\title{NGC 2992}
\author{mvittoria.mavi }
\date{March 2021}
\documentclass[tradiabstract]{aa}  

\usepackage{graphicx}
\usepackage{txfonts}
\usepackage[dvipsnames]{xcolor}
\usepackage{natbib,twoopt}
\usepackage[breaklinks=true]{hyperref} %% to avoid \citeads line fills
\bibpunct{(}{)}{;}{a}{}{,} %% natbib format for A&A and ApJ
\makeatletter
\newcommandtwoopt{\citeads}[3][][]{\href{http://adsabs.harvard.edu/abs/#3}%
{\def\hyper@linkstart##1##2{}%
\let\hyper@linkend\@empty\citealp[#1][#2]{#3}}}
\newcommandtwoopt{\citepads}[3][][]{\href{http://adsabs.harvard.edu/abs/#3}%
{\def\hyper@linkstart##1##2{}%
\let\hyper@linkend\@empty\citep[#1][#2]{#3}}}
\newcommandtwoopt{\citetads}[3][][]{\href{http://adsabs.harvard.edu/abs/#3}%
{\def\hyper@linkstart##1##2{}%
\let\hyper@linkend\@empty\citet[#1][#2]{#3}}}
\newcommandtwoopt{\citeyearads}[3][][]%
{\href{http://adsabs.harvard.edu/abs/#3}
{\def\hyper@linkstart##1##2{}%
\let\hyper@linkend\@empty\citeyear[#1][#2]{#3}}}
\makeatother

\newcommand{\sfr}{$\rm M_{\odot}~yr^{-1}$}
\newcommand{\ergs}{erg s$^{-1}$}
\newcommand{\kms}{km s$^{-1}$}
\newcommand{\ha}{H$\rm \alpha~$}
\newcommand{\hb}{H$\rm \beta~$}
\newcommand{\msun}{$\rm M_{\odot}$}

\begin{document} 

   \title{NGC 2992: The interplay between the multiphase disk, wind and radio bubbles}

   \author{M.V. Zanchettin\inst{\ref{inst1}, \ref{inst2}} 
   \and C. Feruglio\inst{\ref{inst2}, \ref{inst3}} 
   \and M. Massardi\inst{\ref{inst4}}
   \and A. Lapi\inst{\ref{inst1}, \ref{inst3}, \ref{inst4}, \ref{inst5} }
   \and M. Bischetti\inst{\ref{inst6}, \ref{inst2}} 
   \and S. Cantalupo\inst{\ref{inst14}}
   \and F. Fiore\inst{\ref{inst2}, \ref{inst3}} 
    \and A. Bongiorno\inst{\ref{inst7}}
    \and A. Malizia\inst{\ref{inst9}}
   \and A. Marinucci\inst{\ref{inst10}} 
   \and M. Molina\inst{\ref{inst9}}
   \and E. Piconcelli\inst{\ref{inst7}} 
   \and F. Tombesi\inst{\ref{inst11}, \ref{inst7}, \ref{inst12}, \ref{inst13}, \ref{inst15} }
   \and A. Travascio\inst{\ref{inst14}}
   \and G. Tozzi\inst{\ref{inst16}, \ref{inst17}}
   \and R. Tripodi\inst{\ref{inst6}, \ref{inst2}} 
   }

\institute{SISSA, Via Bonomea 265, I-34136 Trieste, Italy \email{mazanch@sissa.it}\label{inst1} \and INAF Osservatorio Astronomico di Trieste, via G.B. Tiepolo 11, 34143 Trieste, Italy\label{inst2} \and IFPU - Institute for fundamental physics of the Universe, Via Beirut 2, 34014 Trieste, Italy \label{inst3} \and INAF, Istituto di Radioastronomia, Italian ARC, Via Piero Gobetti 101, I-40129 Bologna, Italy\label{inst4} \and INFN, Sezione di Trieste, via Valerio 2, Trieste I-34127, Italy\label{inst5} \and Dipartimento di Fisica, Università di Trieste, Sezione di Astronomia, Via G.B. Tiepolo 11, I-34131 Trieste, Italy\label{inst6} \and Department of Physics, University of Milan Bicocca, Piazza della Scienza 3, I-20126 Milano, Italy\label{inst14} \and INAF - Osservatorio Astronomico di Roma, Via di Frascati 33, 00040, Monteporzio Catone, Rome, Italy\label{inst7} \and INAF-OAS Bologna, via Gobetti 101, I-40129 Bologna, Italy \label{inst9} \and ASI - Agenzia Spaziale Italiana, Via del Politecnico snc, 00133 Roma, Italy \label{inst10} \and Dept. of Physics, University of Rome ‘Tor Vergata’, via della Ricerca Scientifica 1, 00133, Rome, Italy\label{inst11} \and Dept. of Astronomy, University of Maryland, College Park, MD, 20742, USA\label{inst12} \and NASA - Goddard Space Flight Center, Code 662, Greenbelt, MD 20771, USA\label{inst13} \and INFN - Roma Tor Vergata, Via Della Ricerca Scientifica 1, 00133, Rome, Italy\label{inst15} \and
Dipartimento di Fisica e Astronomia, Universita` di Firenze, Via G. Sansone 1, I-50019, Sesto Fiorentino (Firenze), Italy\label{inst16} \and 
INAF - Osservatorio Astrofisico di Arcetri, Largo E. Fermi 5, I-50127, Firenze, Italy\label{inst17}
}   

   \date{Received  ; accepted  }

% \abstract{}{}{}{}{} 
% 5 {} token are mandatory
 
    \abstract
   {
   We present an analysis of the gas kinematics in NGC 2992, based on VLT/MUSE, ALMA and VLA data, aimed at characterising the disk, the wind and their interplay in the cold molecular and warm ionised phases. NGC 2992 is a changing look Seyfert known to host both a nuclear Ultra Fast Outflow, and an AGN driven kpc-scale ionised wind. CO(2-1) and \ha arise from a multiphase disk with inclination 80 deg and radii 1.5 and 1.8 kpc, respectively. By modeling the gas kinematics, we find that the velocity dispersion of the cold molecular phase, $\sigma_{\rm gas}$, is consistent with that of star forming (SF) galaxies at the same redshift, except in the inner 600 pc region, and in the region between the cone walls and the disk, where $\sigma_{\rm gas}$ is a factor of 3-4 larger then in SF galaxies for both the cold molecular and the warm ionised phase. This suggests that a disk-wind interaction locally boosts the gas turbulence. We  detect a clumpy ionised wind in both \hb, [O III], \ha, and [N II], distributed in two wide opening angle ionisation cones reaching scales of 7 kpc (40 arcsec). 
  The [O III] wind expands with velocity exceeding $-1000$ \kms~ in the inner 600 pc, a factor of $\sim5$ larger than the previously reported wind velocity. Based on spatially resolved electron density and ionisation parameter maps, we infer an ionised outflow mass of $M_{\rm of,ion} = (3.2 \pm 0.3) \times \, 10^7 \, M_{\odot}$, and a total ionised outflow rate of $\dot M_{\rm of,ion}=13.5\pm1$ \sfr.
  We detected ten clumps of cold molecular gas located above and below the disk in the ionisation cones reaching maximum projected distances of 1.7 kpc, and projected bulk velocities up to 200 \kms. 
  On these scales, the wind is multiphase, with a fast ionised component and a slower molecular one, and a total mass of $M_{\rm of, ion+mol}= 5.8 \times 10^7 \, M_{\odot}$, of which the molecular component carries the bulk of the mass, $M_{\rm of,mol} = 4.3 \times 10^7 \, M_{\odot}$. The dusty molecular outflowing clumps and the turbulent ionised gas are located at the edges of the radio bubbles, suggesting that the bubbles interact with the surrounding medium through shocks, as also supported by the  [O I]/\ha ratio. Conversely, both the large opening angle and the dynamical timescale of the ionised wind detected in the ionisation cones on 7 kpc scales, indicate that this is not related to the radio bubbles but instead likely associated with a previous AGN episode. Finally, we detect a dust reservoir co-spatial with the molecular disk, with a cold dust mass $M_{\rm dust} = (4.04 \pm 0.03) \times \, 10^{6} \, M_{\odot}$, which is likely responsible for the extended Fe K$\alpha$ emission seen on 200 pc scales in hard X-rays and interpreted as reflection by cold dust.
   }


   \maketitle
%
%-------------------------------------------------------------------

\section{Introduction}

%Shimizu+2019 and Garcia-Bernete+2021
Active Galactic Nuclei (AGN) can power massive outflows, potentially impacting on the galaxy interstellar medium (ISM) and altering both star formation and nuclear gas accretion. 
The growth of the super massive black holes (SMBHs) in the galactic center is then stopped, as well as the nuclear activity and winds, until new cold gas accretes toward the nucleus, so starting a new AGN episode \citep{fabian2012,kingpounds}. This is the so called feeding and feedback cycle of active galaxies \citep{tumlinson2017, gaspari2020}. 
Understanding the relation between winds and outflows emerging from accreting SMBHs and their host galaxy ISM is necessary to understand the SMBH and galaxy co-evolution.

Host bulge properties such as velocity dispersion, luminosity, and mass, are tightly correlated with the mass of the SMBH in the galaxy centre \citep{gebhardt,ferrareseford,kormendyho,shankar2016,shankar2017}. These SMBH-host bulge properties relations seem to arise when the black hole reaches a critical mass; at this point the AGN driven winds, the nuclear activity, and the SMBH growth itself are stopped \citep{silkrees,fabian1999, king2003}.
AGN occupy the very centres of galaxies but they can drive gas outflows that reach several kpc away.
These AGN driven winds appear to be ubiquitous, and detected through all gas and dust phases, suggesting that they consist of a mixture of cold molecular gas, warm ionised gas and hot X-ray emitting and absorbing plasma, probably mixed with dust \citep[e.g][and references therein]{tombesi2015, fiore2017, honig2017, smith2019, asmus2019, lutz2020}. 

One of the most outstanding questions regarding AGN driven outflows pertains to the relationships between the different gas phases involved in the winds, their relative weight and impact on the galaxy ISM \citep{bischetti2019, fluetsch2019}. In some cases, molecular and ionised winds have similar velocities and are nearly co-spatial \citep{feruglio2018, alonso-herrero, zanchettin2021}, suggesting a cooling sequence scenario where molecular gas forms from the cooling of the gas in the ionised wind \citep{richings2017, menci2019}. Other AGNs show ionised winds that are faster than the molecular winds \citep{ramosalmeida2022}, suggesting a different origin of the two phases \citep[and references therein]{veilleux2020}. 
Another open question regards the effect of winds on the surrounding ISM where they expand, and whether the properties of disks, their dynamical state and turbulence, are modified by winds.
The interaction between a radio jet, a radiative wind and the ISM has been probed in few nearby Seyfert galaxies, and suggests that the ISM is modified by outflows expanding across the disk \citep[e.g.][]{alonso-herrero2018,feruglio2020,fabbiano2022,garcia-burillo2014,cresci2015, venturi2018, garcia-burillo2019,rosario2019,shimizu2019}.
Therefore observations that cover a large range of wavelengths and spatial scales are needed to quantify the overall mass, momentum, and energy budget of multiphase winds, and their interplay with the ISM, and to reveal the dominant processes that rule the AGN-host galaxy relation. 
To date, detailed studies of the the multiphase wind have been achieved in only a handful of sources \citep[e.g][]{finlez2018, husemann2019,slater2019,herreracamus2019,shimizu2019, garcia-bernete2021}.

In this paper we present an analysis of the gas kinematics in the disk, wind and radio bubble of NGC 2992 based on 
%ALMA, VLT/MUSE and VLA observations.
ALMA, VLT/MUSE and Karl G. Jansky Very Large Array (VLA) observations.
NGC 2992 is a nearby Seyfert galaxy located at $\sim$ 32.5 Mpc \citep[$z = 0.00771$,][projected angular scale of 0.158 kpc/arcsec]{keel1996}, whose disk is seen almost edge-on \citep[i $\sim$ 70 deg,][]{marquez}. 
It hosts an AGN known for its extreme variability, from the near-IR \citep{glass1997} to the X-rays (up to a factor 20) on time-scales of weeks/months \citep{gilli2000,marinucci2018,middei2022,luminari2023}. Its optical classification changes between Seyfert 1.5 and 2 \citep{trippe2008}. It has been suggested that the optical emission line profiles could be affected by the prominent thick dust lane extending nearly along the galaxy major axis, crossing the nucleus \citep{ward1980,colina1987}. 
In the high state, the AGN has a bolometric luminosity of $L_{\rm bol, AGN}=1.2-2.4\times 10^{44}$ \ergs, and shows an Ultra Fast Outflow (UFO) with velocity about 0.21c and total kinetic energy rate of 5\% $L_{\rm bol}$, sufficient to switch on feedback mechanisms on the galaxy host \citep{marinucci2018}, but detected only when the source accretion rate exceeds 2 $\%$ of the Eddington luminosity. UFO components with velocity  up to $\sim$0.4c have been also recently reported by \citet{luminari2023}.
The galaxy is part of the interacting system Arp 245, together with NGC 2993 and the tidal dwarf A245 North \citep{brinks2000}. Tidal features, which connect the three galaxies, have been detected both by infrared \citep{garcia-bernete2015} and optical/near-infrared imaging suggesting that the system is at early stage of interaction \citep{duc2000}.
%Da citare su radio bubbles: \citet{ulvestadwilson,fernandez2022}.
Radio 6 cm observations detected a hourglass-shaped emission extending from north-west to south-east for about 1 kpc, nearly perpendicular to the galaxy disk \citep{ulvestadwilson}. 
%An additional kpc-sized double-lobed radio source has been detected in the linearly polarized radio emission and interpreted as a relic of an earlier episode of AGN activity \citep{irwin2017}.
A simultaneous X-ray and VLBI radio monitoring campaign showed an anticorrelation between the luminosity of the radio core at 6 GHz and X-ray 2-10 keV emission. The X-ray and radio behaviour can be due to flares produced by magnetic re-connection in the accretion disk \citep{fernandez2022}.
The south-eastern radio bubble overlaps with the soft X-ray emission where no bright optical line or continuum emission are detected \citep{colbert2005, xu2022}. \citet{xu2022} proposed that this extended soft X-ray emission near the nucleus could be dominated by hot gas heated by shocks from outflows associated with the radio bubble.
\citet{chapman2000} suggested that the radio morphology is the result of expanding plasma bubbles carried by AGN driven outflows.
SINFONI/VLT and Spitzer data confirm that the most likely driver of the radio 8-shaped structure and the large scale outflow is the AGN \citep{friedrich2010}.
\citet{veilleux2001} found that \ha emission indicates the presence of a ionised wind whose energy source should be a hot, bipolar, thermal wind powered on sub-kpc scale by the AGN. 
GEMINI integral field unit (IFU) data from the inner 1.1 kpc indicates the presence of a blueshifted outflowing gas component and of an arc-shaped [O III] emission, both kinematically decoupled from stellar kinematics, and spatially correlated with the 8-shaped radio structure \citep{guolopereira2021}. This suggests that a relativistic plasma bubble is expanding, compressing the gas along its path and driving an outflow at its boundaries.

In this paper we investigate further the complex gas kinematics in NGC 2992, joining ALMA with VLT/MUSE and VLA radio observations, to constrain the relation between the cold and the warm ionised gas phases, through their different kinematic components, characterise the multiphase winds and assess their interaction with the ISM. 
The paper is organised as follows. Section \ref{sec:obs} presents the observational set-up and data reduction. Section \ref{sec:results} presents the observational results, in particular the radio continuum emission, the gas properties, and kinematics. In Section \ref{sec:discussion} we discuss our results, and in Section \ref{sec:concl} we present our conclusions.

\begin{table*}
     \caption[]{Main properties of the ALMA and VLA data sets used in this work.}
         \label{tab:cube-properties}
\centering                          
\begin{tabular}{c c c c c c c }        
\hline\hline                 
project ID & Obs & Freq.  & beam & continuum rms & line rms & channel width\\
& & [GHz] & [${\rm arcsec}^2$] & [mJy/beam] & [mJy/beam] & [\kms]\\ 
\hline      
2017.1.00236.S & ALMA & 225.4-244.1  & 0.23 $\times$ 0.18 at 87 deg & 0.02 & 0.3 & 10 \\ 
2017.1.01439.S & ALMA & 226.1-244.1  & 0.62 $\times$ 0.54 at 78 deg & 0.05 & 0.4 & 10.4\\
%2017.1.01439.S & ALMA & 226.1-244.1  &$>$ 100 & 0.52 $\times$ 0.45 at 82 deg & 0.5 & 10.4\\
2017.1.01439.S$^{*}$ & ALMA & 226.1-244.1  & 2.18 $\times$ 2.03 at -82 deg & 0.05 &  0.9 & 10.4\\
17B-074 &  VLA & 4.5-6.5  & 0.98 $\times$ 0.4 at -77 deg & 0.14 & - & -\\
\hline
\end{tabular}\\
  \flushleft 
 \footnotesize{{\bf Notes.} *: \texttt{uvrange} < 100 m. The \texttt{uvrange} parameter of the \texttt{tclean} task of CASA software is used to select the baselines in the desired region of the uv-plane. This allows us to separate the CO(2-1) emission on compact and extended physical scales.}
\end{table*}

\section{Observations and data reduction}
\label{sec:obs}

\subsection{VLA Observations}

We analysed archival VLA observations of NGC 2992 in Band C (project code 17B-074, PI Preeti Kharb). The observations were carried out in January 2018 for a total integration time of 3585 seconds in the frequency range 4.5-6.5 GHz.
We used the standard calibrated visibilities provided by the VLA data archive (\href{https://science.nrao.edu/facilities/vla/archive}{https://science.nrao.edu/facilities/vla/archive}) and we used CASA 5.4.1 software \citep{mcmullin} to generate the map of the continuum emission at 6 cm rest-frame. We averaged the visibilities in the 16 spectral windows (spws) and we produced a clean map using task \texttt{tclean} and a Briggs weighting scheme with robust parameter equal to -1.0. We used the \texttt{hogbom} cleaning algorithm with a detection threshold of 3 times the rms sensitivity. The final continuum map has an rms of 0.1 mJy/beam and a beam of 0.98 $\times$ 0.40 arcsec$^2$ with position angle equal to 77 deg (PA, measured from north direction anticlockwise).

\subsection{ALMA observations}\label{sec:alma}

We used two ALMA Band 6 data-set, both covering the CO(2-1) line (observed frequency of 228.776 GHz) and the underlying 1.3 mm continuum emission.
The first observation (program ID 2017.1.00236.S, PI Matthew Malkan) was obtained in December 2017 and spans the frequency range 225.4-244.1 GHz in the configuration C43-6 with 44 antennas, minimum baseline of 15 m, and maximum baseline of 2500 m. This configuration provided an angular resolution of about 0.2 arcsec and a largest angular scale (LAS) of $\sim$ 2.5 arcsec.
The second observation (program ID 2017.1.01439.S, PI Chiara Feruglio) was carried out in March 2018 in the frequency range 226.1-244.1 GHz in the configuration C43-4 with 44 antennas, minimum baseline of 15 m, and maximum baseline of 784 m, reaching an angular resolution of about 0.6 arcsec and LAS$\sim$ 5.7 arcsec.
\newline \noindent We calibrated the visibilities in the CASA 5.4.1 pipeline mode using the default quasar calibrators provided by the observatory: J1037-2934 as bandpass and flux calibrator, J1037-2934 and J0957-1350 as phase-amplitude calibrators for the first data-set, J1037-2934 as bandpass and flux calibrator and J0942-0759 as phase-amplitude calibrator for the second data-set. 
The absolute flux accuracy is better than 10$\%$. 

To estimate the continuum emission, we averaged the visibilities in the four spws, excluding the spectral range covered by the CO(2-1) emission line. In addition, to analyse the continuum emission next to CO(2-1), we modelled the visibilities in the spw covering the line with a first-order polynomial using the \texttt{uvcontsub} task. Using a zero-order polynomial gave consistent results. We subtracted this fit to produce continuum-subtracted CO(2-1) visibilities. 
We imaged the data by adopting a natural weighting scheme and the \texttt{hogbom} cleaning algorithm with a detection threshold of three times the r.m.s. noise, and a spectral resolution of 10.4 and 10 \kms for 2017.1.01439.S and 2017.1.00236.S data set respectively. 
For the 2017.1.01439.S data set we produced also clean data cubes with different resolution by selecting different baselines (uv ranges). We produced a map of the compact (diffuse) emission using the \texttt{uvrange} parameter of task \texttt{tclean} to select visibilities corresponding to distance in the uv plane < 100 m (>100 m), that is corresponding to a $\sim$ 2.5 kpc physical scale (Table \ref{tab:cube-properties}). 
The final map of the diffuse component has been obtained by subtracting the image with \texttt{uvrange} $>100$ m from that with \texttt{uvrange} $<100$ m with task \texttt{immath}. This ensures that power from compact sources does not affect the diffuse component. 
The properties of the data cubes obtained are summarised in Table \ref{tab:cube-properties}.

\subsection{VLT/MUSE observations}

NGC 2992 was observed with MUSE for 5 nights between January 22 and February 13 2015 (program 094.B-0321, PI Alessandro Marconi). The observations were acquired in seeing-limited, wide field mode (WFM) covering 1 $\times$ 1 $\rm arcmin^2$, spanning the spectral range 4750–9350 $\AA $.  This allows us to cover the main optical emission lines such as $\rm H\beta \lambda$4861$\AA $ , [O III]$\rm \lambda$5007$\AA $ doublet, $\rm H\alpha \lambda$6563$\AA $, and the [NII]$\lambda 6548-6583\AA$ doublet.
MUSE spectral binning is 1.25 $\AA $/channel ($\approx$ 70 \kms) and its resolving power in WFM is 1770 at 4800 $\AA $, and to 3590 at 9300 $\AA $.
The field of view covers the central part of NGC 2992 corresponding to a square region of about 9 kpc on the side.
The data includes five Observing Blocks (OBs), each one divided in four 500s exposures, together with an equal number of 100s sky exposures. Each sky exposure was employed in the data reduction to create a model of the sky lines and sky continuum to be subtracted from the closest science exposure in time. The average seeing during the observations was $\sim$ 0.9 arcsec.
The data reduction was performed using the MUSE Instrument Pipeline v2.8.5 and the ESO Recipe Execution Tool, version 3.13.3 \citep{weilbacher2014}. 
We used the MUSE pipeline standard recipes (\textit{scibasic} and \textit{scipost}) to remove instrumental signatures from the data applying a bias and flat-field correction, the wavelength and flux calibrations and the pipeline sky subtraction following the standard procedure.
In order to remove the stellar continuum, we first applied a Voronoi tessellation \citep{cappellaricopin2003} to achieve an average signal-to-noise ratio (S/N) equal to 120 per wavelength channel on the continuum under 5530 $\AA$.  
We then performed the continuum fit by using the Penalized Pixel-Fitting \citep[pPXF, ][]{cappellari2004} code on the binned spaxels. The stellar continuum was modeled using a linear combination of \citet{vazdeki2010} synthetic spectral energy distributions for single stellar population models in the wavelength range 4700-7200 $\AA$. 
We fitted the continuum together with the main emission lines in the selected wavelength range ($\rm H\beta \lambda$4861$\AA $, [O III]$\rm \lambda$4959,5007$\AA $, $\rm H\alpha \lambda$6563$\AA $, [OI]$\rm \lambda$6300,6364$\AA$, [NII]$\rm \lambda$6548,6584$\AA$ and [SII]$\rm \lambda$6716,67313$\AA$) to better constrain the underlying stellar continuum. 
We then subtracted the fitted stellar continuum derived in each Voronoi cell spaxel by spaxel, after rescaling the modeled continuum to the median continuum in each spaxel.

\begin{figure*}[ht]
  \resizebox{\hsize}{!}{ \includegraphics{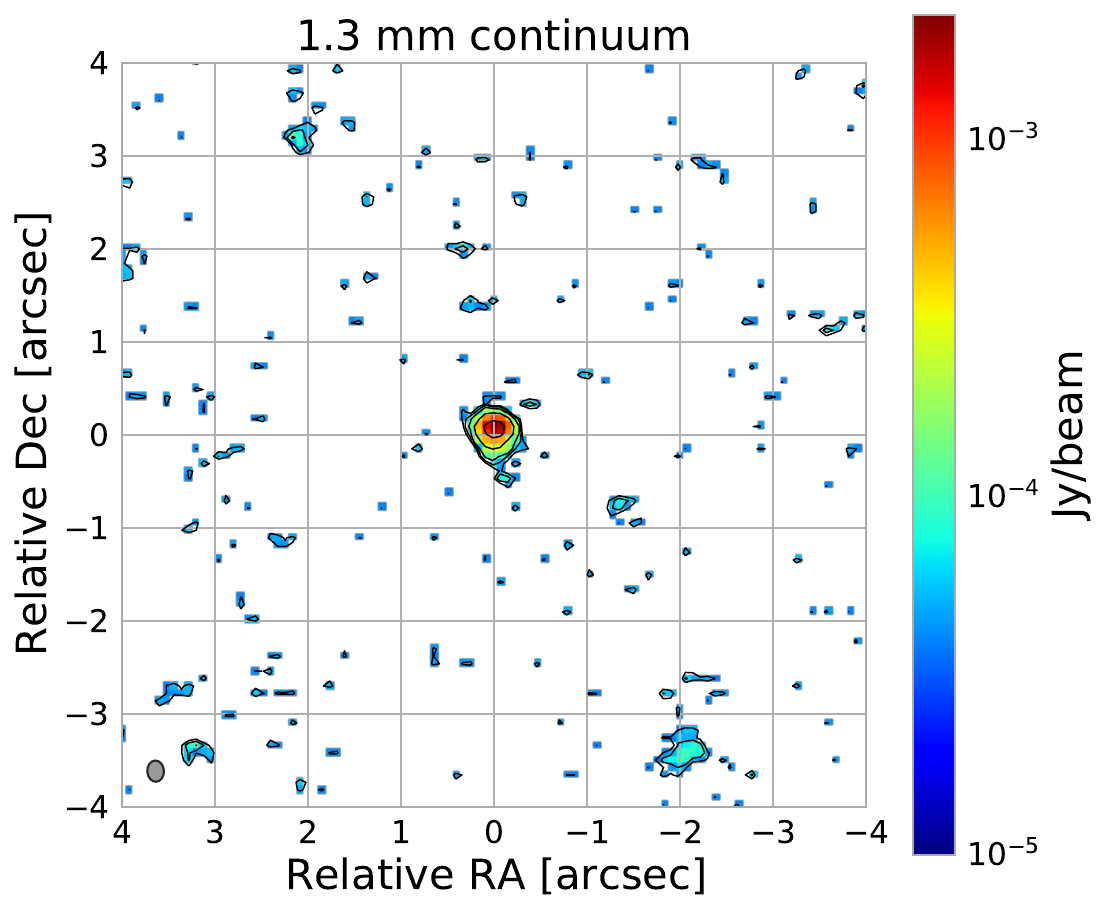} \includegraphics{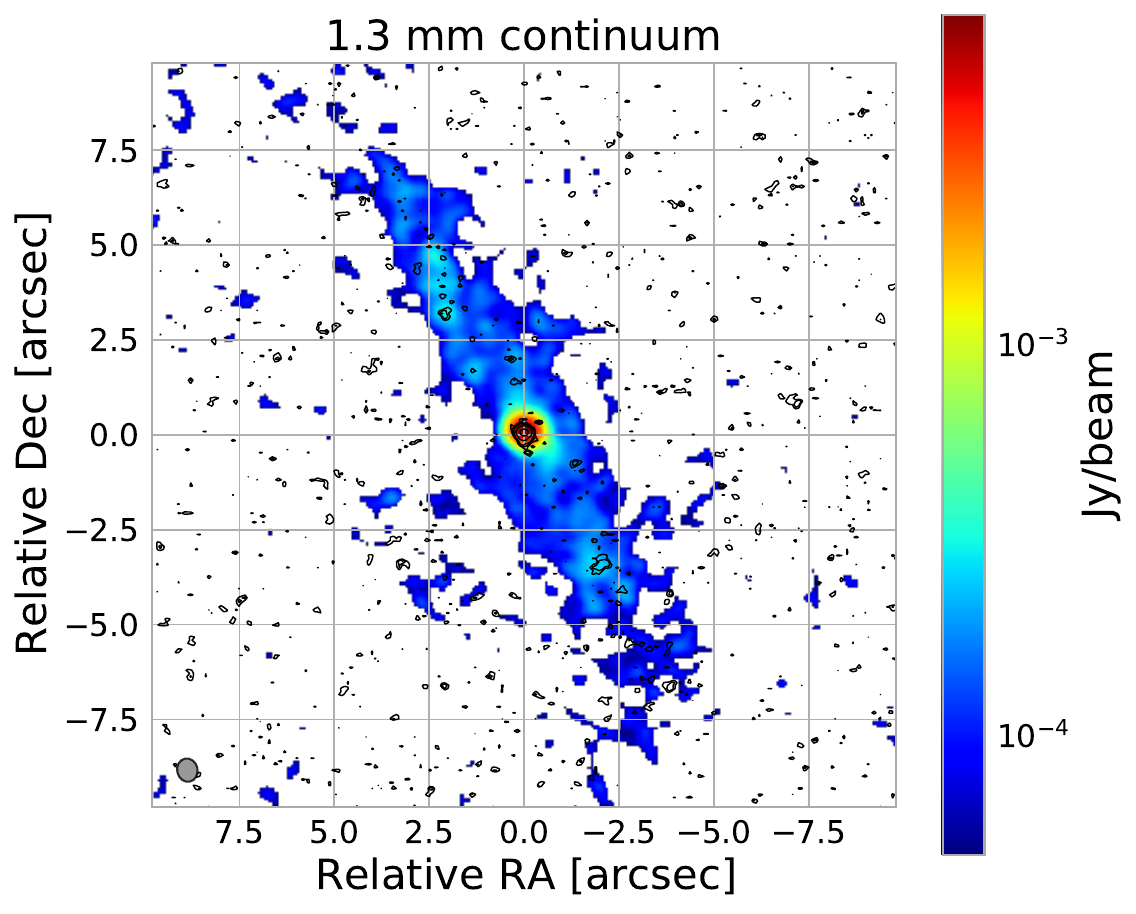} \includegraphics{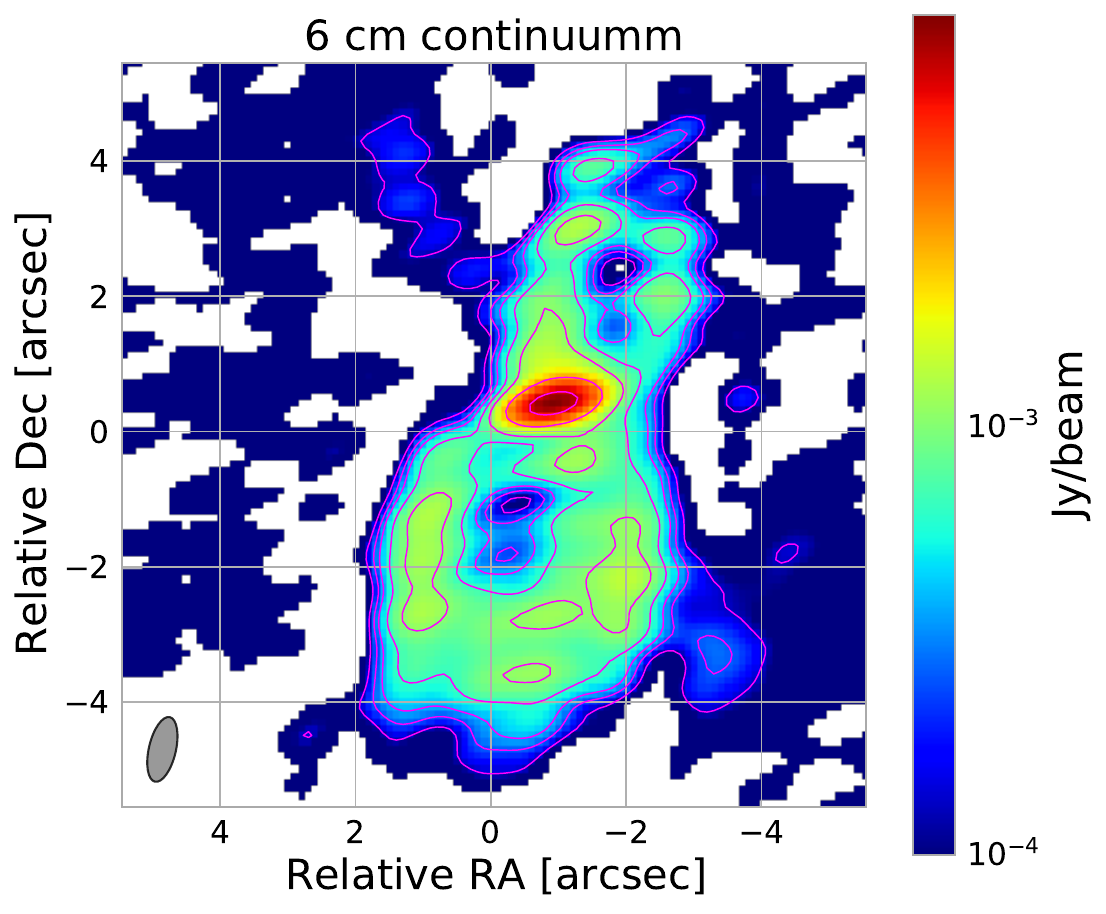}}
 \caption{Left panel: map of the 1.3 mm continuum emission at $\sim$0.2 arcsec resolution with a cut-out at 2$\sigma$. Contours are drawn at (2, 3, 6, 15, 60)$\sigma$, $\sigma = 0.02$ mJy/beam. Central panel: ALMA map of the 1.3 mm continuum seen at $\sim0.6$ arcsec resolution with a cut-out at 1$\sigma$, $\sigma = 0.05$ mJy/beam. Compared to the left panel, in this map both the nuclear emission and the dusty disk are detected. Contours as in left panel. Right panel: VLA 6 cm map of radio continuum. Contours are drawn at (1, 2, 3, 5, 8, 20, 50, 100)$\sigma$, $\sigma = 0.14$ mJy/beam. The grey ellipses show the ALMA beam in the left-middle panels and the VLA beam in the right panel.}
  \label{continuo}
\end{figure*} 

\section{Results}
\label{sec:results}

\subsection {Radio continuum emission}

\begin{table}
     \caption[]{Main properties of the VLA 6 cm and ALMA 1.3 mm extended continua.}
         \label{tab:continuo}
\centering                          
\begin{tabular}{c c c c c}        
\hline\hline                 
Obs & Peak Flux & Total Flux density & size \\
 & [mJy/beam] & [mJy] & [$\rm arcsec^2]$ \\
 (a) & (b) & (c) & (d) \\
\hline      
VLA & 8.5 $\pm$ 0.2 & 55.8 $\pm$ 0.1 & 44 \\
%2017.1.00236.S & 2.26 $\pm$ 0.05 & 2.71 $\pm$ 0.09 & 6 \\
ALMA & 6.9 $\pm$ 0.1 & 20.2 $\pm$ 0.5 & 58 \\
\hline  
\end{tabular}\\
  \flushleft 
 \footnotesize{ {\bf Notes.} The table reports the main properties of the continuum emission at 6 cm and at 1.3 mm shown in Figure \ref{continuo}. (a) The observatory (top row: VLA project ID 17B-074, bottom row: ALMA project ID 2017.1.01439.S), (b) the flux at the peak position (RA, DEC = 09:45:41.94, -14:19:34.6), (c) the total flux density in a region of size reported in (d).}
\end{table}
  
Figure \ref{continuo} (left panel) shows the 1.3 mm continuum map with 0.2 arcsec resolution. Performing a 2D gaussian fitting, we measure a continuum flux density $S_{\rm 1.3 mm} = 2.71 \pm 0.09$ mJy and a peak flux $S_{\rm peak} = 2.26 \pm 0.05$ mJy/beam at RA, DEC = 09:45:41.9448, -14:19:34.6072. According to our 2D fit the continuum emission is consistent with a point source.
The 1.3 mm continuum emission at lower resolution (0.6 arcsec beam, project ID 2017.1.01439.S) is shown in Figure \ref{continuo}, middle panel. We detect peak flux $S_{\rm peak} = 6.9 \pm 0.1$ mJy/beam at the position of the AGN, and a clumpy extended component with an approximate size of 15 arcsec extending from north-east to south-west along PA $\sim$ 30 deg. 
The latter component aligns well with the inclined gaseous disk traced by both CO(2-1) and optical emission lines (see Sect. \ref{sect:moldisk} and \ref{ioniseddisk}), suggesting that the extended continuum emission may be due to cold dust in the galactic disk (see Discussion).
We measure a total flux density at 1.3 mm of $S_{\rm 1.3 mm} = 20.2 \pm 0.5$ mJy above 3$\sigma$ threshold extended on a 58 $\rm arcsec^2$ area.
The extended 1.3 mm continuum is detected only in the 0.6 arcsec resolution data, whereas it may be resolved-out in the high resolution data.
Figure \ref{continuo} (right panel) shows the 6 cm radio continuum map. The peak position of the continuum, obtained through 2D fitting in the image plane, is consistent with that of the 1.3 mm continuum and has a flux at the peak $S_{\rm peak, 6 cm} = 8.5 \pm 0.2$ mJy/beam. Both the 6 cm and the 1.3 mm continua peaks are consistent with the AGN optical position from NED \citep[RA, DEC  = 09:45:42.05, -14:19:34.98,][]{argyle1990}.
The total radio emission above a 3$\sigma$ threshold covers a region of about 44 $\rm arcsec^2$ with an integrated flux density of $55.8 \pm 0.1$ mJy.
The radio emission shows a double lobe 8-shaped structure of size about 9.4 arcsec ($\sim$1.4 kpc) extended from north-west to south-east ($\rm PA=158$ deg), tracing the radio bubbles first detected by \citet{ulvestadwilson}. 
We detect an additional fainter and elongated radio emission extended $\sim$ 10 arcsec (i.e. 1.5 kpc) from north-east to south-west along $\rm PA = 213$ deg, with a flux of $1.2\pm0.1$ mJy, which partly overlaps with the 1.3 mm extended continuum.
We measure a total flux density above a 3$\sigma$ threshold for the two lobes and nucleus of $54.6\pm 0.1$ mJy, corresponding to a radio power $P_{\rm radio,5 GHz} = 3.89 \times 10^{38} \mathrm{erg \, s^{-1}}$.
The main properties of the radio continuum emission from VLA and ALMA data are summarized in Table \ref{tab:continuo}.

\begin{figure*}[ht]
 \resizebox{\hsize}{!}{\includegraphics{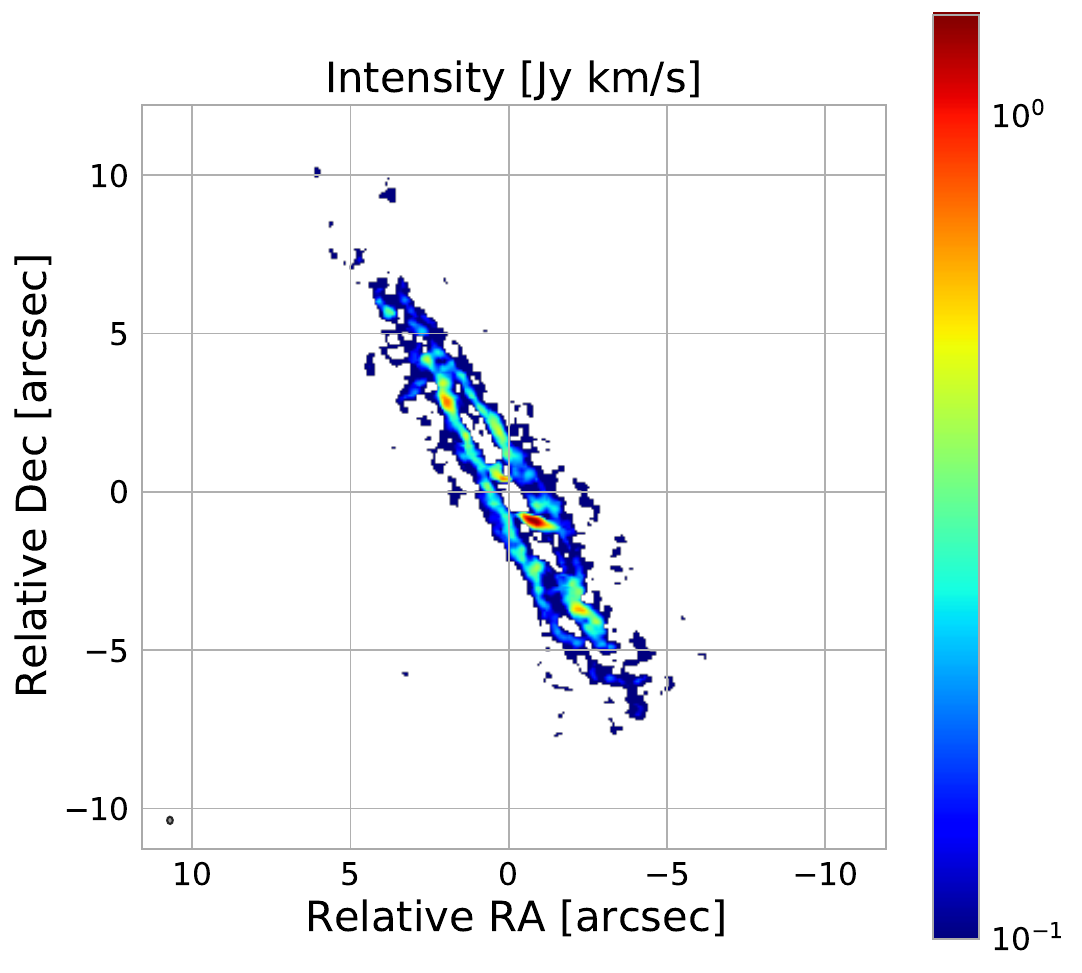} \includegraphics{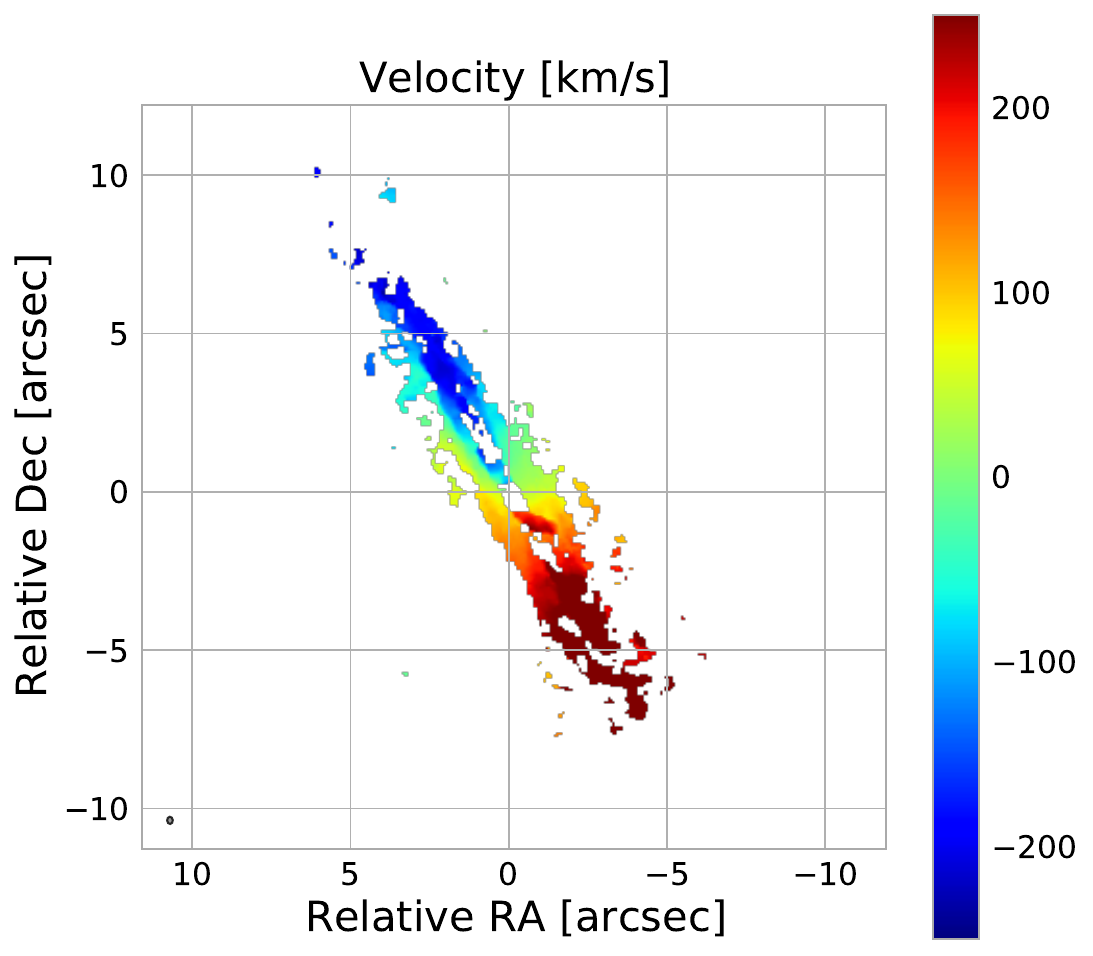} \includegraphics{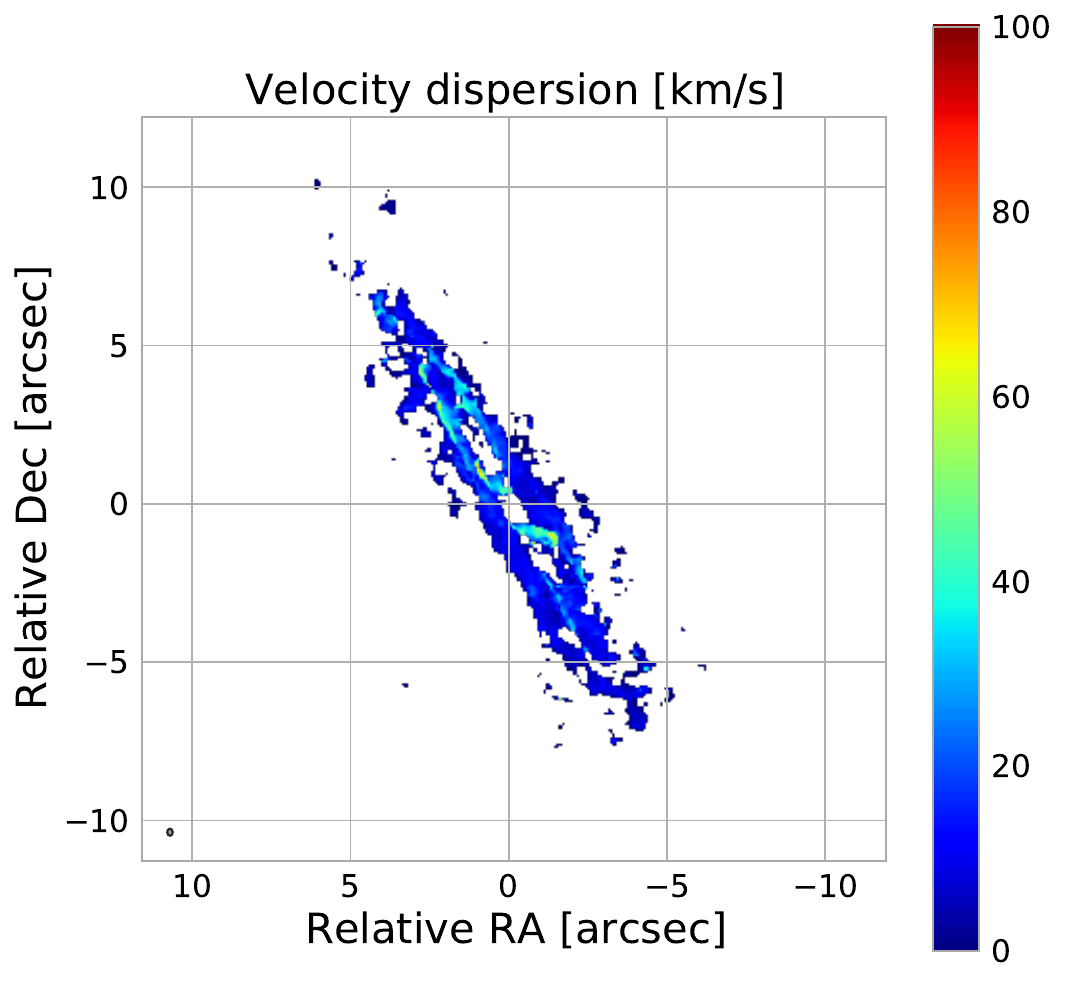}} 
    \resizebox{\hsize}{!}{ \includegraphics{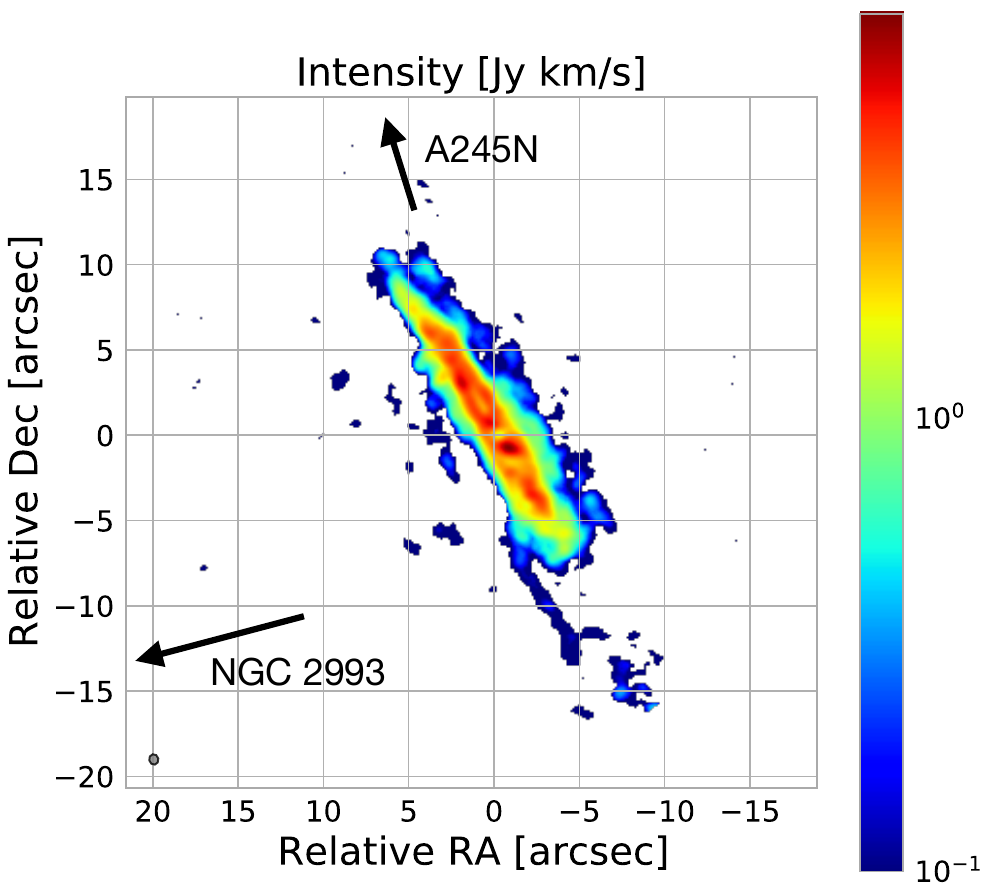} \includegraphics{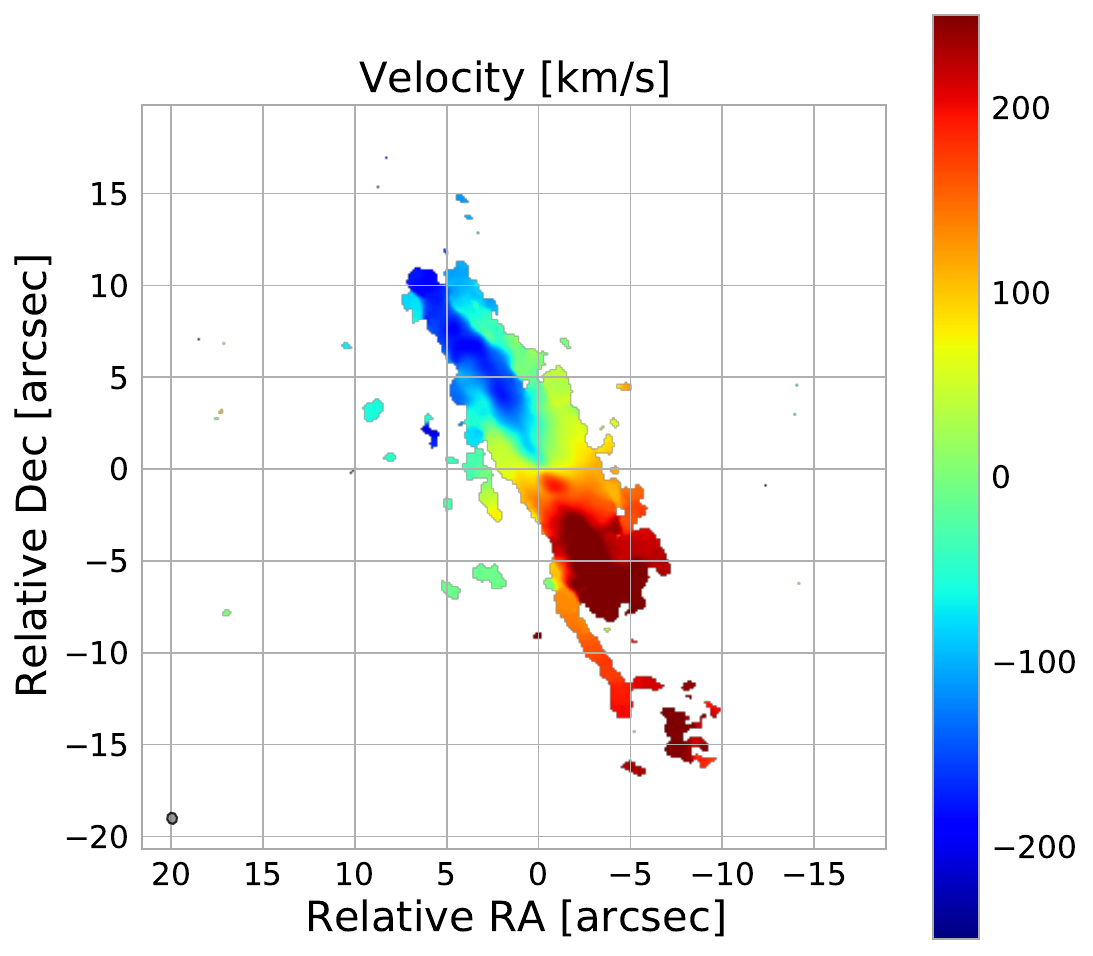} \includegraphics{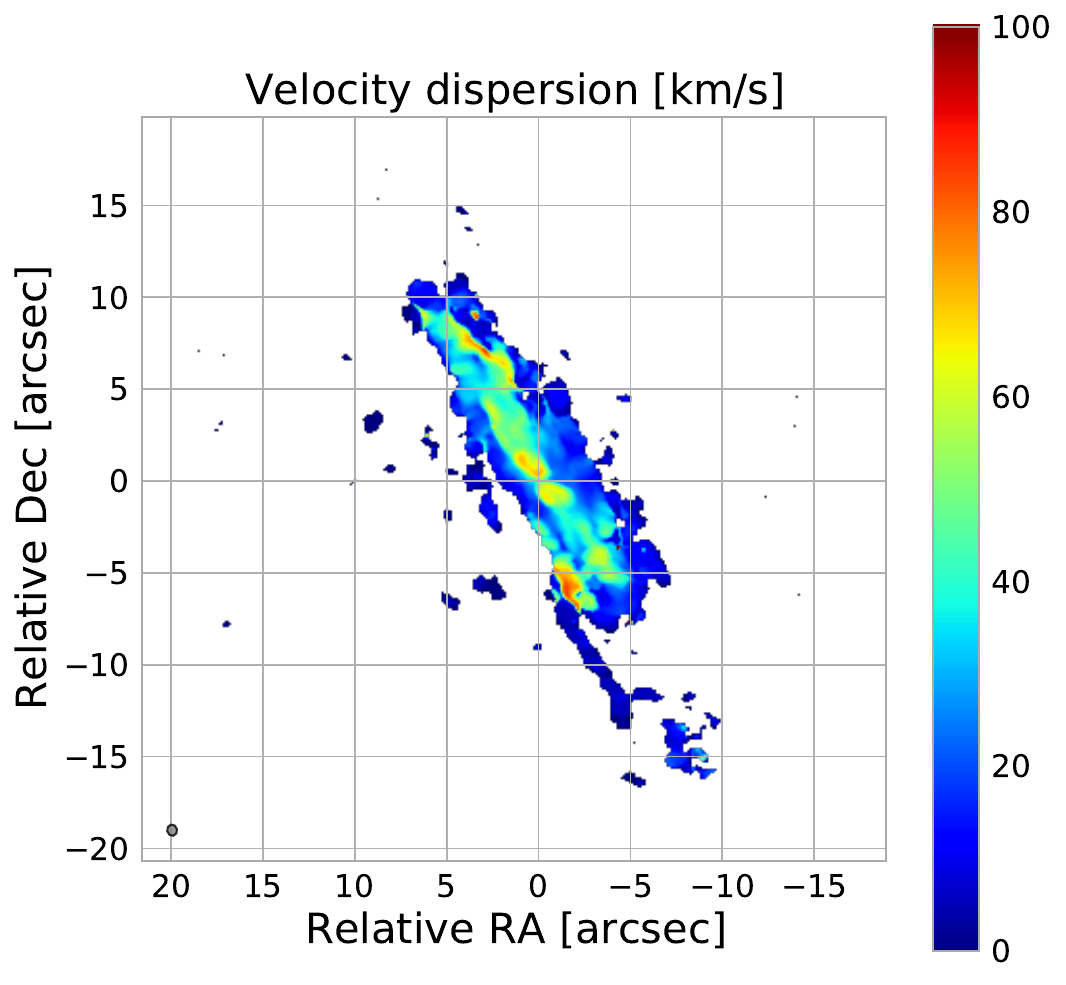}} 
 %\resizebox{\hsize}{!}{\includegraphics{Figure2.pdf}}
 \caption{From left to right: CO(2-1) integrated flux (moment-0), mean velocity (moment-1) and velocity dispersion maps (moment-2). Top panels show data with 0.2 arcsec angular resolution, lower panels data with 0.6 arcsec resolution. Regions with emission below 3$\sigma$ have been blanked. The synthesised beam of each data-set is shown by the grey filled ellipses in the lower-left corner of the map. In the bottom-left panel arrows indicate the direction towards NGC 2993 and Arp 245 North.}
  \label{momentmaps}
\end{figure*}

\begin{table*}
     \caption[]{Parameters of the dynamical models of the multiphase disk.}
         \label{tab:barolomodel}
\centering                          
\begin{tabular}{c c c c c c c c}        
\hline\hline                 
Ang. Res.  & $v_{\rm rot,CNR}$ & $v_{\rm rot, disk}$  & $\sigma_{\rm gas, CNR}$ & $\sigma_{\rm gas, disk}$ & $\sigma_{\rm gas, CNR}/v_{\rm rot, CNR}$ & $\sigma_{\rm gas, disk}/v_{\rm rot, disk}$ & $M_{\rm dyn}(r<1\ {\rm kpc})$ \\ 
$\rm [arcsec]$ & [\kms] & [\kms] & [\kms] & [\kms] &  &  & [$10^{10} M_{\odot}$] \\  
(a) & (b) & (c) & (d) & (e) & (f) & (g) & (i)\\
\hline
\multicolumn{8}{c}{\textit{Cold Molecular gas}} \\
0.2 & 164$\pm$14 & 236$\pm$11 & 38$\pm$12 & 16$\pm$10 & 0.2 & 0.1 & 0.7 \\
0.6 & 139$\pm$13 & 260$\pm$10 & 60$\pm$11 & 25$\pm$10 & 0.4 & 0.1 & 0.9\\
\hline \hline
seeing & $v_{\rm rot,nucl}$ & $v_{\rm rot, disk}$  & $\sigma_{\rm gas, nucl}$ & $\sigma_{\rm gas, disk}$ & $\sigma_{\rm gas, nucl}/v_{\rm rot, nucl}$ & $\sigma_{\rm gas, disk}/v_{\rm rot, disk}$ & $M_{\rm dyn}(r<1\ {\rm kpc})$\\ 
$\rm [arcsec]$ & [\kms] & [\kms] & [\kms] & [\kms] &  &  & [$10^{10} M_{\odot}$] \\  
 (a) & (l) & (c) & (m) & (e) & (n) & (g) & (i) \\
\hline 
\multicolumn{8}{c}{\textit{Ionised gas}} \\
0.9 & 91$\pm$36 & 216$\pm$35 & 117$\pm$36 & 55$\pm$35 & 1.3 & 0.3 & 0.6 \\
\hline
\end{tabular}\\
  \flushleft  \footnotesize{ {\bf Notes.} (a) angular resolution or seeing;  (b) rotation velocity at the CNR, and (c) in the disk; (d) gas velocity dispersion at CNR, and (e) in the disk; (f) ratio between the velocity dispersion and the rotational velocity at CNR, and (g) in the disk; (i) dynamical mass (derived from the relation $M_{\rm dyn} = rv^2_{rot}/2$G) enclosed within a 1 kpc radius; (l) the rotational velocity, and (m) velocity dispersion at nucleus (r$<$350 pc), and (n) their ratio.}
\end{table*}

\begin{figure*}[htb]
 \resizebox{\hsize}{!}{
 \includegraphics{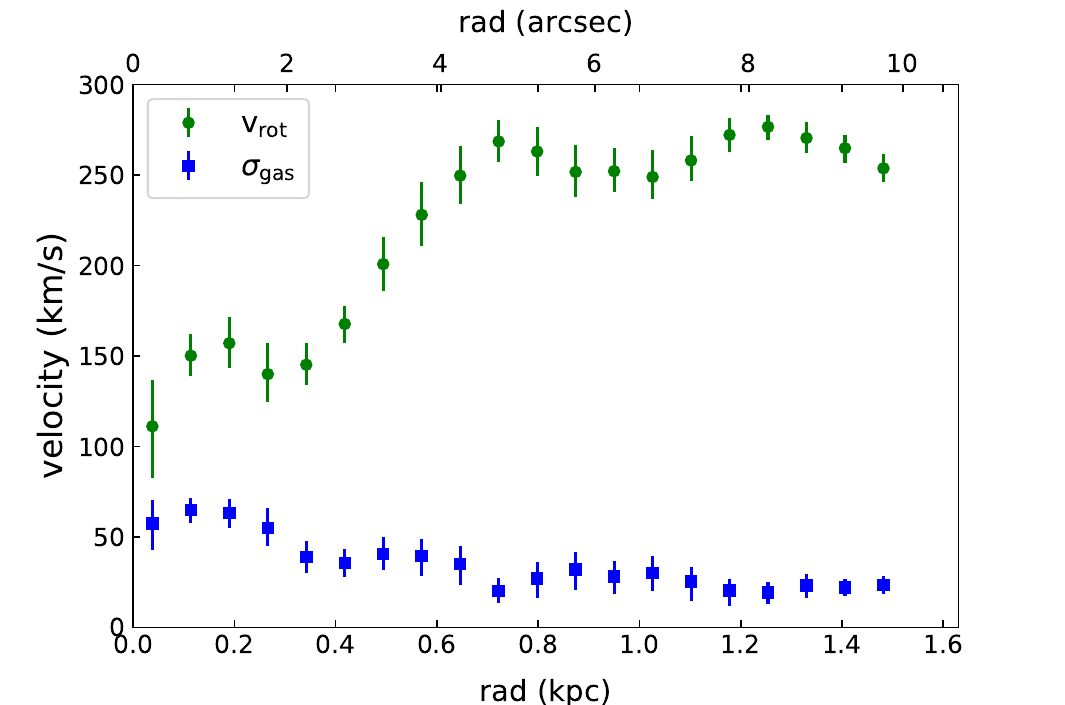}
 \includegraphics{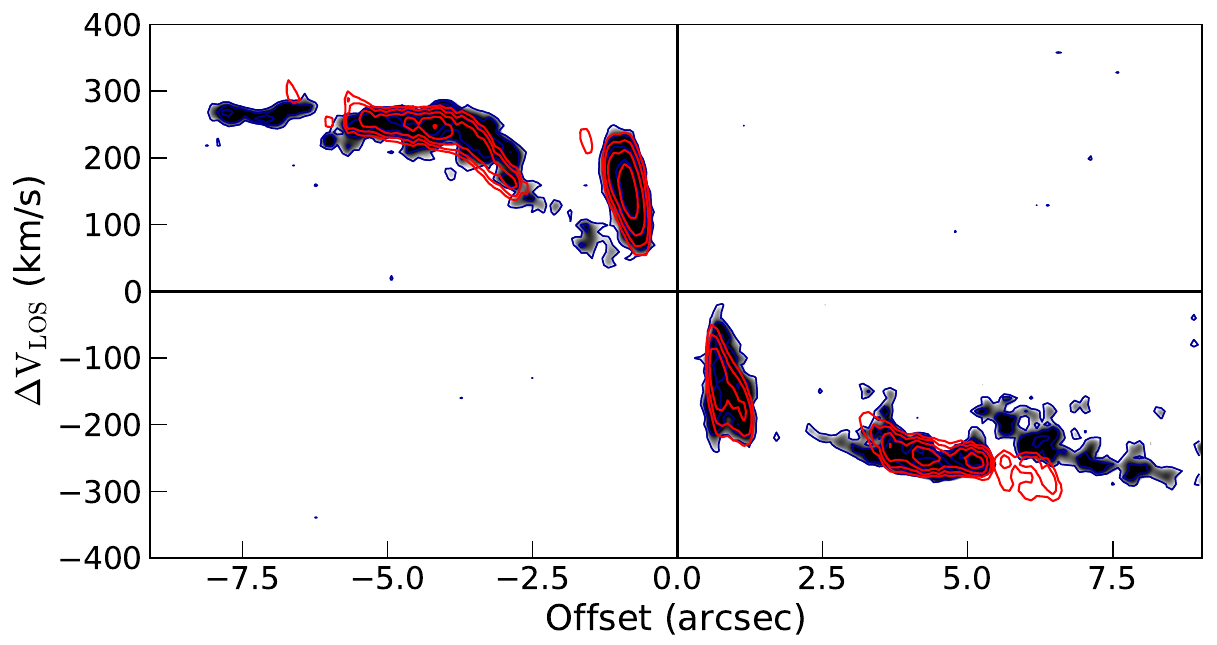} \includegraphics{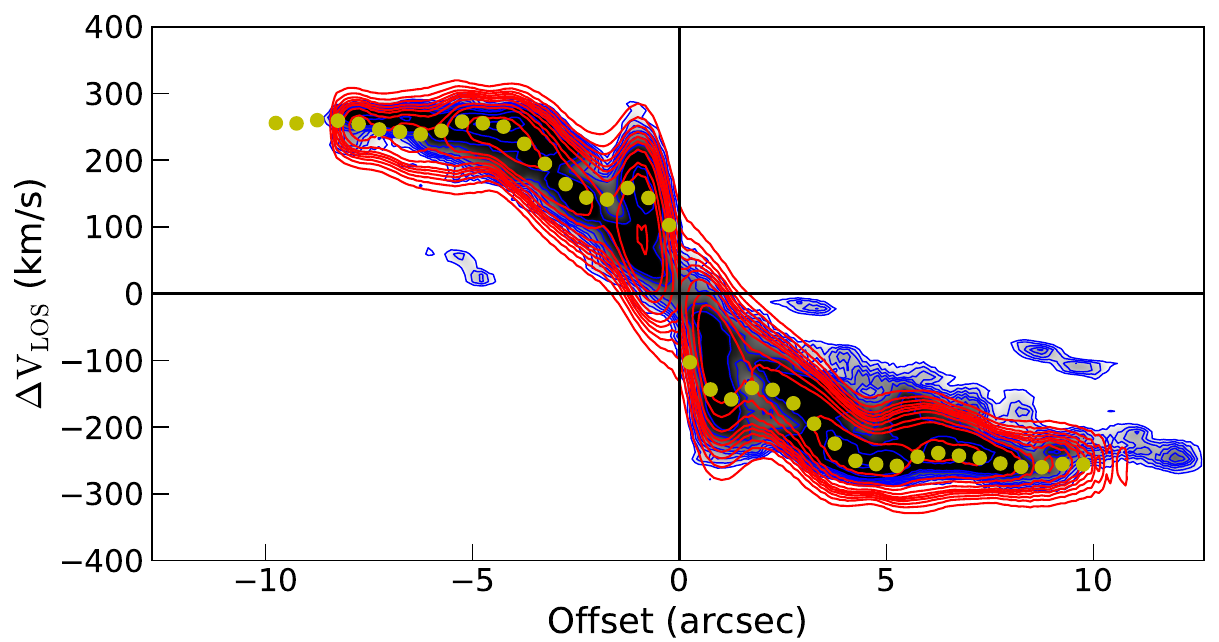}}
 \caption{Left panel: rotation velocity (green symbols) and velocity dispersion (blue symbols) as a function of the radius obtained from the best-fit disk dynamical model of the CO(2-1) data with 0.6 arcsec angular resolution. Middle/right panels: position-velocity diagrams along the kinematic major axis for the 0.2 arcsec and 0.6 arcsec resolution observations, respectively. The slit widths are set to the FWHM size of the synthetic beam major axis for each respective observation. Red contours and yellow circles represent the disk model, while blue contours represent the data. Contours are drawn at (1, 2, 4, 8, 16, 32, 64) $\sigma$.}
  \label{barolomodel}
\end{figure*} 

\subsection{Cold Molecular disk}
\label{sect:moldisk}

\noindent Figure \ref{momentmaps} shows the CO(2-1) integrated intensity (moment-0), the velocity (moment-1) and the velocity dispersion (moment-2) at 0.2 arcsec (top row) and 0.6 arcsec (bottom row) resolutions. The emission extends from north-east to south-west approximately along the inclined disk direction, that is close to edge-on \citep[i$\approx$ 80 deg][and this work]{marquez}. The maps with higher resolution detect clumpy structures in the host galaxy disk, clearly visible in the CO(2-1) intensity. 
In the lower resolution observation we detect the cold molecular tail towards south-east that connects NGC 2992 to its merging companion NGC 2993 (not included in the ALMA field of view). 
The CO(2-1) line profile from each data-cube considering regions above a 3$\sigma$ threshold in the intensity maps are reported in Figure \ref{app:spectraCO}. We integrate the spectra to derive the flux, FWHM and line luminosity $L^\prime$ of CO(2-1) line (Table \ref{tab:spectra}), the latter obtained using the relation of \citet{solomon}. 
To estimate the cold molecular mass we adopt a conversion factor $\alpha_{\rm CO} = 3.82 \rm M_{\odot} (K\,km \, s^{-1} \,pc^{2})^{-1} $ derived as in \citet{accurso}, for a stellar mass $\log(M_*/M_{\odot}) = $ 10.31 \citep{koss}. We find a total cold molecular reservoir of $M({\rm H_2})=9.7\times 10^8$ \msun, derived from the 0.6 arcsec resolution data. 
The highest resolution data filter out about 60\% of the CO(2-1) flux (Fig. \ref{app:spectraCO}). 
The mean-velocity maps (Figure \ref{momentmaps}) show a gradient oriented north-east to south-west along PA$\sim$ 210 deg, with range from $-250$ to 250 \kms (PA in degrees is measured anti-clockwise from north from the receding side of the galaxy).
The CO(2-1) velocity dispersion maps show values in the range from 10-20 \kms in the outer regions to 60-80 \kms in the inner regions. We detect two gas clumps with enhanced velocity dispersion, located in the inner 2$\times$4 $\rm arcsec^2$ region (see Section \ref{sect:moldisk}). 
%that we can also identified in the position-velocity (PV) diagrams along the kinematic major axis (i.e. $\sim$210 deg) reported in the bottom panels of Figure \ref{barolo}.
The velocity dispersion map in the lower resolution data-set shows enhanced values also in the region in which the tidal tail connects to the galaxy disk, and at the north-west border of the CO emitting region.

\begin{figure}[htb]
 \resizebox{\hsize}{!}{\includegraphics{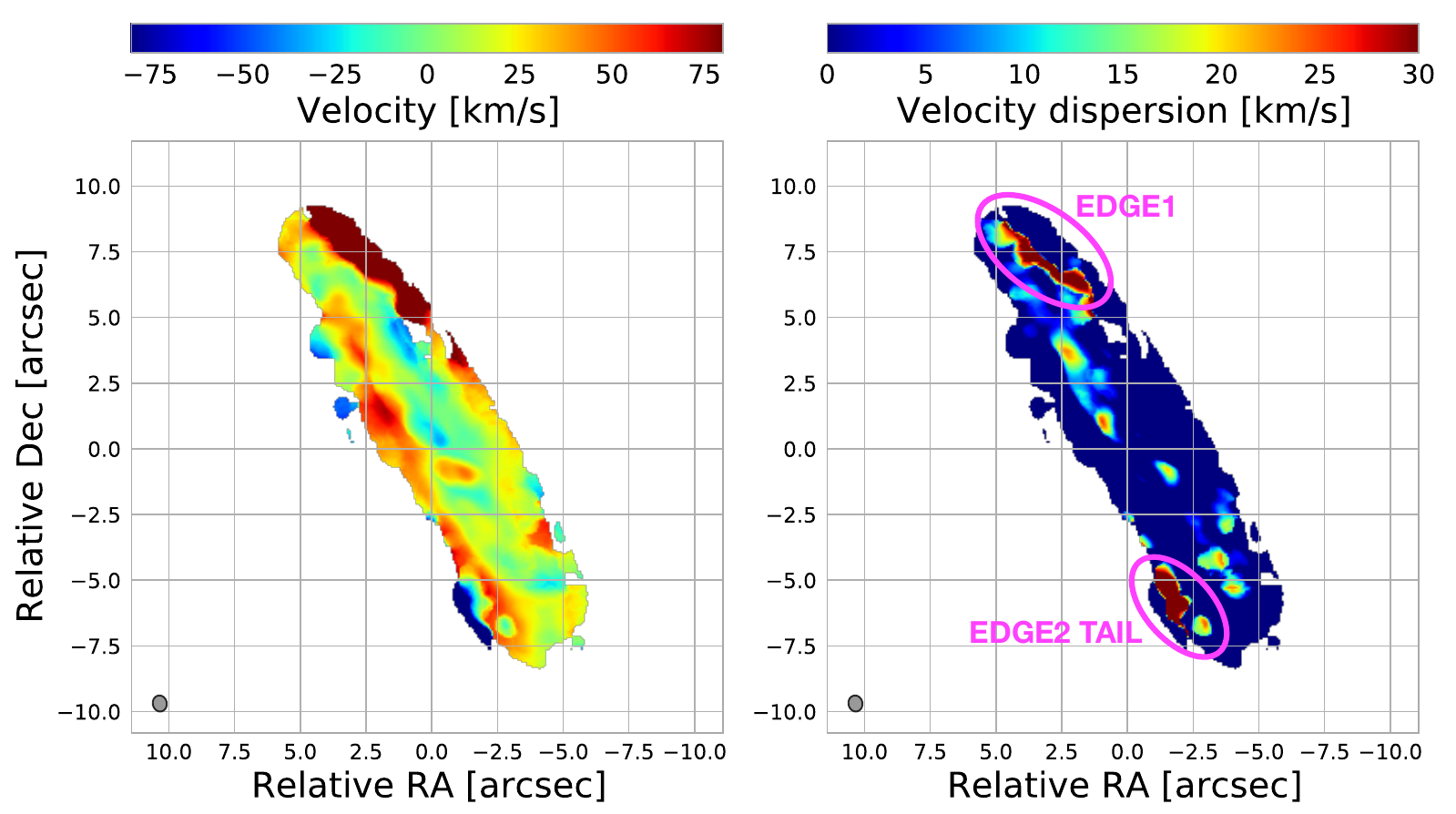}}
\caption{Left panel: residual mean velocity map. Right panel: residual velocity dispersion map, magenta circles mark the position of the cold molecular perturbations described in section \ref{sec:molpert}. Residual maps are obtained by subtracting the best fit disk model maps to the observed ones.}
  \label{fig:COres}
\end{figure}

We build a dynamical model of the system fitting the observed CO(2-1) data cubes with the 3D-Based Analysis of Rotating Objects from Line Observations \citep[$^{3D}\rm BAROLO$,][]{diteodoro}.  
We fit a 3D tilted-ring model to the 0.6 arcsec resolution data using 0.5 arcsec wide annuli. In the first run we allow four parameters to vary: rotation velocity, velocity dispersion, disk inclination, and position angle. We fix the kinematic centre to the peak position of the 1.3 mm continuum. A second run with three free parameters (rotation velocity, velocity dispersion and position angle) and inclination fixed to 80 deg, produces lower amplitude residuals; therefore, we adopt the latter as best-fit disk model. The inclination is fixed at the mean value found in the first run and the disk models return a position angle of $210\pm5$ deg. 
Figure \ref{barolomodel} shows the rotation velocity, $v_{\rm rot}$, and velocity dispersion, $\sigma_{\rm gas}$, versus radius of the best fit disk model, the former ranging from a central value of 100 \kms to 250 \kms at radii 0.6 out to 1.5 kpc.  
The (beam smearing corrected) gas velocity dispersion is 40-60 \kms in the nucleus, and about 25 \kms in the outer parts of the disk. 
The disk model provides a good description of the data, as demonstrated by the position velocity diagram taken along the kinematic major axis (i.e. $\sim$210 deg, Figure \ref{barolomodel}, right panel), in which data are consistent with the inclined rotating disk described above (red contours and yellow filled circles). 
Figure \ref{fig:COres} shows the residual maps obtained by subtracting the mean velocity and velocity dispersion model maps from the data. The main features detected in the residual velocity map are red-shifted regions at the north-west edge of the mask, where also the velocity dispersion is larger than 30 \kms. 
Molecular clumps with modest deviations from disk-like kinematics, about 25 \kms blue- and redshifted with respect the the rotation velocity at that position, are detected across the disk as regions with enhanced $\sigma_{\rm gas}$ (regions from cyan to orange in right panel of \ref{fig:COres}). These are likely due to the CNR in the central region, and to projection effects due to the high disk inclination. 
The same methodology applied to the higher resolution (0.2 arcsec) data delivers a dynamical model that is consistent with the previous one (Table \ref{tab:barolomodel}). These data resolve a circum-nuclear ring (CNR), with inner radius of about 65 pc, and with a slightly different PA with respect to the one of the outer disk (Figure \ref{barolomodel}).
%\textbf{The dynamical ratios, $v_{rot}/\sigma_{gas}$, are in the range 2-10, consistent with a rotationally supported molecular disk.}
The dynamical mass enclosed within a radius of 1 kpc, and computed as $M_{\rm dyn} = rv^2_{\rm rot}/2$G, is $M_{\rm dyn}=7-9\times 10^{9} \rm M_{\odot}$, corresponding to an escape velocity $v_{\rm esc} = \sqrt{\frac{{\rm G}M_{\rm dyn}}{r}} = $ 170-180 \kms at 1 kpc.

\begin{figure}[htb]
 \resizebox{\hsize}{!}{\includegraphics{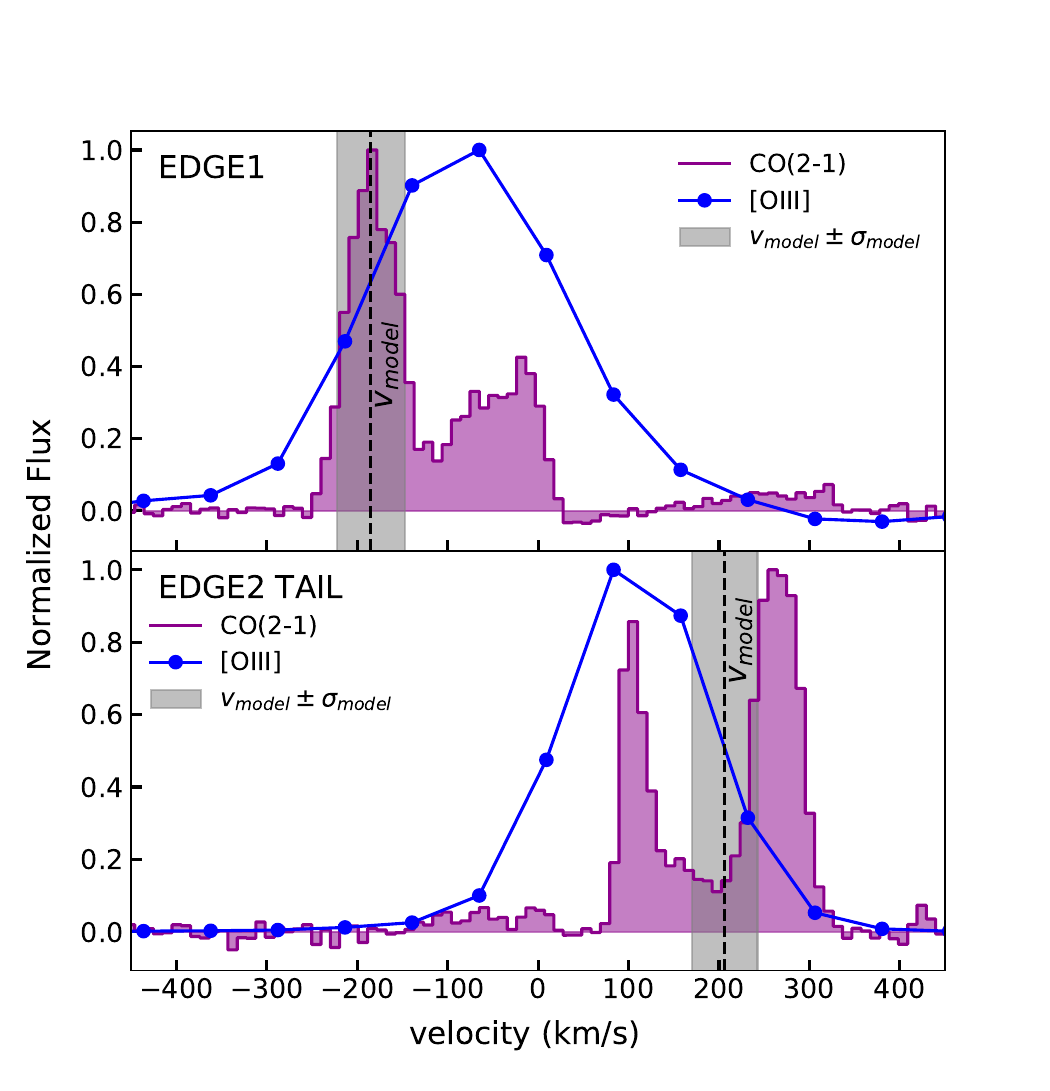}}
  \resizebox{\hsize}{!}{\includegraphics{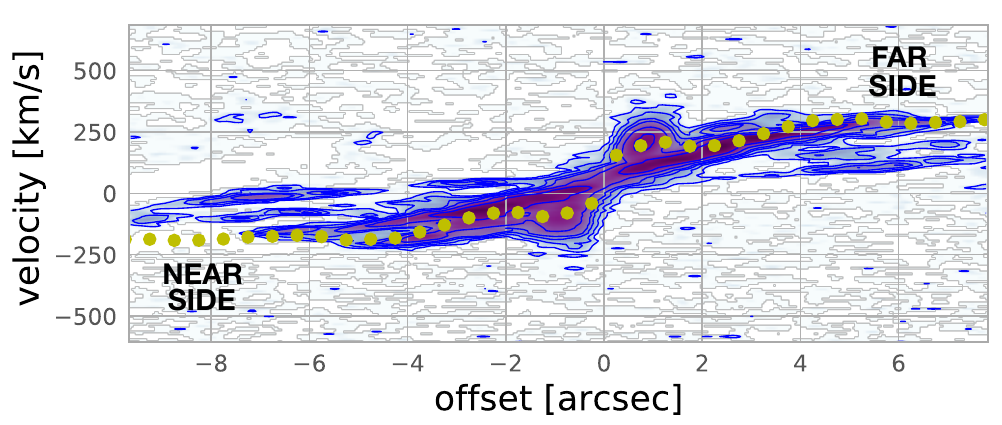}}
 \caption{Normalised CO(2-1) and [O III] spectra at region  EDGE1 (upper panel) and EDGE2 TAIL (middle panel). Purple histograms = CO(2-1) line.  Blue symbols and solid line = [O III]. The dashed black line and grey shaded area represent the LOS velocity of the best fit disk model  and the [$v_{\rm model}$-$\sigma_{\rm model}$,$v_{\rm model}$+$\sigma_{\rm model}$] range, where ($v_{\rm model}$, $\sigma_{model}$) are (-185, 37) \kms and (206, 36) \kms at at EDGE1 and EDGE2 TAIL respectively. Bottom panel: CO(2-1) position-velocity plot (from 2017.1.01439.S) cut through EDGE1, EDGE2 TAIL and the disk centre (along PA = 193 deg). The slit width is set to 0.6 arcsec (equal to the FWHM size of the synthetic beam major axis). Yellow symbols represent the disk model while blue contours represent the data, contours are drawn at (2, 5, 10, 15, 30, 60)$\sigma$.}
  \label{fig:molperturbations}
\end{figure}

\subsection{Cold molecular wind} \label{sec:molpert}

\begin{figure*}[ht]  
 \resizebox{\hsize}{!}{\includegraphics{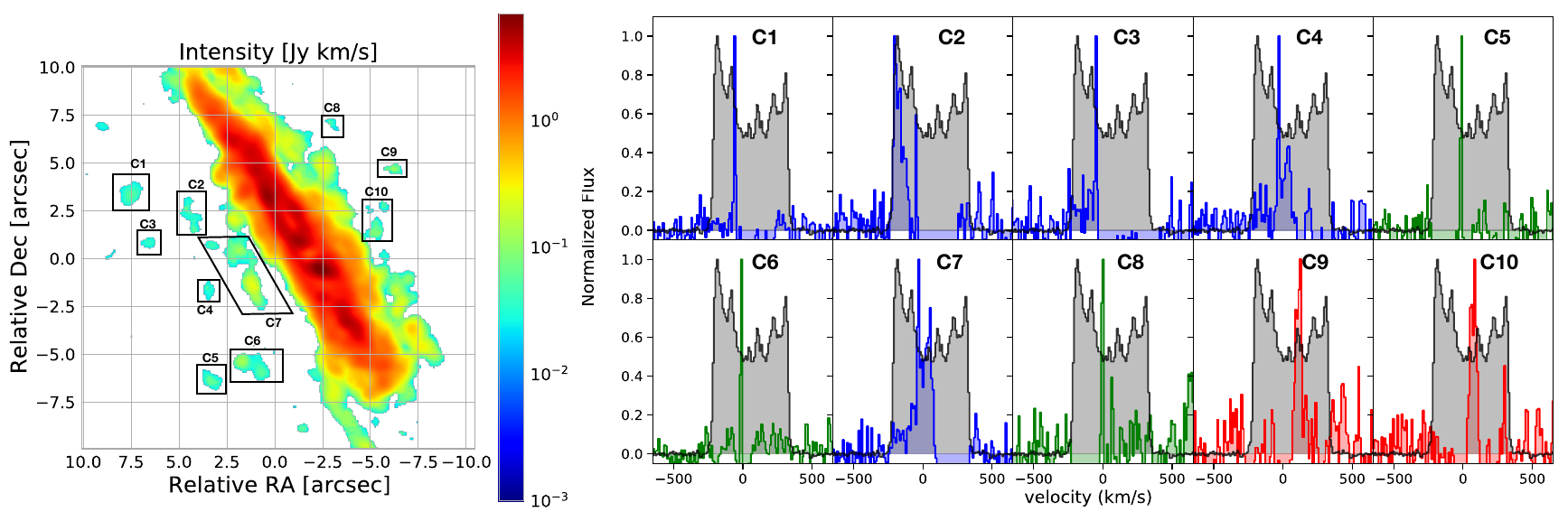}}
 \caption{Left panel: CO(2-1) intensity map with a cut at $>3\sigma$. The labelled black polygons show the CO(2-1) clumps identified as cold molecular wind, i.e. not participating to the disk rotation. Right panel: the blue, green and red spectra are extracted from the each molecular clump identified in the left panel, as labelled. Spectra are coloured according to the velocity shift of their peak: red with $v_{\rm peak} > $10 \kms,  green with  -10 $< v_{\rm peak} < $10 \kms,  blue with $v_{\rm peak} < -10$ \kms.
 The grey shaded histograms show the total CO(2-1) emission above a 3$\sigma$ threshold, extracted from the coarser resolution data-cube.}
 \label{fig:clumps}
\end{figure*} 

The largest residuals in Figure \ref{fig:COres} are detected as a redshifted emission ($v \sim$80 \kms) at the north-western edge of the disk (EDGE1), and as a blue-shifted one ($v \sim$-80 \kms) at the onset of the tail, at the south-eastern edge of the disk (EDGE2-TAIL). In the latter position, the velocity dispersion map shows residual emission of amplitude about 40 \kms. 
The CO and [O III] spectra extracted at EDGE1 and EDGE2-TAIL are shown in Figure \ref{fig:molperturbations}. Both show two bright components peaking at different velocities. 
At EDGE1 our disc best fit dynamical model has a projected velocity $v(model) \sim-185$ \kms and $\sigma (model) \sim$40 \kms, that is consistent with the strongest CO component peaking at -180 \kms. The additional emission components, spanning velocities of [-100,0] \kms and [150,300] \kms, thus trace off planar motion of the gas, associated to a wind.
The [O III] emission line at EDGE1 is offset by 100 \kms with respect to the CO disk component, suggesting that in this region the ionized gas does not partake in the disk rotation but instead traces perturbed kinematics, due to a wind. \ha and \hb emission lines show consistent line profile and velocity offset as  [O III].
EDGE2-TAIL also shows a CO spectrum with two emission components. At this location the best fit dynamical model requires a velocity of $\sim$205 \kms and $\sigma (model)\sim$40 \kms, which is in between the two main CO peaks. Off-planar motions at EDGE1 (EDGE2-TAIL) are seen in the PV plot shown in Figure \ref{fig:molperturbations}, as regions with offset [-9,-5] arcsec ([4,9] arcsec) and line of sight velocity that is smaller than that of the disk. 
EDGE2-TAIL is further complicated by the tidal tail due to the interaction of NGC 2992 with NGC 2993 \citep{duc2000}, so it is hard to identify and separate tidal interaction from any wind.
For this reason, we compute the wind parameters only at EDGE1 and do not include EDGE2-TAIL in the following analysis of the cold molecular wind.
At EDGE1 we derive a cold molecular mass $M(\rm H_2)= 2.7 \times 10^7 \, \frac{\alpha_{\rm CO}}{3.82 ~\mathrm{M_{\odot} (K\,km \, s^{-1} \,pc^{2})^{-1}}}$ \msun, 
and for EDGE2-TAIL  $M(\rm H_2)= 0.9 \times 10^7 \, \frac{\alpha_{\rm CO}}{3.82 \, \mathrm{M_{\odot} (K\,km \, s^{-1} \,pc^{2})^{-1}}}$ \msun.

\begin{table*}
     \caption[]{Measured properties of CO(2-1) at at different spatial frequencies.}
         \label{tab:spectra}
\centering                          
\begin{tabular}{c c c c c c c }        
\hline\hline                 
ID & uvrange & FWHM & $S_{\rm CO}$  & $L'_{\rm CO(2-1)}$ & $M(\rm H_2)$ & Area \\ & [m] & [\kms] & [Jy \kms] & [$\rm 10^7 \, K \, km \, s^{-1} \, pc^2$] & [$\rm 10^8 \, M_{\odot}$] & [arcsec$\rm ^2$] \\
(a) & (b) & (c) & (d) & (e) & (f) & (g)\\
\hline      
00236.S & all & 470$\pm$10 & 132.54$\pm$0.03 & 10.8 & 4.1 & 33 \\
\hline
01439.S & all & 4640$\pm$10 & 311.56$\pm$0.02  & 25.4 & 9.7 & 132 \\
        & $>$100 & 4840$\pm$10 & 103.58$\pm$0.03 & 8.4 & 3.2 & 97 \\
        & $<$100 & 4640$\pm$10 & 307.02$\pm$0.01 & 25.0 & 9.6 & 30\\
\hline  
\end{tabular}\\
  \flushleft 
 \footnotesize{ {\bf Notes.} (a) the project code, (b) the \texttt{uvrange} parameter, (c) the full width half maximum (FWHM) of the line (spectra are shown in Figure \ref{app:spectraCO}):, (d) the integrated flux, (e) the line luminosity, (f) the cold molecular gas mass obtained using a conversion factor $\alpha_{\rm CO}\,= 3.82 \,\mathrm{M_{\odot} (K\, km \, s^{-1} \,pc^{2})^{-1}}$, and (g) the area of the considered region.}
\end{table*}

\begin{table*}
     \caption[]{Properties of the cold molecular outflowing clumps.}
         \label{tab:molof}
\centering    
\label{table:molof}
\begin{tabular}{l c c c c c c c c}        
\hline\hline                 
ID & $S_{\rm CO}$ & $\Delta v$  & $L'_{\rm CO(2-1)}$ & $M_{\rm of}$ & $r_{\rm of}$ & $v_{\rm of}$ & $\dot{M}_{\rm of}$ & $\dot{E}_{\rm kin}$ \\ 
 & [mJy \kms] & [\kms] & [$\rm 10^5 \, K \, km \, s^{-1} \, pc^2$] & [$\rm 10^6 \, M_{\odot}$] & [$\rm kpc$] & [$\rm km \, s^{-1}$] & [$\rm M_{\odot} \, yr^{-1}$] & [$\rm erg \, s^{-1}$]\\
(a) & (b) & (c) & (d) & (e) & (f) & (g) & (h) & (i)\\
\hline 
C1  & 221.9$\pm$0.2  & 41  & 1.81  & 0.69 & 1.71 & -59  & 0.07  & 8.4$\times \rm 10^{37}$\\
C2  & 667.1$\pm$0.1  & 165 & 5.44  & 2.08 & 1.15 & -204 & 1.13  & 1.5$\times \rm 10^{40}$\\
C3  & 200.4$\pm$0.1  & 93  & 1.63  & 0.62 & 1.49 & -49  & 0.06  & 4.9$\times \rm 10^{37}$\\
C4  & 382.6$\pm$0.1  & 144 & 3.12  & 1.19 & 0.96 & -29  & 0.11  & 2.9$\times \rm 10^{37}$\\
C5  & 210.1$\pm$0.2  & 41  & 1.71  & 0.65 & 1.35 & -8   & 0.01  & 2.8$\times \rm 10^{35}$\\
C6  & 659.6$\pm$0.1  & 124 & 5.38  & 2.05 & 1.08 & -8   & 0.05  & 1.1$\times \rm 10^{36}$\\
C7  & 2040.9$\pm$0.1 & 288 & 16.64 & 6.36 & 0.59 & -29  & 0.95  & 2.5$\times \rm 10^{38}$\\
C8  & 135.9$\pm$0.2  & 41  & 1.11  & 0.42 & 1.19 & 2    & 0.002 & 2.5$\times \rm 10^{33}$\\
C9  & 234.1$\pm$0.1  & 82  & 1.91  & 0.73 & 1.14 & 126  & 0.25  & 1.2$\times \rm 10^{39}$\\
C10 & 501.2$\pm$0.1  & 93  & 4.09  & 1.56 & 0.61 & 84   & 0.66  & 1.5$\times \rm 10^{39}$\\
\hline  
\end{tabular}\\
  \flushleft 
 \footnotesize{ {\bf Notes.} Measured properties of the outflowing CO(2-1) clumps detected Figure \ref{fig:clumps}. (a) ID, (b) integrated CO(2-1) flux, (c) line width, (d) CO(2-1) luminosity, (e) clump mass obtained using a conversion factor $\alpha_{\rm CO}\,= 3.82 \,\mathrm{M_{\odot} (K\,km \, s^{-1}\,pc^{2})^{-1}}$, (f) projected distance from the AGN, (g) the clump velocity defined as the peak velocity, (h) the mass rate at radius $r_{\rm of}$, and (i) the wind kinetic power.}
\end{table*} 

In addition, we identify as cold molecular wind the regions which do not participate to the disk rotation as described by our best fit model, and detected as molecular clumps located above and below the disk. 
The clump spectra are shown in Figure \ref{fig:clumps} (right panel) and coloured according to their peak velocity with respect to the rest frame, red for $v_{\rm peak} > $10 \kms, green for -10 \kms$< v_{\rm peak} < $10 \kms and blue for $v_{\rm peak} < $-10 \kms. We note that the clumps on the western side of the disk are mainly redshifted, and those on the eastern side mainly blueshifted. We detect 10 molecular clumps with $>$ 3$\sigma$ significance (Table \ref{tab:molof}). 
For each clump we measure the wind velocity, $v_{\rm of}$, as the peak velocity, the line luminosity, the cold molecular mass, $M_{\rm of}$, and the projected distance of each clump from the AGN, $r_{\rm of}$.
We find a total cold molecular mass in the clumps of 
$M(\rm H_2)_{\rm clumps}= 1.64 \times 10^7~ (\alpha_{\rm CO}/3.82)$ \msun.
The wind mass outflow rate is computed at each radius $r_{\rm of} $ by applying the relation from \citet{fiore2017}, assuming a spherical sector:
\begin{equation} \label{eq1}
\dot{M}_{\rm of} = f \frac{M_{\rm of} \, v_{\rm of}}{r_{\rm of}}. 
\end{equation}

A factor $f=3$ is used with the assumption of a conical wind with constant density, as previously adopted \citep[e.g.][]{vayner2017, feruglio2017, brusa2018}. However, in other works \citep[e.g][]{bischetti2019,bischetti2019b,cicone2015,veilleux2017, marasco2020,tozzi2021} the mass outflow rate is rather computed with $f$ equal to 1, by assuming a steady mass-conserving flow with constant velocity, in which the density at the outer radius is only 1/3 that of the average \citep[see][for a detailed discussion]{veilleux2017}. The resulting values, assuming $f=3$, relative to each component are reported in in column (e) of Table \ref{tab:molof}.

\begin{figure*}[ht]
 \resizebox{\hsize}{!}{\includegraphics{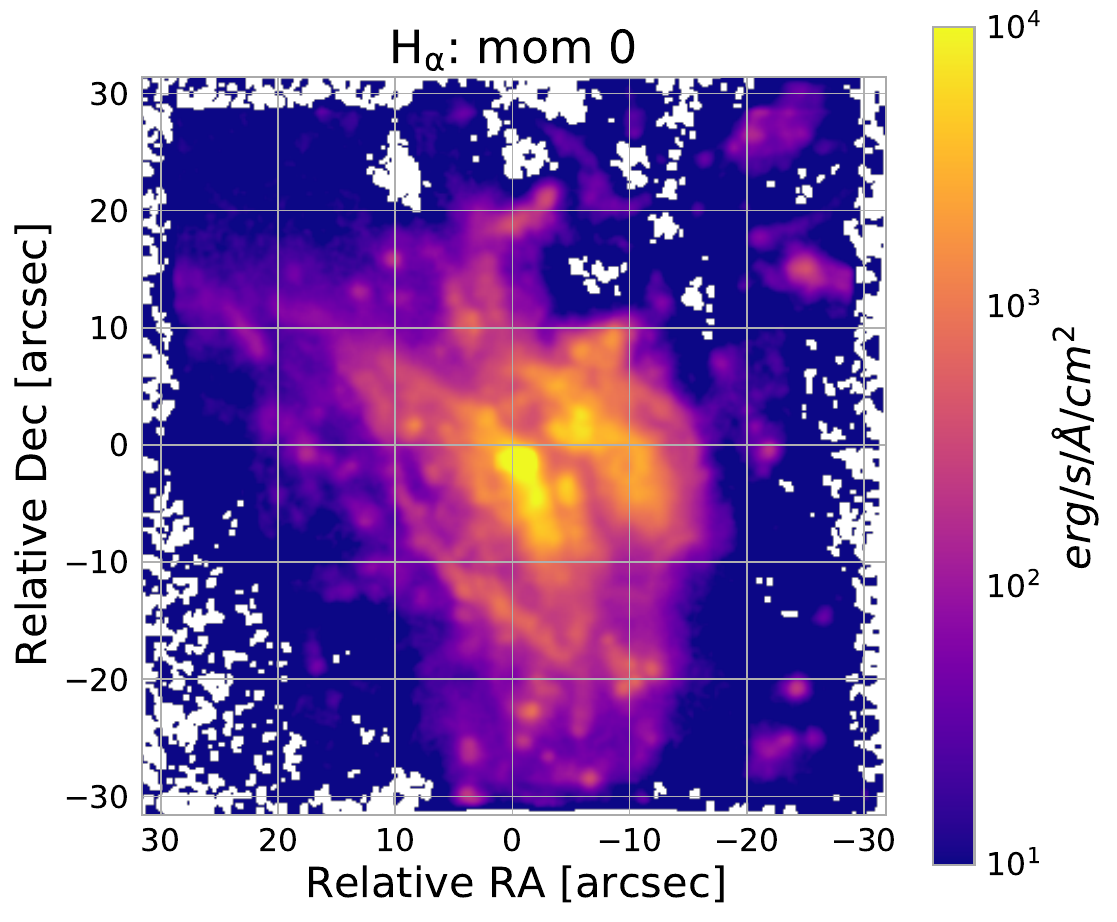} \includegraphics{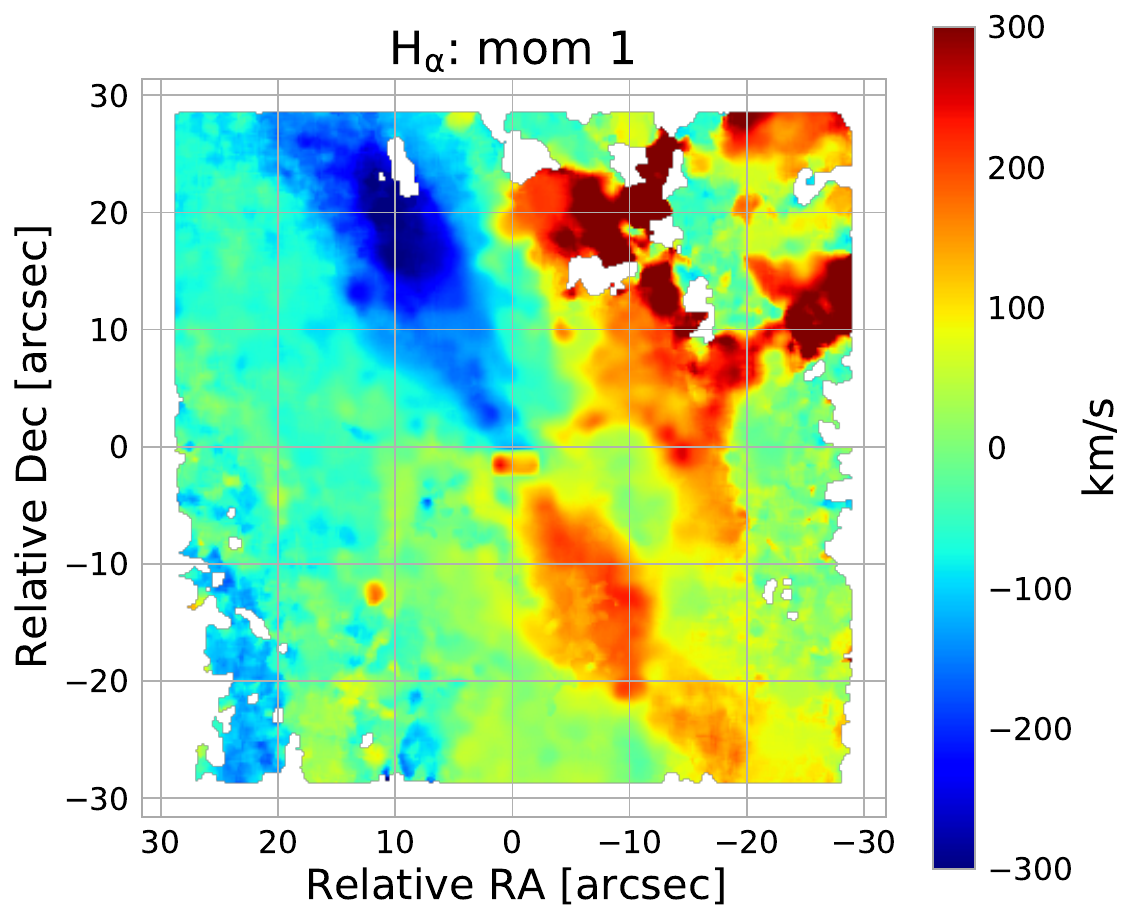} \includegraphics{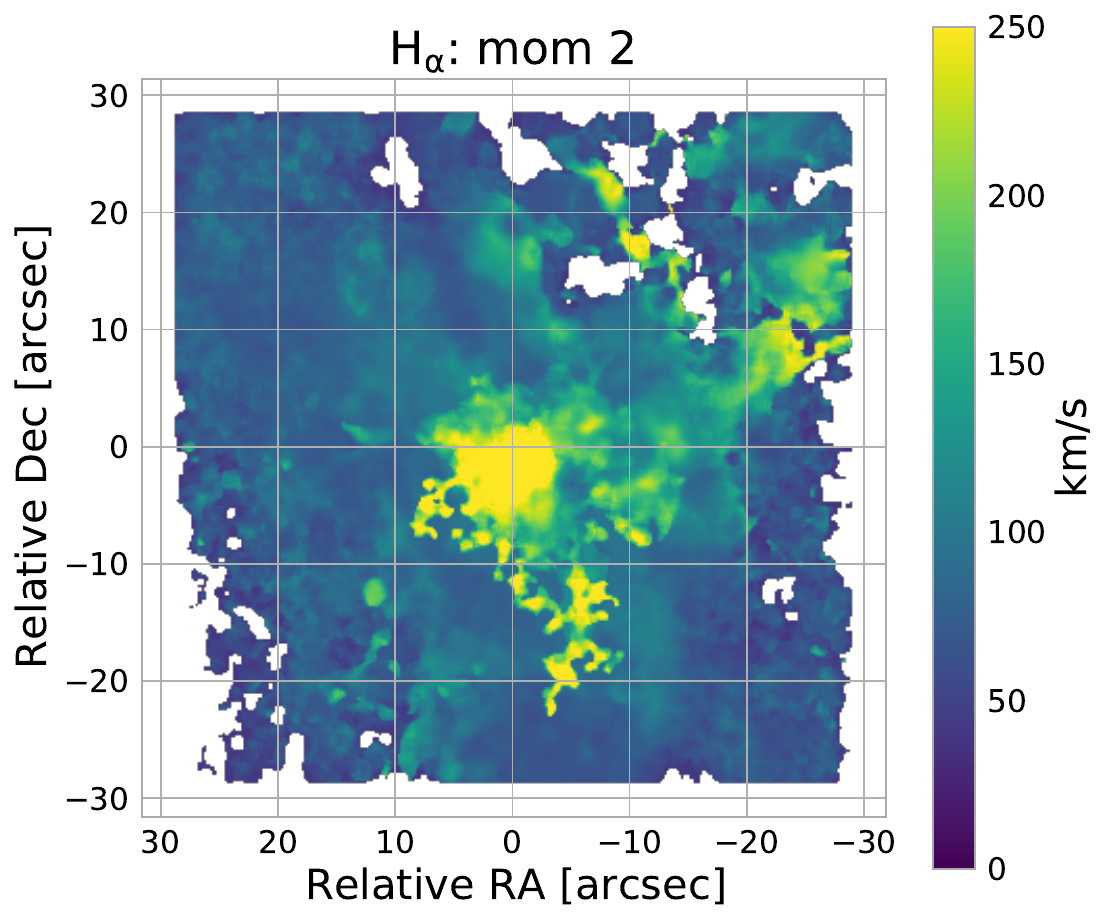}} 
 \resizebox{\hsize}{!}{ \includegraphics{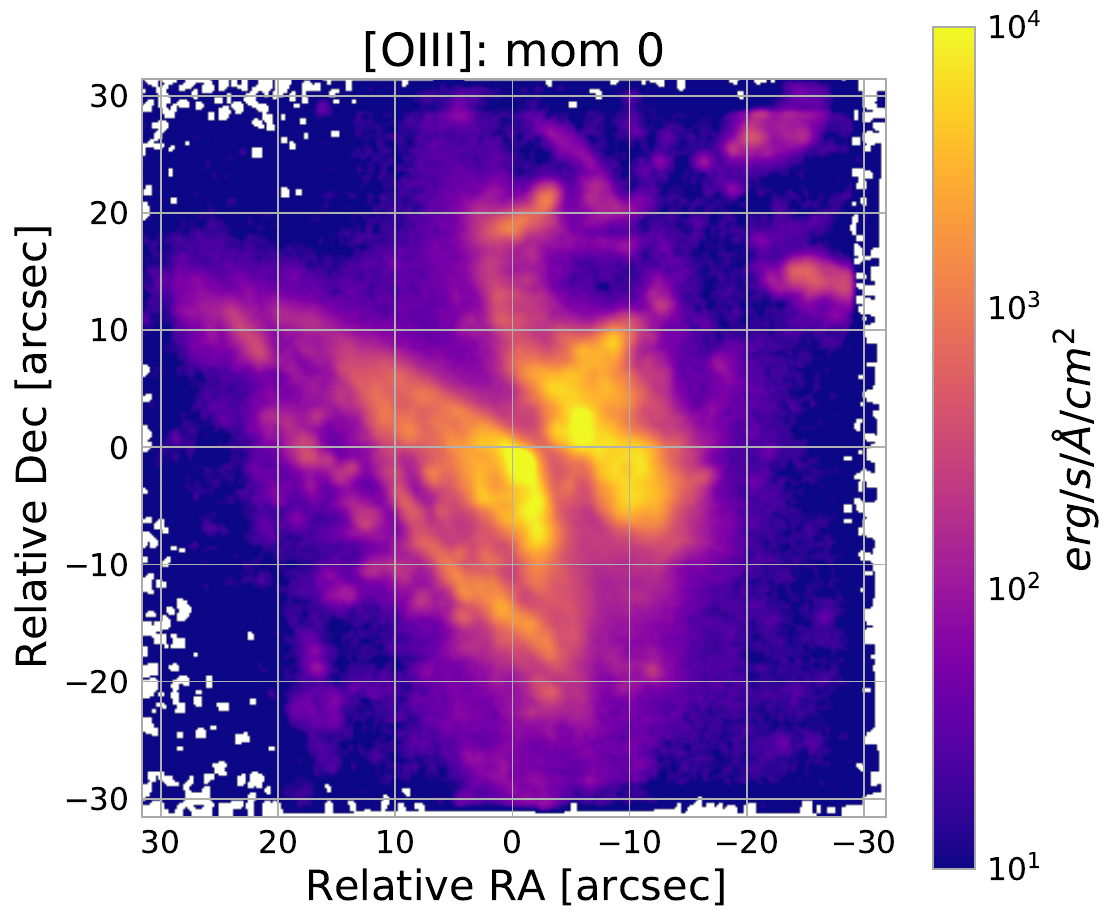} \includegraphics{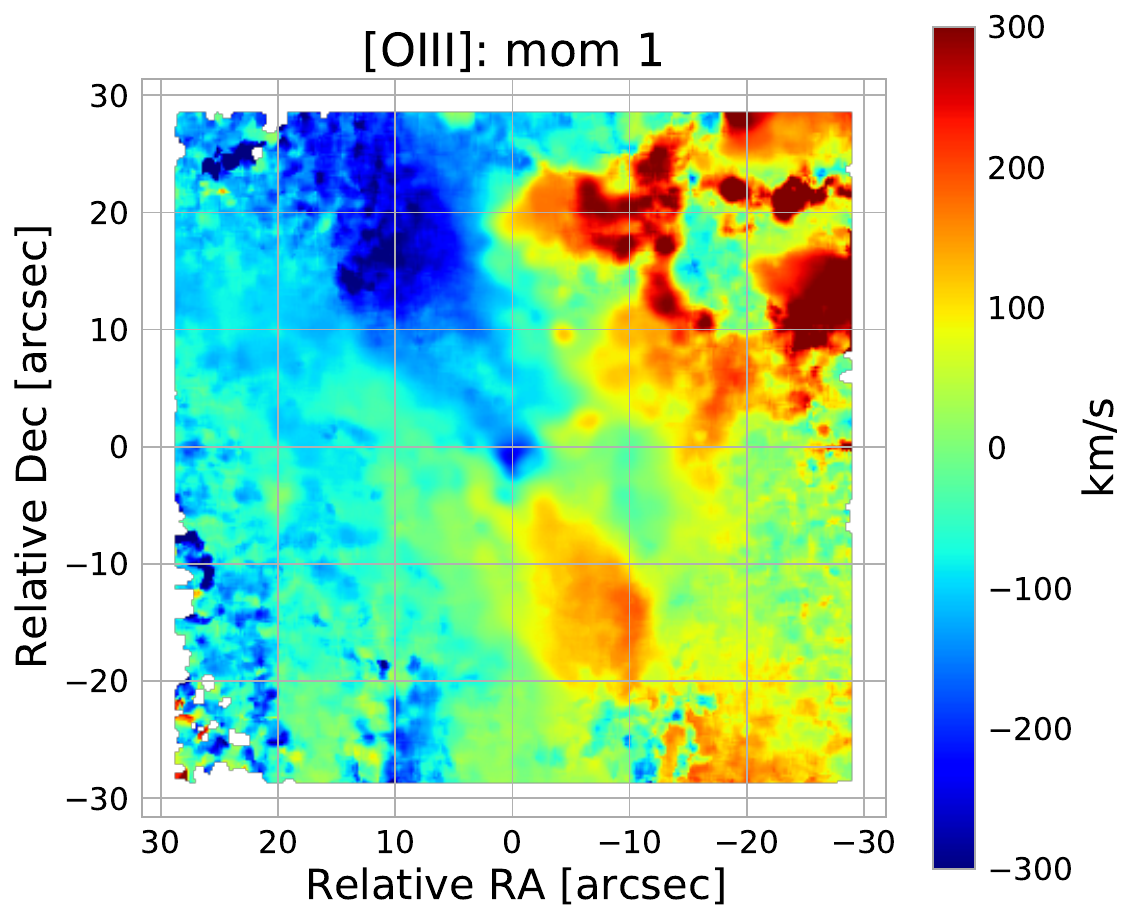} \includegraphics{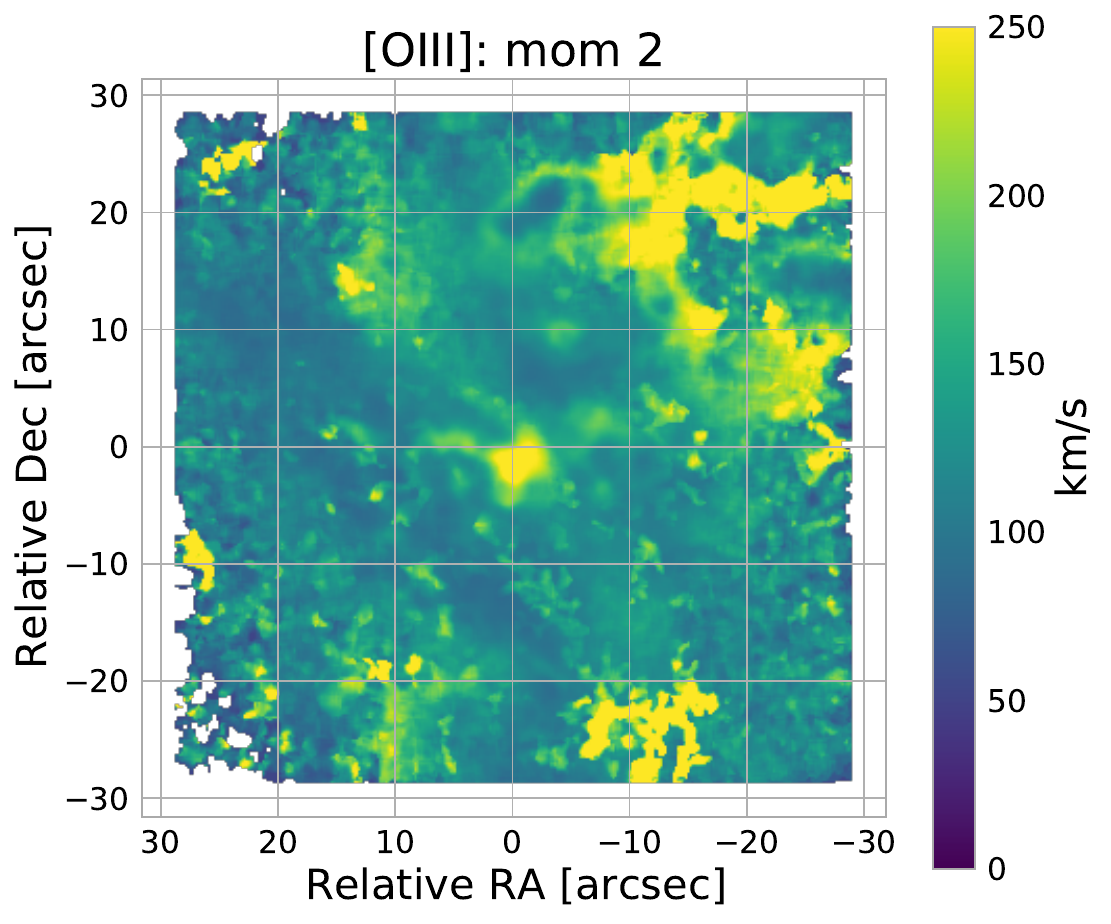}} \resizebox{\hsize}{!}{ \includegraphics{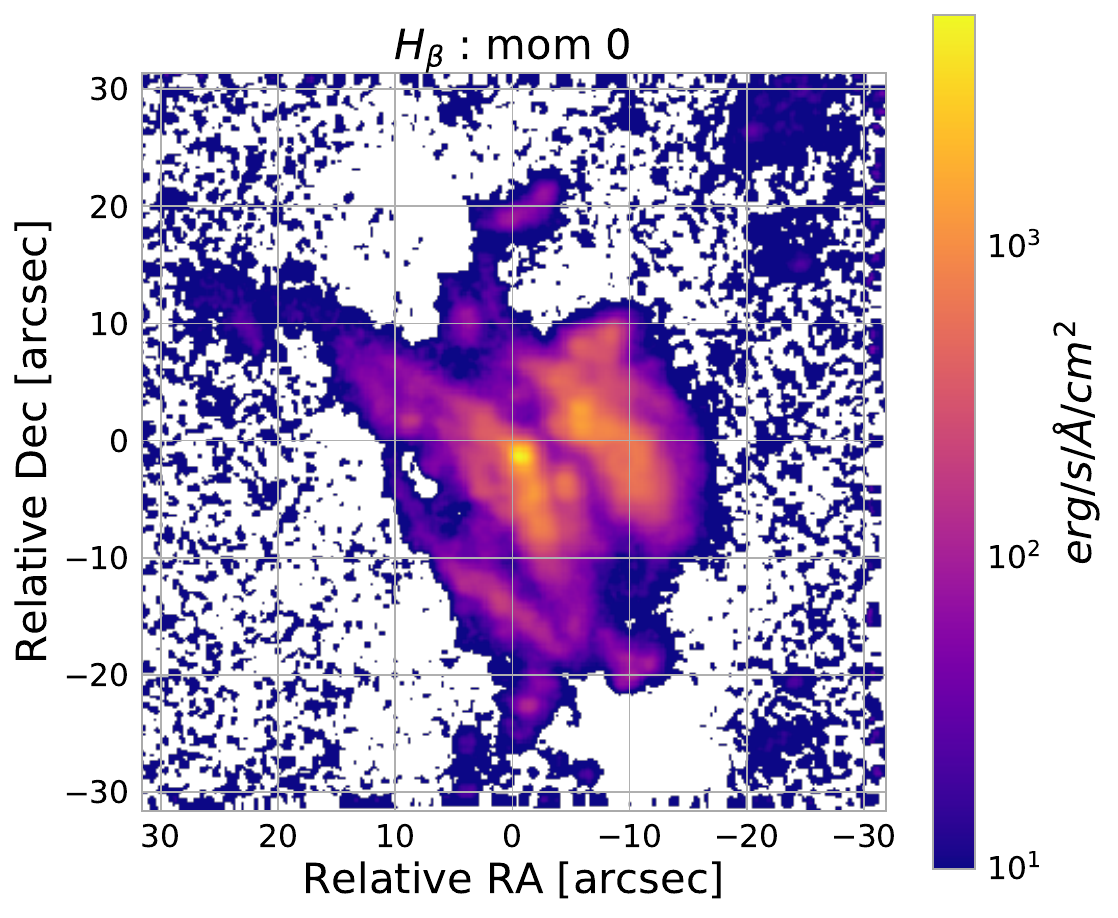} \includegraphics{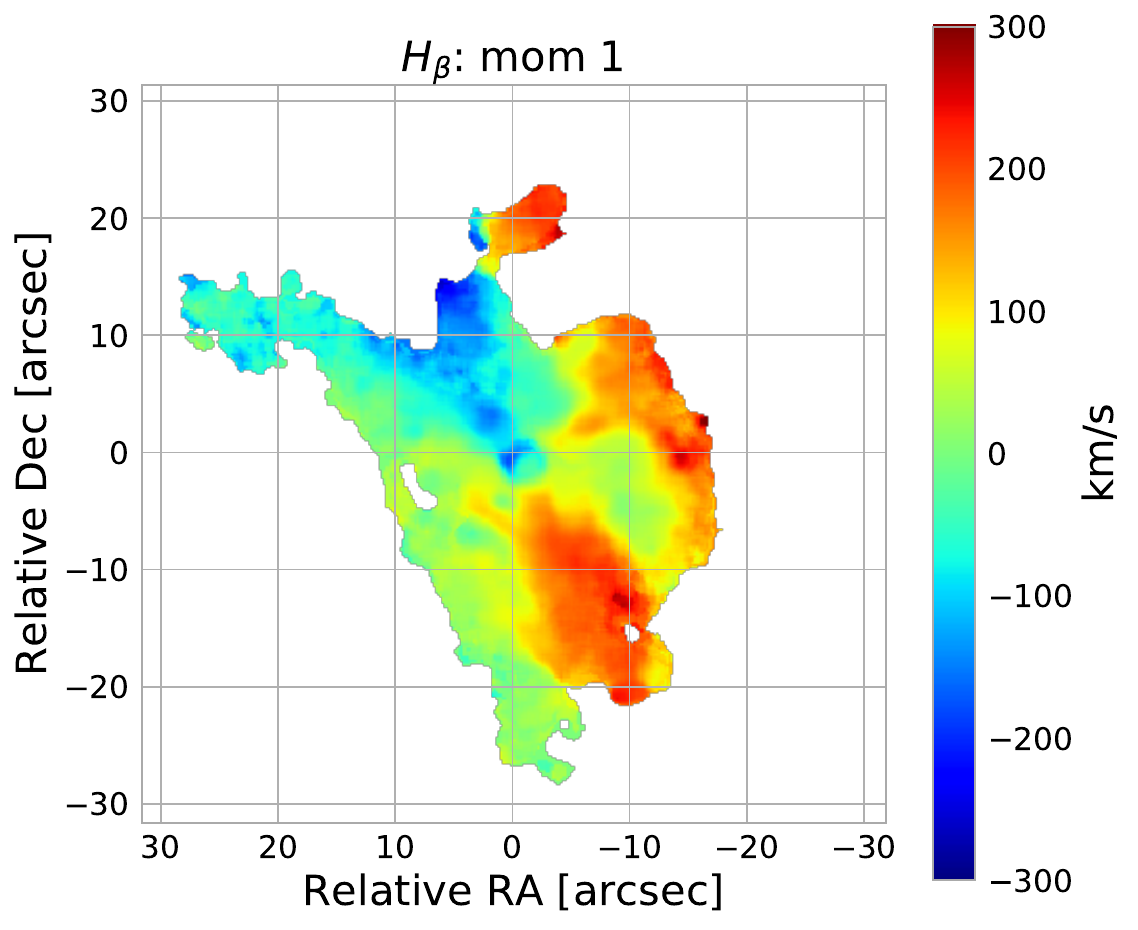} \includegraphics{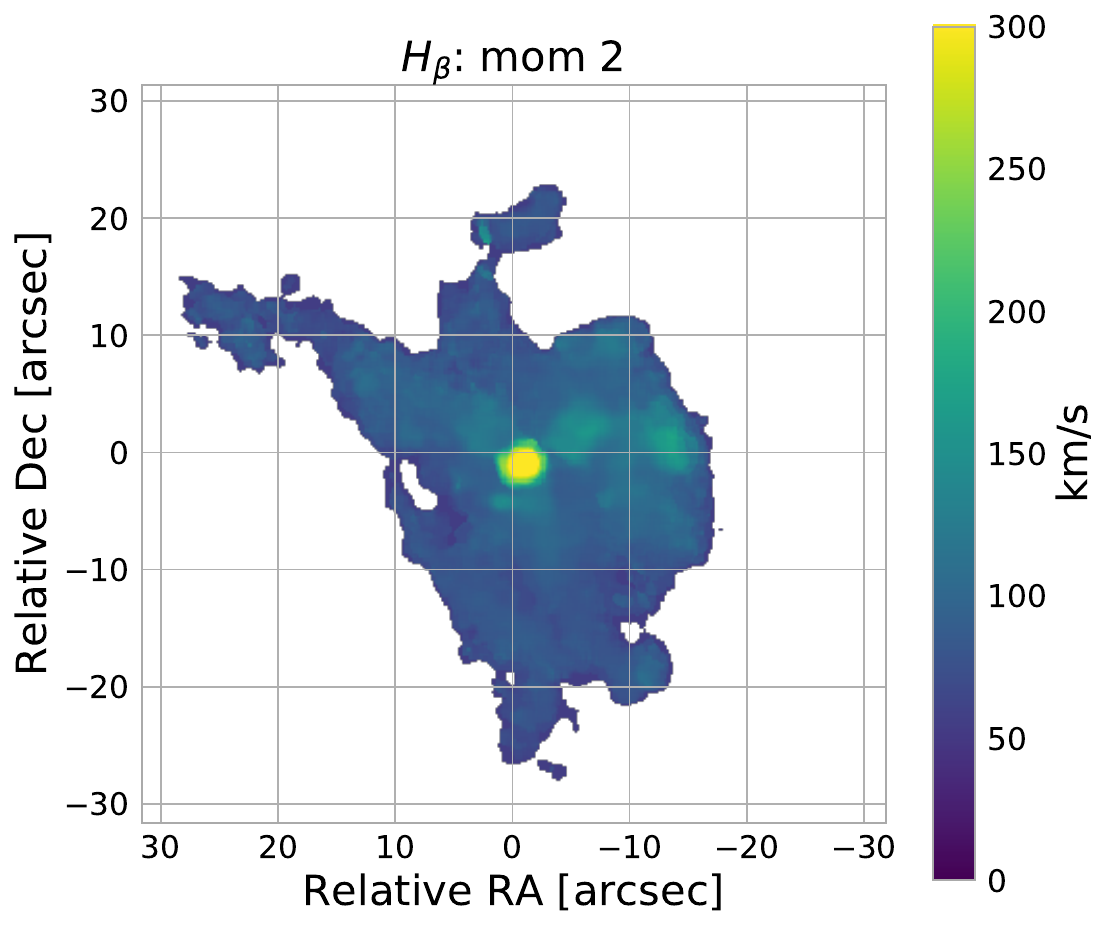}}  \resizebox{\hsize}{!}{ \includegraphics{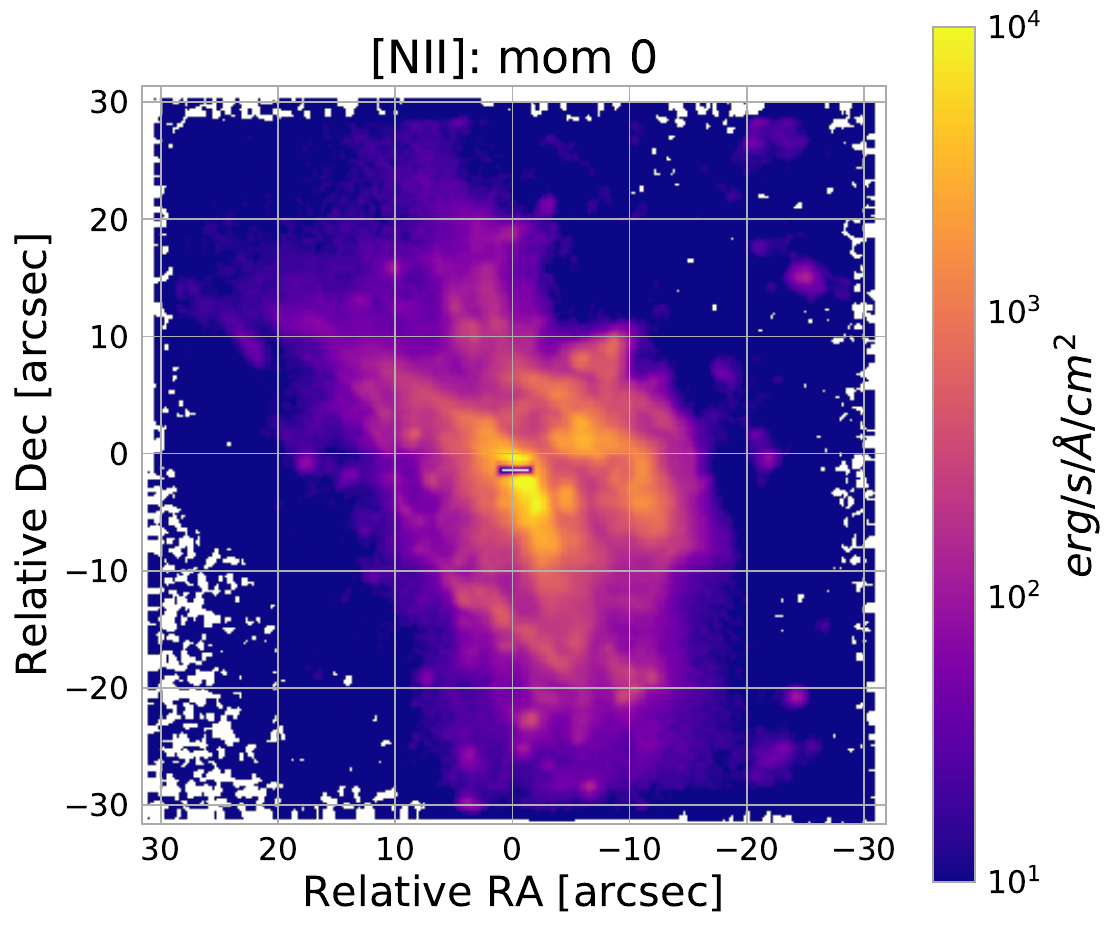} \includegraphics{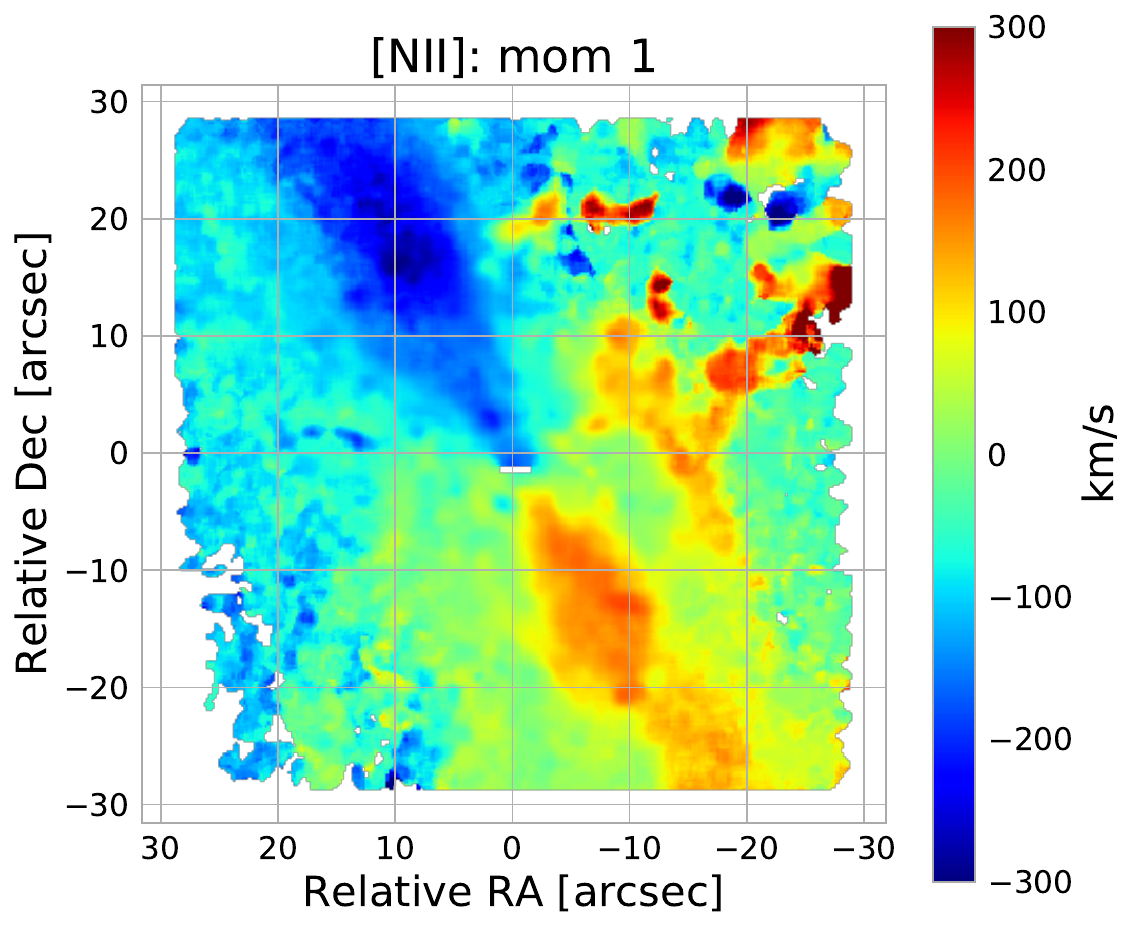} \includegraphics{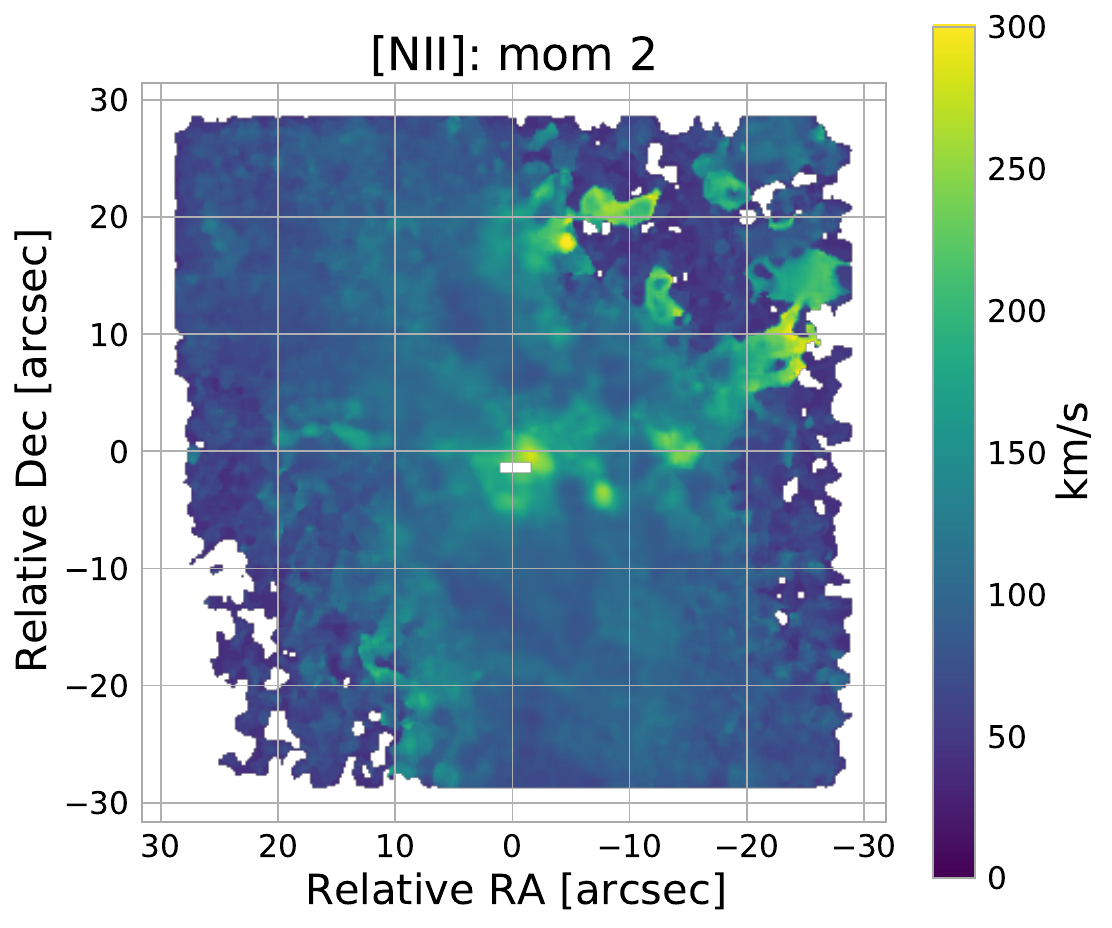}} \caption{Moment maps of (top to bottom) H${\alpha}, [O III], {\rm H}{\beta}$, [NII]$\lambda 6583 \AA$ data-cubes. From left to right: integrated flux (moment-0), mean velocity (moment-1) and velocity dispersion maps (moment-2). The maps were obtained using the CUBEXTRACTOR package by \citet{cantalupo2019}, using detection threshold that optimise the signal-to-noise  (3$\sigma$ threshold for \ha, 2$\sigma$ threshold for [O III], 5$\sigma$ for H${\beta}$ and 4$\sigma$ for [NII]).}
  \label{fig:Cubexmoments}
\end{figure*} 

\subsection{Ionised gas distribution}\label{sec:ionisedgas}

We analysed the \hb, [O III]$\rm \lambda$5007$\AA $, \ha, and [NII]$\lambda 6583\AA$ emission lines, extracted from the MUSE data-cube by using the CUBEXTRACTOR package \citep{cantalupo2019}. We used the procedure described below for each emission line, following the approach described in \citet{borisova2016, cantalupo2019}.
%Differently we did not apply PSF subtraction because we are studying a local Seyfert and not extended emission around high red-shift quasars.
First, we used the \texttt{CubeSel} tool to produce three z-direction cut data-cubes around the [O III] line and the \hb line with a spectral window of 100$\AA $, while \ha and and [NII] doublet lines are all included in a spectral window of 300$\AA $
We removed any possible continuum emission for each spaxel within the MUSE FOV by applying a fast median-filtering approach using the \texttt{CubeBKGSub} task \citep{cantalupo2019}.
This tool allowed us to apply a median sigma clipping algorithm with a filter of 30 spectral layers ($\approx$ 40 $\AA $), carefully masking the emission lines.
Then we smoothed the spectrum with a median filter with radius equivalent to two bins to minimize the effect of line features in individual spectral regions. Finally we subtracted from each voxel in the cube the estimated continuum from the corresponding spaxel and spectral region. 
In this way we obtained the continuum subtracted cubes. 
In order to perform the following analysis, we produce a data-cube for each emission line of interest.
For what concerns the \ha and [NII] doublet line complex, due to their blending, we modelled the emission lines performing a pixel-by-pixel Gaussian fit with a minimum of one and a maximum of two components for each emission line.
To produce the \ha cube, we subtracted the model [NII] doublet emission in each voxel, and analogously to obtain the [NII]$\lambda 6583\AA$ cube we subtracted the previously model \ha and [NII]$\lambda 6548\AA$ profile.
%Then we subtracted the modelled emission lines in order to produce the \ha cube with [NII] doublet emission subtracted and [NII]$\lambda 6583\AA$ cube with \ha and [NII]$\lambda 6548\AA$ contribution subtracted.
\texttt{Cube2Im} task of CUBEX tool produces 3D masks of the extended emission lines in each data-cube, by applying a signal-to-noise (SN) threshold of 3 for \ha, of 2 for [O III], of 5 for H${\beta}$ and of 4 for [NII].
We applied different thresholds in order to optimise the signal-to-noise ratio.
By using only the voxels of the continuum-subtracted datacube defined in the 3D-mask, the \texttt{Cube2Im} task allowed us to obtain the following three data products: (i) the surface brightness map, that is, an "optimally extracted Narrow Band image"; 
(ii) a map of the velocity distribution obtained as the first moment of the flux distribution; and (iii) a map of the gas velocity dispersion $\sigma_{\rm gas}$, which was derived as the second moment of the flux distribution. 
All these data products shown in Figure \ref{fig:Cubexmoments} were obtained by assuming rest wavelengths, corrected according to the AGN redshift derived from CO(2-1) data and smoothed with a Gaussian-kernel of $\sigma = 2$ pixels. 

\begin{figure*}[ht] \resizebox{\hsize}{!}{\includegraphics{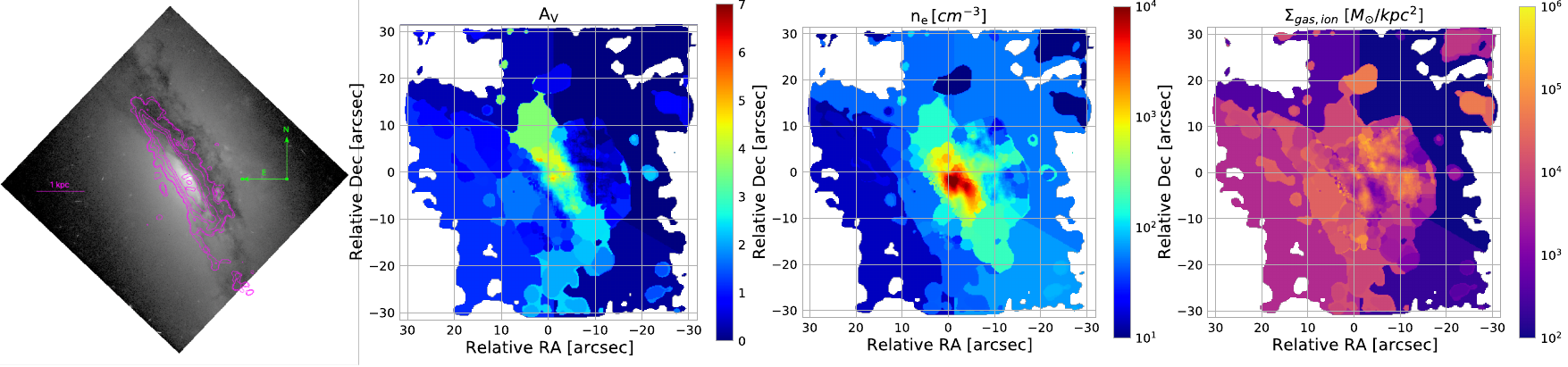}}
 \caption{Left panel: HST image of NGC 2992 in the FQ606W filter (from the Mikulski Archive for Space Telescopes, Proposal 05479), showing the dust lane. Magenta contours are from the CO(2-1) intensity map (Figure \ref{momentmaps}). Left-centre panel: Voronoi map of the total extinction in V band, $A_{V}$. Right-centre panel: electron density map, $n_{\rm e}$, derived from the ionisation parameter (Section \ref{sec:ionisedgas} and Appendix \ref{appendixc}). Right panel: mass surface density of the ionised gas.}
  \label{fig:gasproperties}
\end{figure*}

The intensity maps (Figure \ref{fig:Cubexmoments} ) show similar morphology in the four tracers. The emission peaks are consistent with the AGN position from ALMA. The emission is reduced along the cold molecular disk due to extinction by dust in the galaxy disk and dust lane \citep{trippe2008}, mostly affecting the H${\beta}$ emission (Figure \ref{fig:gasproperties}). 
Two ionised cones emerge approximately perpendicular to the galactic disk in the north-west and south-east directions, with opening angle of about 120 deg, consistent with the value estimated by \citet{garcia-lorenzo2001}. The edges of the ionisation cones are well defined in all tracers, as is the clumpy gas distribution within the cones.
The mean velocity maps show a velocity gradient oriented from north-east to south-west, with range from -300 to 300 \kms, which likely traces the ionised disk. The kinematic major axis is consistent with that identified by CO(2-1). The disk velocity gradient is detected in all emission lines. %\ha and [N II] emission clearly shows broadening due to the presence of the broad line region (BLR) in correspondence of the AGN position. 
Within the ionisation cones, the mean velocity maps show redshifted velocities exceeding 300 \kms in the north-western cone, and enhanced blueshifted velocities in the south-eastern cone, meaning that the west cone is mainly receding, while the eastern cone is mainly approaching, although different kinematic components are present along any line of sight. 
The $\sigma_{\rm gas}$ maps show the average velocity dispersion of the gas, which is limited by the MUSE resolution of about 70 \kms and therefore appears saturated around this value. \ha shows enhanced velocity dispersion values mainly at the center, where the Gaussian FWHM exceeds 300 \kms. Here we expect residual contamination from the BLR \citep[e.g.][]{guolopereira2021}. 
%after the PSF subtraction. 
\ha shows  $\sigma_{\rm gas}$ up to 250 \kms in the north-western cone where the average velocity is mainly redshifted. [O III] shows large dispersion with $\sigma_{\rm gas}\sim 250$ \kms both at the nucleus and in the ionisation cones. \hb shows enhanced $\sigma_{\rm gas}$ only at the AGN position.

\begin{figure*}[ht]
 \resizebox{\hsize}{!}{\includegraphics{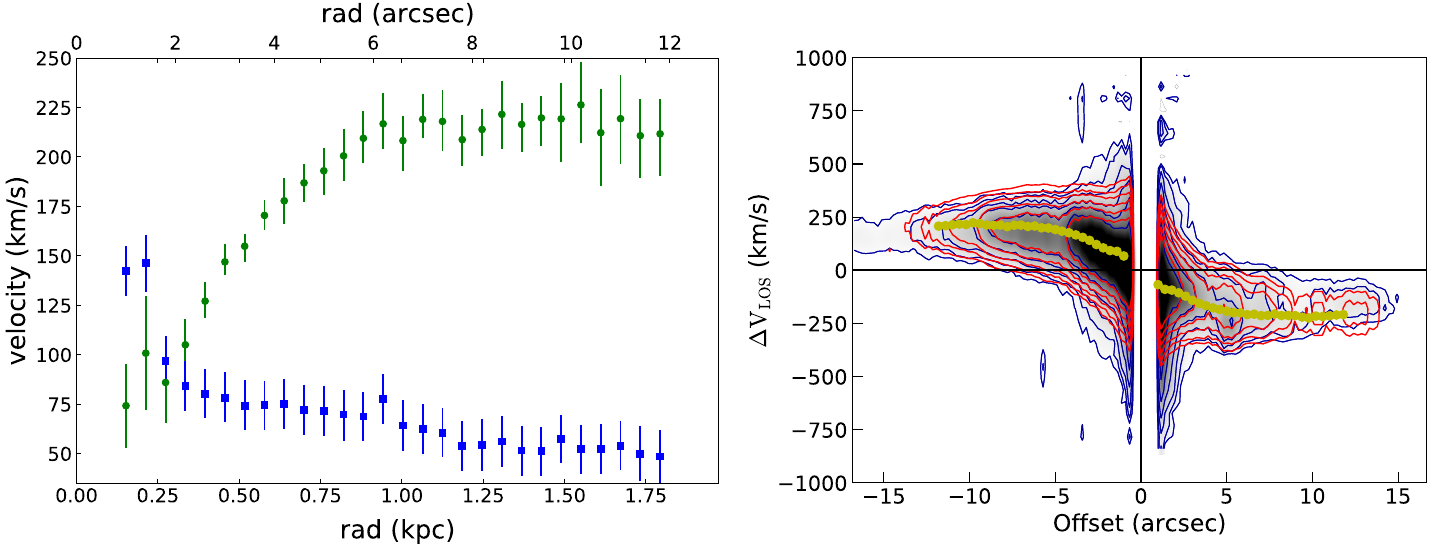}}
 \caption{$^{3D}$BAROLO dynamical model of the ionised gas disk obtained by fitting the \ha data cube. Left panel: rotation curve (green symbols) and velocity dispersion (blue symbols), corrected for beam smearing, as a function of the radius obtained from the best-fit disk model. Right panel: position-velocity diagram along the kinematic major axis, PA = 210 deg. Red contours and yellow filled circles represent the disk model, blue contours represent the data. Contours are drawn at (1, 2, 4, 8, 16, 32, 64)$\sigma$. }
  \label{fig:barolo-ionised}
\end{figure*} 

\subsection{Ionised gas disk}\label{ioniseddisk}

\noindent We built a 3D tilted-ring model of the ionised gas disk by fitting the \ha data cube with $^{3D}$BAROLO,
%, allowing three parameters to vary: rotation velocity, velocity dispersion and systemic velocity. We fixed the kinematic centre to the AGN position (RA = 9:45:41.943 DEC = -14:19:34.584). The inclination and the position angle are fixed to the output values found for molecular gas disk, $i = 80$ deg and PA $= 210$ deg. We apply a 3$\sigma$ threshold for the detection and 
following the same approach adopted for CO(2-1), fixing the size of the annuli to 0.4 arcsec, and excluding the central region of radius 0.6 arcsec, that is dominated by the residual contamination from the BLR. Figure \ref{fig:barolo-ionised} shows the \ha rotation curve (90 \kms at the center, increasing to 220 \kms at 1.5 kpc). The velocity dispersion decreases from $\sim$120 \kms near the nucleus down to $\sim$55 \kms at larger radii. The \ha rotation curve is globally consistent with the CO(2-1) one, but it extends further out to 1.8 kpc.
We derived the dynamical mass as a function of the radius, $M_{\rm dyn} = rv^2_{\rm rot}/2$G and we found a total dynamical mass of $\sim 10^{10} \, \rm M_{\odot}$ enclosed in a 1.5 kpc region.
Figure \ref{fig:barolo-ionised} (right panel) shows the position velocity diagram along the kinematic major axis (PA=210 deg). The disk model well describes the data at projected radii $>4$ arcsec, while the nuclear region shows gas with very high red(blue)-shifted velocity up 500 \kms (-750 \kms). This is not due to the BLR as the latter is unresolved, but rather traces the ionised wind. The same pattern is seen also in the PV diagram taken for the [O III]$\lambda 5007 \AA$ emission line (Figure \ref{fig:ionisedoutflow}). 
%The ionised outflow will be discussed in Sec. \ref{sec:ionisedof}.

\subsection{Ionised gas densities and masses}\label{sec:ionprop}

\begin{figure*}[ht] 
\resizebox{\hsize}{!}{\includegraphics{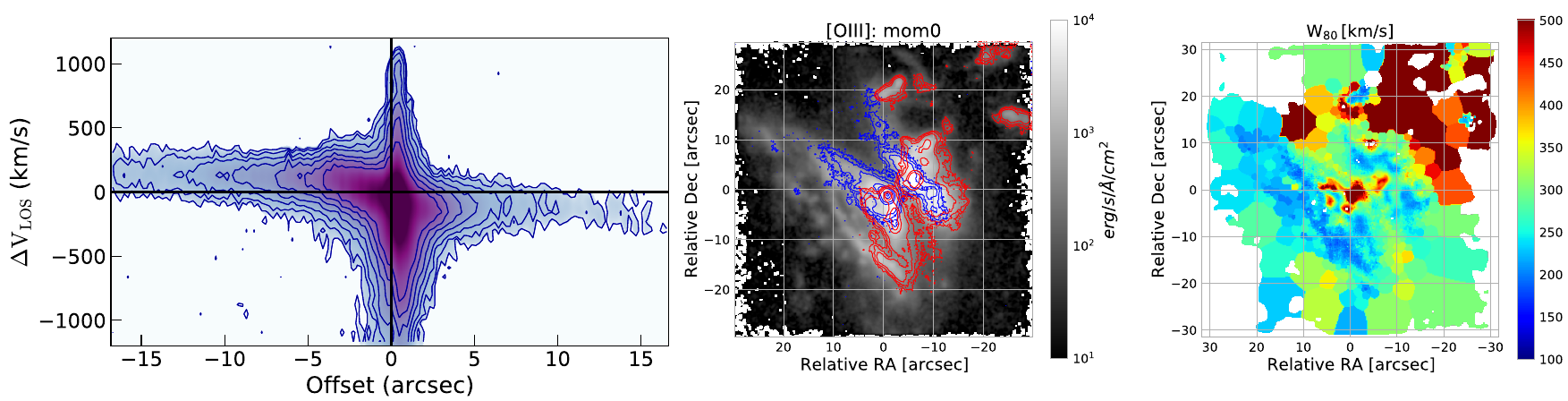}}
 \caption{Left panel: position-velocity diagram of the [O III] emission along the kinematic major axis, PA = 210 deg. Central panel: [O III] intensity map (grey scale), red and blue contours show the high velocity [O III] emission between [200, 1000] and [-1000,-200] \kms, respectively. Right panel: map of the $W_{80}$ of the [O III] line.}
  \label{fig:ionisedoutflow}
\end{figure*}

To assess the ionised gas contribution to the global baryon budget, we calculate the gas mass in the warm ionised phase from the luminosity of [O III], assuming recombination in a fully ionised medium with a gas temperature of $10^{4}$ K and by using the following equation \citep{carniani2015,bischetti2017}:
\begin{equation}\label{eq:ionisedgasmass}
    M_{\rm [O III]} = 0.8 \times 10^8 M_{\odot} \bigg( \frac{C}{10
^{{\rm [O/H]}-{\rm [O/H]}_{\odot}}} \bigg) \bigg( \frac{L_{\rm [O III]}}{10^{44}} \bigg)  \bigg( \frac{n_{\rm e}}{500} \bigg)^{-1}
\end{equation}
where [O/H] is the metallicity, $n_{\rm e}$ is the electron density in unit of $\rm cm^{-3}$, and  $L_{\rm [O III]}$ is the [O III] luminosity in $\rm erg \, s^{-1}$. 
We correct $L_{\rm [O III]}$ for dust extinction, because the dust lane seen in the HST image (Figure \ref{fig:gasproperties} left panel) is expected to produce significant attenuation along the disk \citep{trippe2008}.
We calculate the extinction map in V band, $A_V$, through the Balmer decrement H${\alpha}/H{\beta}$, assuming a \citet{calzetti2000} attenuation law.
\newline We apply a Voronoi tessellation \citep{cappellaricopin2003} based on H${\beta}$, that is the faintest optical line, to derive both the extinction and the ionization parameter (see below), requiring a SNR $\ge$10 per wavelength channel in each cell.\footnote{The noise is computed as the r.m.s. extracted from the \hb cube masking the lines emission. An initial smoothed 3D mask is applied, obtained by including the voxels with emission above 1 r.m.s. for at least 3 connected spectral channels}. To account for the outer regions, where H$\beta$ is undetected but [O III] is detected, we calculate upper limits for the H${\beta}$ flux. In the following we derive all parameters in Eq. \ref{eq:ionisedgasmass} using this Voronoi tassellation, except for the  metallicity, assumed constant and equal to the solar value. 
\newline The highest values of $A_V$ are found at the dust lane and inclined disk, while the ionisation cones show modest reddening, on average $A_V < 1$ (Figure \ref{fig:gasproperties} left-central panel).
We calculate the electron density (Figure \ref{fig:gasproperties}, central-right panel) using the relation by \citet{baronnetezer2019}:
\begin{align}
    n_{\rm e} = & 3.2 \frac{L_{\rm bol}}{10^{45} {\rm erg \, s^{-1}}} \biggl(\frac{r}{1\ {\rm kpc}}\biggr)^{-2} \frac{1}{U}~ \rm cm^{-3}
\end{align}
\noindent
where $L_{\rm bol}$ is the bolometric luminosity ($10^{44.13}~ \rm erg \, s^{-1}$) and $U$ is the ionisation parameter derived using the [O III]/$\rm H{\beta}$ and [NII]/$\rm H{\alpha}$ line ratios, see Appendix \ref{appendixc}.
The typical uncertainty associated with $n_{\rm e}$ is 0.6 dex  \citep{baronnetezer2019}. The highest $n_{\rm e}\sim 10^3-10^4 \, \rm cm^{-3}$ is found in the nuclear region. Further out across the disk $n_{\rm e}$ is $100-1000 ~ \rm cm^{-3}$, while in the ionisation cones  $n_e$ is  $\sim 10-100~ \rm cm^{-3}$.
By applying equation \ref{eq:ionisedgasmass} we derive the mass surface density of ionised gas (Figure \ref{fig:ionisedoutflow}, right panel). We derive a total ionised gas mass $M_{\rm gas,ion} = (4.5 \pm 0.3)\times \rm 10^7 \, M_{\odot}$, of which $M_{\rm gas, ion} = (1.27 \pm 0.03) \rm 10^7 \, M_{\odot}$ (or 27\% of the total) is in the host galaxy disk, while the remaining fraction is located off the galactic plane, mainly in the ionisation cones.

\subsection{Ionised wind}\label{sec:ionisedof}

\begin{table*}
     \caption[]{Properties of the ionised gas wind derived from [O III] emission.}
         \label{tab:ionisedof}
\centering    
\label{table:ionisedof}
\begin{tabular}{l c c c c c c c c}        
\hline\hline                 
 & $L_{\rm [O III]}$ & $A_V$  & $n_{\rm e}$ & $M_{\rm of}$ & $r_{\rm of}$ & $v_{\rm of}$ & $\dot{M}_{\rm of}$ & $\dot{E}_{\rm kin}$\\
 & [$\rm 10^{42} erg \, s^{-1}$] &  & [$\rm cm^{-3}$] & [$\rm 10^{6} M_{\odot}$] & [$\rm kpc$] & [\kms] & [$\rm M_{\odot} \, yr^{-1}$] & [$\rm 10^{40} erg \, s^{-1}$] \\
  & (a) & (b) & (c) & (d) & (e) & (f)  & (g) & (h) \\
\hline 
Nucleus (N) & 22.9$\pm$3 & 3.89 & 2325 & 0.39$\pm$0.05 & 0.3 & -895$\pm$5 & 3.6$\pm$0.5 & 91$\pm$13 \\
West Cone 1 (W1) & 6.2$\pm$1 & 1.38 & 259 & 10$\pm$2 & 1.5 & 217$\pm$5  & 4.2$\pm$0.7 & 7$\pm$1\\
West Cone 2  (W2)& 1.1$\pm$0.2 & - & 54 & 8$\pm$1 & 6.7 & 217$\pm$5  & 0.8$\pm$0.2 & 1.3$\pm$0.4\\
East Cone 1 (E1) & 2.2$\pm$0.3 & 1.49 & 176 & 5.0$\pm$0.7 & 1.5 & -375$\pm$5  & 3.8$\pm$0.5 & 16.8$\pm$3\\
East Cone 2 (E2) & 0.5$\pm$0.1 & - & 22 & 9$\pm$2 & 5.6 & -227$\pm$5  & 1.1$\pm$0.2 & 1.8$\pm$0.6\\ 
\hline  
\end{tabular}\\
  \flushleft 
 \footnotesize{ {\bf Notes.} The table reports the main properties of the outflowing ionised gas detected through [O III] emission line in different regions: N = nuclear region; W1 (W2) = west cone at $\rm r < 1.5  kpc$ ($\rm r > 1.5\  kpc$); E1 (E2) = east cone at $\rm r < 1.5  kpc$ ($\rm r > 1.5\  kpc$). (a) Extinction-corrected [O III] luminosity (b) Extinction, assumed negligible in W2 and E2, (c) Electron density with a typical uncertainty of 0.6 dex \citep{baronnetezer2019}, (d) wind mass, (e) wind radius, (f) wind velocity defined as $v_{98}$, (g) outflow rate at radius $r_{\rm of}$, (h) outflow kinetic power.}
\end{table*}

In section \ref{ioniseddisk} we showed that, in addition to disk rotation and BLR, H$\alpha$ kinematics traces a wide opening-angle ionised wind. Here we examine the [O III] emission. This is not contaminated by BLR and, therefore, best traces the ionised wind. The [O III] PV plot (Figure \ref{fig:ionisedoutflow}) shows that the ionised gas disk is detected also in [O III]. The [O III] wind is detected in the nuclear region, where its (projected) velocity exceeds $v_{\rm [O III]}= | 1000 |$ \kms, and in the ionisation cones, where $v_{\rm [O III]}$ gradually decreases with increasing distance from the nucleus. 
A representation of the [O III] kinematics within the ionisation cones is given in the middle panel of Figure \ref{fig:ionisedoutflow}, where we plot red and blue contours, tracing the integrated [O III] emission in the velocity ranges of [200,1000] \kms and [-1000,-200] \kms respectively. These contours highlight ionised gas moving at high velocities, both approaching and receding, within the cones. In several positions, the gas has both an approaching and receding component, likely tracing the walls of the wind cone filled by clumpy gas along the line of sight \citep{veilleux2001,friedrich2010,mingozzi2019,guolopereira2021}.
Figure \ref{fig:ionisedoutflow}, right panel, shows the [O III] $W_{80}$ map  \footnote{ $W_{80} \, = v_{90} \, - v_{10}$ represents the width of the line that contains 80\% of the wind flux, calculated as the difference between the 90th and 10th percentiles velocities.}, a proxy of the gas turbulence, derived using the Voronoi tessellation (see Sec. \ref{sec:ionprop}). The ionised gas distribution shows regions of enhanced $W_{80}$, reaching up to 500 \kms both at nucleus and in the cones out to kpc distance from the center, highlighting the presence of broad components due to the outflowing gas.

\begin{figure}[ht]
 \resizebox{\hsize}{!}{ \includegraphics{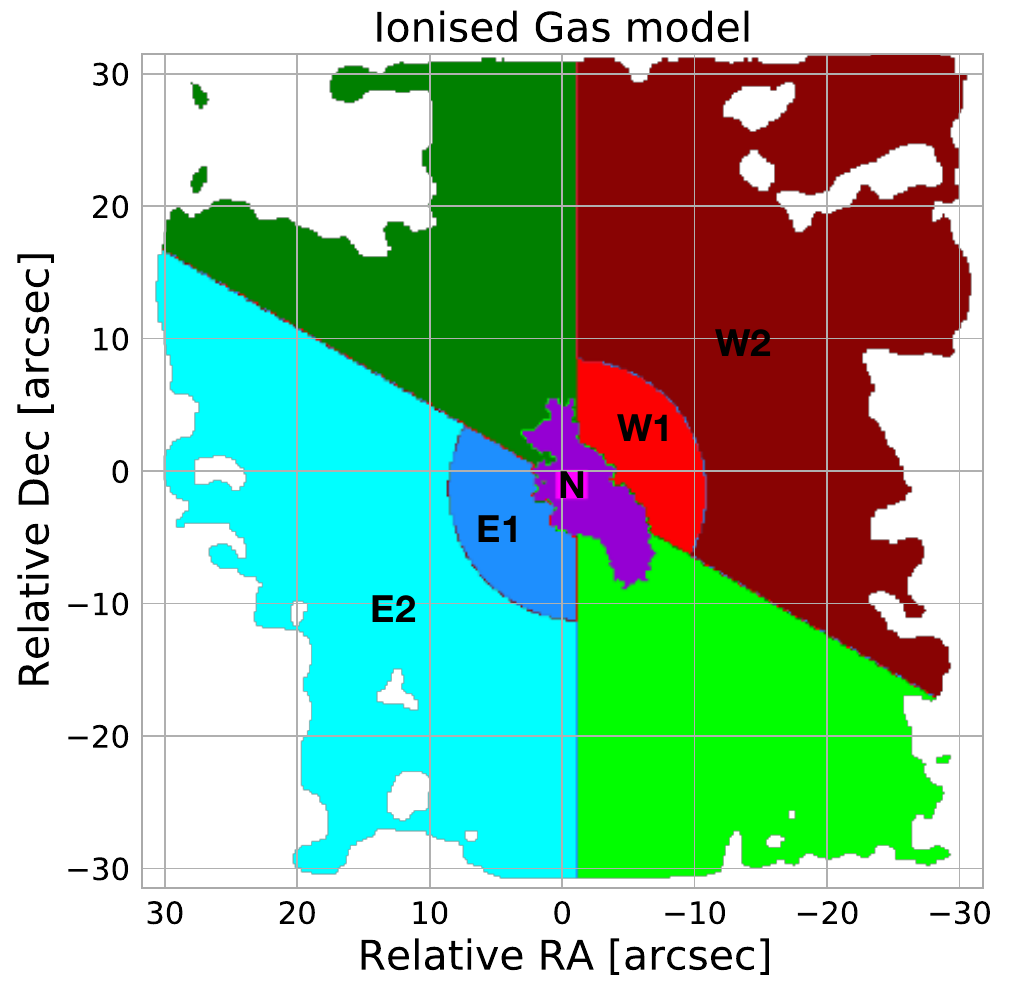}}
\resizebox{\hsize}{!}{\includegraphics{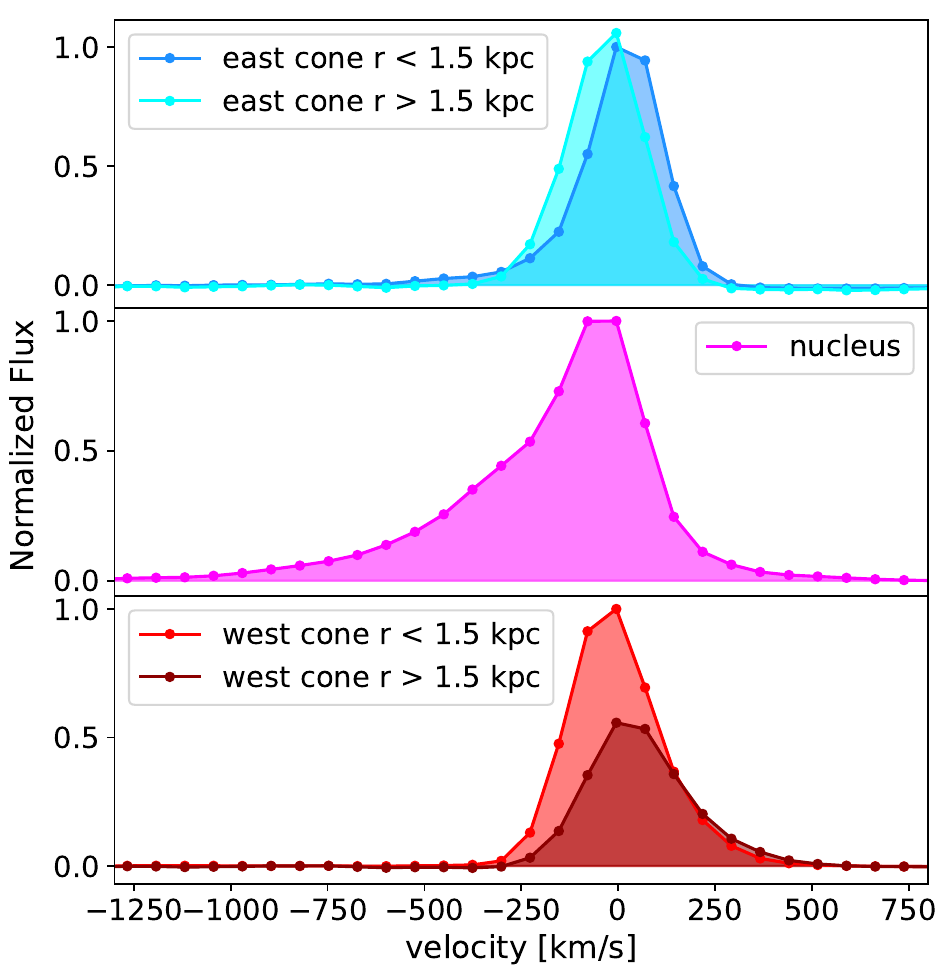}}
\caption{Top panel: simplified model of the source, based on the ionised gas kinematics. Magenta = nucleus (N); Purple = inner disk; Light/dark green: outer disk;  Red/Burgundy = west ionisation cone (W1/W2); Blue/Cyan = east ionisation cone (E1/E2). Bottom panel: normalised [O III] line profiles at the wind regions, spectra are coloured accordingly to the model map in left panel.}
  \label{fig:outflow}
\end{figure}

\section{Discussion}
\label{sec:discussion}

    \subsection{The multiphase disk}  
    The cold molecular and ionised gas disks have approximately the same size, 1.5 and 1.8 kpc (radius), respectively. In CO(2-1) we resolve a disk and an inner CNR, the latter with an inner radius of 65 pc. The rotation velocity of the cold molecular disk is $v_{\rm rot,mol}=260\pm 10$ \kms, that of the ionised disk is $v_{\rm rot}=216\pm 35$ \kms, suggesting that the cold molecular and ionised disks are co-spatial and trace the same kinematics.
    The velocity dispersion of the cold molecular gas within the disk (from the highest resolution data) is $16\pm10$ \kms, which agrees with the average value found for local SF galaxies \citep[][and references therein]{ubler2019}. For the ionised component we find $\sigma_{\rm gas}=55\pm35$ \kms, which is affected by the limited spectral resolution of MUSE ($\sim$70 \kms) and therefore cannot be compared with the average value in local SF galaxies \citep[$20$ \kms, e.g.][]{epinat2010}. 
    Conversely, in the CNR, we find that $\sigma_{\rm gas}$ is enhanced with respect to local star-forming galaxies \citep{ubler2019}, $\sigma_{\rm gas}=38\pm12$ \kms for the cold molecular and $117\pm36$ \kms for the ionised phases, respectively. These velocity dispersions are a factor of 3-4 larger than those found in the disks of SF galaxies \citep{ubler2019}.  The [O III] line, associated to the ionised component, in this location is broadened by the effect of the ionised wind.
    We conclude therefore that we find enhanced $\sigma_{\rm gas}$ only in the central kpc of NGC 2992, where the wind interacts with the ISM, but not on larger scales across the host galaxy disk (Figure \ref{barolomodel} and \ref{fig:barolo-ionised}). 
   The dynamical ratio for the cold molecular component is $v_{\rm rot}/\sigma_{\rm gas}=10$ at 1 kpc radii, consistent with a rotationally supported disk. At EDGE1 and EDGE2-TAIL, we measure $v_{\rm rot}/\sigma_{\rm gas}\sim 5.4$. For the cold molecular phase we do not find dynamical rations smaller than unity, which are observed in mergers \citep{cappellari2017}. However, we do find dynamical ratios smaller than unity in the ionised gas phase, which at nucleus shows  $v_{\rm rot}/\sigma_{\rm gas}\sim0.8$, likely owing to the effect of the fast ionised wind detected in this region.

    %%%%% NON CANCELLARE
   % \red{\citet{garcia-bernete2015} derived from SED fitting: for the circumnuclear region $M_{dust}$ = (7.6 $\pm$ 1.3) $10^6$ \msun ($T_{dust}$ = 29 $\pm$ 1 K) and SFR = (2.5 $\pm$ 0.4) \sfr, for the disk $M_{dust}$ = (19.6 $\pm$ 2.7) $10^6$ \msun ($T_{dust}$ = 21 $\pm$ 1 K) and SFR = (0.7 $\pm$ 0.1) \sfr.}

    \subsection{The multiphase wind} 
    
    The projected velocity of the ionised wind reaches its maximum ($v_{98}$ about $-900$ \kms) close to the nucleus, at the apex of the eastern cone, where the [O III] profile shows a prominent asymmetric blueshifted wing exceeding $-1000$ \kms (Figure \ref{fig:outflow}). At this side of the galaxy, the sightline reaches the nucleus and the BLR, seen through broad \ha emission \citep{veilleux2001,guolopereira2021}. $v_{98}$ decreases further out in the ionisation cones, reaching few hundred \kms (Figure \ref{fig:outflow}). The latter values of projected velocity in the outer ionisation cones are in agreement with those reported  for \ha and NIR emission lines \citep{veilleux2001,guolopereira2021,friedrich2010}, which however did not detect the very fast [O III] component with velocity of $-1000$ \kms. Owing to the wide opening angle of the cones and the projection effects (the cones orientation is mainly on the plane of the sky), these [O III] velocity estimates represent a lower limit. 
    
     In order to derive the ionised wind mass and mass loss rate we built a simplified source model, where we identified regions with similar kinematic properties, and analysed them individually. As the NGC 2992 disk is viewed close to edge-on and the ionisation cones are close to the plane of the sky, we can separate disk and wind relatively easily. In the source model (Figure \ref{fig:outflow}), we considered a disk inclination of 80 deg based on the best fit dynamical model, and an aperture of 120 deg for the cones.
We identified 8 regions that are the nucleus (including BLR), the galactic disk, which in turn has a high density portion around the nucleus at r $<$ 1 kpc, and two lower $n_{\rm e}$ portions further out on the north and south side (green regions in Figure). 
%The west ionisation cone is behind the galactic disk, whereas the east cone is in front of the disk, with respect to the observer. 
Each ionisation cone is further divided into two sub-regions, the inner and the outer cones, whose boundary is set at $r = 1.5$ kpc (Figure \ref{fig:gasproperties}). 
The [O III] line at nucleus (N, magenta region) is characterised by a very broad, asymmetric profile, with a blue-shifted wing exceeding -1000 \kms from the rest frame ($v_{98}$ \footnote{$v_{98}$ is defined as the wind velocity enclosing the 98$\%$ of the cumulative velocity distribution of the outflowing gas.} $=-895\pm5$ \kms, Table \ref{tab:ionisedof}), that traces the nuclear ionised wind. 
The disk inclination of 80 deg, allows a line of sight to the nucleus, whereas the rear part of the nuclear wind is obscured by the dusty disk. In the east cone the [O III] profile is mainly blueshifted decreasing from $v_{98}\sim-375$ \kms (E1) to $v_{98}\sim-227$ \kms (E2).
While in the west cone, accordingly to the [O III] velocity map of Figure \ref{fig:Cubexmoments}, the [O III] profile is mainly redshifted with $v_{98}=217\pm5$ \kms (W1, W2).
We used eq. \ref{eq:ionisedgasmass} to derive the mass in the ionised wind. The extinction is assumed negligible in the outer part of the ionisation cones, where no dust lane is detected (see left panel of Figure \ref{fig:gasproperties}). 
We calculated the mass outflow rate in each region using eq. \ref{eq1}, where the wind radius is the outer radius of each region and the wind velocity is the maximum in absolute value between $v_{02}$ \footnote{$v_{02}$ is defined as the wind velocity enclosing the 2$\%$ of the cumulative velocity distribution of the outflowing gas, analogous to $v_{98}$ but on the blue-shifted side of the line profile.} and $v_{98}$. The wind velocity so derived is a lower limit owing to the low inclination of the ionisation cones.
Table \ref{table:ionisedof} summarizes the [O III] wind parameters. 
We derived a total ionised wind mass of $M_{\rm of,ion} = (3.2\pm 0.3) \times  \, 10^7 \, \rm M_{\odot}$, and a total ionised outflow rate of $\dot{M}_{\rm of, ion} = (13.5 \pm 1) \rm M_{\odot} \, yr^{-1}$.
By using \hb and the same procedure, we found a ionised gas mass of $6.4 \times 10^7 \, \rm M_{\odot}$, that is a factor of 2 larger than the mass estimated from [O III], in agreement with previous results \citep{liu2013, harrison2014, carniani2015}. This indicates that the wind mass based on [O III] is a robust lower limit.

    The total ionised wind mass is larger than the estimate reported by \citet{veilleux2001}. The discrepancy is likely due to the assumption of uniform electron density (100 cm$^{-3}$) in \citet{veilleux2001}, and to the fact that they did not consider the very fast nuclear wind component.   
   Most studies assume a constant $n_{\rm e}$ \citep[e.g.][]{veilleux2001,liu2013b,husemann2016,tozzi2021}, or use the ratio of [S II] emission lines to derive the electron density \citep[e.g.][]{nesvadba2006, nesvadba2008,mingozzi2019,marasco2020,kakkad2022}.
Typical [SII]-based estimates are in the range 200-1000 $\rm cm^{-3}$, but this method underestimates the true densities by one to two orders of magnitude \citep{baronnetezer2019, revalski2022}.
Moreover the combination of [O III] or \ha line luminosity with [SII]-based electron density may result in inconsistent gas mass estimates, because [O III] and \ha emission lines are emitted throughout most of the ionised cloud, while the [SII] lines are primarily emitted close to the ionization front in the cloud. Thus, the [O III] and \ha lines trace higher electron density regions, compared to [SII]. 
Based on the \citet{baronnetezer2019} prescription, we derived a spatially resolved $n_{\rm e}$, which spans a broader density range, 100-$10^4$ cm$^{-3}$ (Figure \ref{fig:gasproperties}), compared to \citet{mingozzi2019,kakkad2022}. Based on these $n_{\rm e}$ and the [O III] kinematics, we derived ionised outflow rates at three characteristic radii, 300 pc, 1.5 and $\sim6$ kpc (Table \ref{table:ionisedof}).
 At radius 300 pc we found a outflow rate $\dot M_{\rm of}=3.6\pm0.5$ \sfr, which at face value is consistent with the one derived by \citet{guolopereira2021} based on \ha. However, both our estimates of electron density and of wind velocity are an order of magnitude and a factor of 5 larger, respectively, than in \citet{guolopereira2021}, suggesting a faster and tenuous wind in the nuclear region. This cannot be probed by \ha due to BLR contamination.
 If we examine the large scale ionised wind through the cones, the [O III] wind velocity is consistent with that reported by \citet{veilleux2001} for \ha. For an electron density of 40 $\rm cm^{-3}$ the [O III] outflow rate, $3.6\pm0.5$ \sfr (Table \ref{table:ionisedof}),is in agreement with that derived by \citet{veilleux2001}.
  The ionised wind is certainly AGN driven as demonstrated by multiple authors \citep[e.g.][]{veilleux2001,friedrich2010}, and multiphase on kpc scales, as both the ionised and the cold molecular phases participate in the wind. 

  Near the northern cone wall we detected a region with high velocity dispersion in CO that shows a wind component (EDGE1, Figure \ref{fig:molperturbations}). At this location both CO and [O III] trace a multiphase wind with approximately the same velocity (about -50 \kms), and a cold molecular mass of $2.7\times 10^7$ \msun, with [O III] showing a larger $\sigma_{\rm gas}$ than CO. A similar configuration is found in the southern cone (EDGE2-TAIL), but here the kinematics is complicated by a tidal tail connecting to NGC 2993.  
  
  Ten cold molecular outflowing clumps are detected below the disk (in the eastern side) out to  projected distances of $\sim 0.8$ kpc (Table \ref{table:molof}), and above the disk (western side, Figure \ref{fig:clumps} ). We did not detect cold molecular gas at projected distances larger than $\sim$ kpc from the disk.
    The velocity of the cold molecular clumps is of the order tens \kms, reaching a maximum of -200 \kms (clump C2). Their line width (FWZI) appears very narrow for some clumps (40 \kms), while for other clumps FWZI is in the range 144 to 288 \kms, with average $\rm FWZI=110$ \kms. This suggests that this is disk material that has been entrained in the ionised wind and lifted up. For this reason we adopted for the cold molecular wind the same conversion factor of the disk, obtaining a cold molecular wind mass of $M_{\rm of,mol}=4.3 \times 10^7$ \msun, that is about twice as massive as the ionised mass on the same scale (1.5 kpc radius, see Table \ref{table:ionisedof}).
     Should the conversion factor be lower, as in the case of optically thin outflows \citep{morganti2015,dasyra2016}, $\alpha_{\rm CO}\,= 0.3 \,\mathrm{M_{\odot} (K\, km \, s^{-1} \,pc^{2})^{-1}}$. This would translate into a $\times 10$ smaller cold molecular gas mass in outflow, with $M_{\rm of,mol}= 0.13 \times 10^7$ \msun
     in the clumps, and $M_{\rm of,mol}= 0.21 \times 10^7$ \msun in EDGE1, with a total cold molecular gas mass of $M_{\rm of,mol}= 0.34 \times 10^7$ \msun.
 
  In the inner part of the cones (W1 and E1) the ionised wind is faster ($v_{\rm of} \sim  -220 $ \kms and 375 \kms) then the cold molecular one, which spans a slower velocity range on these scales (1-1.5 kpc radius, Table \ref{tab:molof}). Therefore at the kpc scale the multiphase wind has a fast ionised component, and a slower cold molecular one which carries the bulk of the mass. 
On scales larger than kpc, the wind is only in the ionised phase.
Comparing the results of this work with the correlation between the mass outflow rate and the AGN bolometric found by \citet{fiore2017}, we note that the ionized and X-ray winds are broadly consistent with the correlations, whereas the cold molecular outflow rate is significantly below the best-fit correlation for cold molecular winds. This result is consistent with other cold molecular outflows in nearby objects \citep[e.g.][]{feruglio2020, zanchettin2021, ramosalmeida2022, lamperti2022}.
%(e.g.  Zanchettin+21 for Mrk509, Feruglio+20 for ESO428-G14, Lamperti+22 for a sample of  ultra-luminous infrared galaxy systems and Ramos-Almeida+22 for a sample of radio-quiet type-2 quasars). 
The correlation found by \citet{fiore2017} could be biased towards positive outflow detections and, therefore, to high values of outflow rate, because derived for a sample of high luminous objects. This highlights the importance of performing statistical studies on unbiased AGN samples to derive scaling relations.
\newline By comparing the energetic of the UFO and the nuclear portion of the [O III] wind, we find that the momentum boost $\dot{P}_{\rm of}/\dot{P}_{\rm rad}$\footnote{The wind momentum boost is the ratio between the outflow momentum flux, $\dot{P}_{\rm of} = \dot{M}_{\rm of} \cdot v_{98}$, and the radiative momentum flux, $\dot{P}_{\rm rad} = L_{\rm bol}/$c.}, is of the order $10^{-2}$, smaller than what expected for a radiatively driven momentum conserving wind. The [O III] wind that we observe on scales of 300 pc may be related to a previous UFO episode, given the different timescales implied by the UFO and [O III] wind kinematics.

%  By adopting our standard disk conversion factor, and summing all components, the total molecular outflow rate is $\rm \dot{M}_{OF} \sim 3.29$ \sfr, and energy rate $\dot{E}_{kin} = 1.79 \times \rm 10^{40}$ \ergs. The molecular outflow rate is therefore smaller than the ionised one.

\begin{figure}[ht]
\resizebox{\hsize}{!}{\includegraphics{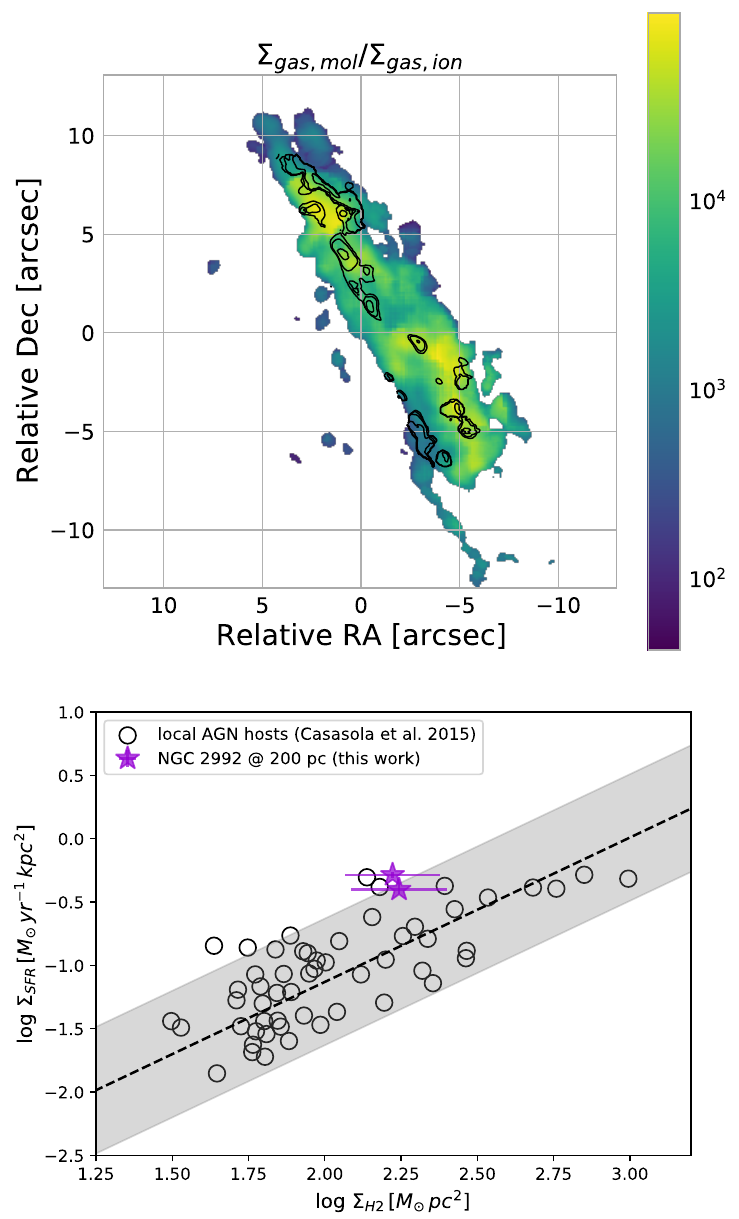}}
 \caption{Upper panel: map of the cold molecular to ionised ([O III])} gas surface densities ratio. The CO map has been degraded to the same angular resolution of the [O III] map. Black contours are from the residual velocity dispersion map (right panel Figure \ref{fig:COres}) and are drawn at (5, 10, 20, 40) \kms.
 Bottom panel: the spatially resolved star formation law on scales of $\sim200$ pc. Purple symbols are the two sides of the disk where radio continuum is detected (this work). Black symbols are data derived for local AGN host galaxies, along with their best fit relation (dashed line) and dispersion (shaded area) from \citet{casasola2015}. 
 \label{fig:SK}
\end{figure}

    \subsection{The radio bubbles and their interplay with the ISM}
     The 8-shape radio feature with limb-brightened prominent loops has been interpreted as two expanding gas bubbles. \citet{chapman2000} suggested that a compact radio outflow may power the bubbles and produce the nuclear outflow seen in the NIR.  The radio outflow is unresolved in our VLA data and should be confined within 150 pc from the AGN. The power of the radio outflow would be 
      $P_{\rm of, radio} \approx 5.8\times10^{43}(P_{\rm radio}/10^{40})^{0.70}\rm erg \, s^{-1} =  6 \times 10^{42} \, \mathrm{erg \, s^{-1}}$, 
         by using the power in the radio bubbles and the empirical relation by \citet{cavagnolo2010} derived at 1.4 GHz\footnote{The 5 GHz radio power is consistent with the measurements at 1.4 and 8.4 GHz \citep{melendez2010}}. However this approach has large uncertainties and may be considered a lower limit to the outflow power \citep{morganti2021,mukherjee2018}, but such a outflow would be able to power the ionised wind seen on sub-kpc scales. 
  
\begin{figure*}[ht]
 \resizebox{\hsize}{!}{ \includegraphics{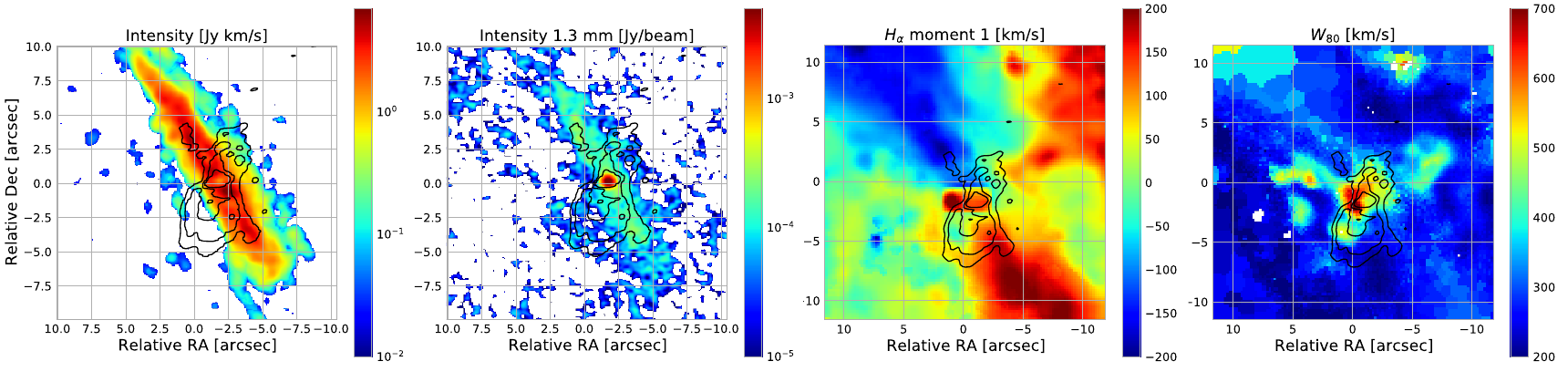}} 
 \caption{From left to right: CO(2-1) mean intensity map (same as Figure \ref{momentmaps} bottom-left panel), ALMA 1.3 mm continuum (same as Figure \ref{continuo} right panel), \ha moment 1 map (same as Figure \ref{fig:Cubexmoments} top-central panel) and [O III] $W_{80}$ map (same as Figure \ref{fig:ionisedoutflow} right panel). Black contours are from EVLA radio continuum map at 6 cm (Figure \ref{continuo} left panel), drawn at (1, 5, 10)$\sigma$. }
  \label{radio_contours}
\end{figure*}

    The radio bubbles are not parallel nor perpendicular to the disk, but probably have an intermediate PA so that are misaligned with respect to the ionisation cones seen in optical lines. For this reason it is unlikely that the ionised gas wind on large scales is directly related to the radio bubble and outflow \citep{veilleux2001}, but rather associated with previous AGN episodes. We found that the bubbles are spatially related to interesting ISM features (Figure \ref{radio_contours}). The outflowing cold molecular clumps are located at the edges of the southern bubble, and the 1.3 mm continuum emission from cold dust is detected outside the disk plane and coincident with the position of the southern bubble, at its southern rim. For the northern bubble the association with cold dust is less clear. This spatial overlap suggests that both cold gas and dust have been uplifted by the bubble. At the rim of the southern bubble, there is an arch of ionised gas (seen in \ha) with red-shifted velocity \citep[see also][]{guolopereira2021}, spatially coincident with dust emission. The rightmost panel of Figure \ref{radio_contours} shows that both bubbles overlap with regions where the ionised gas has large line width ($W_{80}$). These pieces of evidences indicate that the bubbles interact with the surrounding ISM including all gas phases and dust, and are physically related to the multiphase wind on sub-kpc scale. As proposed by \citet{xu2022}, it is possible that the hot gas probed by soft X-ray emission at the bubble location is heated by wind shocks \citep{zakamska2014,giroletti2017, zubovas2014, nims2015}.
    In order to verify whether there is evidence of shocks, we investigated the forbidden [OI]$\lambda 6003 \AA$ line, which is included in the MUSE data and may be used to trace shock-heated ISM.    
   High [OI]/\ha ratios (log[OI]$\lambda 6003 \AA$/\ha > - 1.6) are expected where shocks are present \citep{monrealibero2010, rich2011, rich2014, rich2015}. \citet{veilleux1987} suggest that shock heating should be considered as an alternative ionization mechanism in Low-Ionization Nuclear Emission-line Regions (LINERs).  
   \citet{monrealibero2006, monrealibero2010} also support the idea that a relative increase in the velocity dispersion and shift of the strong emission line ratios in the diagnostic BPT-like diagram toward the LINERs region is indicative of gas excited by slow shocks.
   We computed the [OI]/\ha ratio for NGC 2992 (see Appendix \ref{appendix_OI} and the right panel of Figure \ref{fig:OIHa}). We found a ratio in the range of -1.7 < log[OI]$\lambda 6003 \AA$/\ha < -0.8, which is consistent with values reported for Luminous infrared galaxies (LIRGs) and ultraluminous infrared galaxies (ULIRGs) \citep{veilleux1987, monrealibero2006, monrealibero2010, rich2011, rich2014, rich2015}. Inside the radio bubble, we found log[OI]/\ha $\sim$ -1.1, whereas towards the edge of the radio bubble, the value is about  log[OI]/\ha $\sim$ -1.4.  These values are  consistent with shock regions, and, together with the evidence of increased gas velocity dispersion within the radio bubble (Figure \ref{radio_contours}), support the scenario in which the hot gas in the bubble may be heated by wind shocks.   
   The shocks could be provided either by the fast [O III] wind \citep{zakamska2014}, that close to the nucleus exceeds $-1000$ \kms, or by the UFO, which has kinetic power in the range $7\times 10^{42}-1.1\times 10^{44}$ erg/s, depending on the state of the highly variable AGN \citep{marinucci2018}, or both.

   \subsection {The star formation law across the disk} 
    We computed the spatially resolved ratio between the cold molecular and ionised gas mass surface densities, $\Sigma_{\rm gas, mol}/\Sigma_{\rm gas, ion}$ (Figure  \ref{fig:SK}). The disk is dominated by a clumpy cold molecular distribution, while at EDGE1 and EDGE2-TAIL the cold molecular to ionised mass ratio decreases.
    From the 6 cm radio emission detected within the inclined disk, and by applying the radio - FIR correlation \citep{shao2018}, we derived a $\rm SFR = 0.4$ \sfr, where we considered only the radio emission located at the disk, and excluded the AGN and the radio bubble emission. This SFR is smaller than the value obtained from FIR SED decomposition \citep[$\rm SFR=3.6\pm0.3$ \sfr, ][]{gruppioni2016}. 
    Assuming that 6 cm emission arises from the same region where cold dust is detected through the 1.3 mm continuum (Figure \ref{continuo}), and scaling from the 1.3 mm flux, we would obtain a $\rm SFR = 1.13$ \sfr, which is still a factor of $\times 3$ smaller that the \citet{gruppioni2016} value. The discrepancies could be related to the sensitivity limit of the 6 cm observations, which probe a smaller region of the disk compared to both FIR Herschel \citep{gruppioni2016} and ALMA observations. 
    Based on the 6 cm radio continuum and the CO(2-1) emission, we derived $\Sigma_{\rm SFR} = 0.4$ and 0.5 \sfr $\rm kpc^{-2}$ and $\Sigma_{\rm H2}= 168$ and 175 \msun $\rm pc^{-2}$ (corrected for the disk inclination, i=80 deg), for the two sides of the disk, respectively, and we used them to study the spatially resolved cold molecular star formation law on scales of 200 pc.  
    Figure \ref{fig:SK} shows the star formation law on scales of 200 pc for NGC 2992 and a compilation of local AGN host galaxies from \citet{casasola2015}. We found that NGC 2992 is consistent with the correlation found for local AGN hosts. 
    
    \subsection{The cold dust} 
ALMA continuum data show a dust reservoir co-spatial with the galaxy cold molecular disk in a region that is 1.8 kpc$^2$ wide. A mid-IR extended emission at 11.2 $\rm \mu m$, likely produced by dust heated by SF, was detected on approximately the same scales \citep{garcia-bernete2015}.
    By assuming a grey body model with dust temperature $T_{\rm dust}=30$ K and $\beta=1.5$ \citep[e.g.][]{skibba2011}, we derived a cold dust mass of $M_{\rm dust}= (4.04 \pm 0.03)\times 10^6 \rm M_{\odot}$, consistent with the value obtained from {\it Herschel} unresolved data \citep{garcia-bernete2015}. We thus derived a gas-to-dust ratio GDR$\sim$80 across this region, broadly consistent with the Milky Way value \citep{bohlin1978} and with nearby galaxies with similar metallicity \citep{remy-ruyer2014,galliano2018, devis2017b,devis2019}.
    We argue that this dust reservoir is responsible for the extended Fe K$\alpha$ emission seen on 200 pc scales around the nucleus in hard X-rays and interpreted as reflection by cold dust by \citet{xu2022}. This seems a common feature of Compton thick AGN \citep[e.g.][]{feruglio2020}, and can be probed also in Compton thin AGN if they are in a low state, as is the case of NGC 2992 \citep{xu2022}.
    \citet{malizia2020} argued that NGC 2992 is an example of a Seyfert 1 disguised as a Seyfert 2 due to galactic obscuration from material located in the host galaxy on scales of hundreds of parsecs and not aligned with the putative absorbing torus of the AGN, therefore contributing significantly to the AGN obscuration.
    %therefore contributing to the total amount of column density. 
    Based on our $M(\rm H_2)$, we estimated a line-of-sight $\rm H_2$ column density of $\rm N(H_2) \approx 1.14 \times 10^{22} \, cm^{-2} $ toward the dusty disk, broadly consistent with the $N_{\rm H}$ derived in optical and X-rays \citep{yaqoob2007, malizia2012}, and in the range reported by \citet{malizia2020} for mis-classified type 1 AGN. In NGC 2992 the modest absorption is likely related to the dust on the host galaxy rather than to the dusty torus. This is in good agreement with the small torus radius of about 1.4 pc, typical of Seyfert 1 galaxies, derived for NGC 2992 by \citet{garcia-bernete2015}.

   % However these winds are launched in the innermost regions of the accretion disk and, despite their kinetic power, their feedback effect should have a small duty cycle \citep{fiore2017}.    
    %Therefore can derive the accretion rate necessary to power the AGN at the nucleus of NGC 2992, calculated as:
   %\begin{equation}
       %\dot{M}_{BH} = \frac{L_{bol}}{c^2 %\eta}.
    %\end{equation}
    %where $L_{bol}$ is the bolometric luminosity ($10^{44.13}~ \rm erg/s$), c is the speed of light and $\eta$ is the mass–energy conversion efficiency, which for Seyfert galaxies is usually assumed to be $\eta = 0.1$ \citep{frank2002}. We use these values to derive an accretion rate of $\dot{M}_{BH} \sim 0.02$ \sfr. This value is about 2 orders of magnitude lower then the mass outflow rates of the nuclear multiphase winds. This implies that most of the outflowing gas does not originate in the AGN, but in the surrounding ISM, leading to two scenarios. One in which the plasma bubbles are expanding, pushing gas away from the nuclear region as proposed by \citet{guolopereira2021}. Or the one in which the thermally dominated wind is diverted along the galaxy minor axis by the pressure gradient of the ISM and is powered on a sub-kpc scale by low-energy radio jets, as already suggested by \citet{veilleux2001}.

\section{Conclusions}
\label{sec:concl}

We presented an analysis of the multiphase gas in the disk and wind of NGC 2992, based on ALMA, VLA and VLT/MUSE data. 
We found that CO(2-1) and \ha emission lines trace an inclined (i=80 deg) multiphase disk with radius 1.5 and 1.8 kpc, respectively.
The rotation curves of the cold molecular and ionised phases are consistent, with $v_{\rm rot,mol}=260\pm 10$ \kms and $v_{\rm rot, ion}=216\pm 35$ \kms, therefore the cold molecular and ionised disks are co-spatial and trace the same kinematics.
The velocity dispersion of the cold molecular phase, $\sigma_{\rm gas}$, is consistent with that of SF galaxies at the same redshift \citep{ubler2019}, with the exception of some peculiar regions where the wind interacts with the host galaxy disk. 

We found a fast, clumpy ionised wind traced by [O III] within a wide opening angle bi-conical configuration, extending from the nucleus out to 7 kpc, with velocity exceeding 1000 \kms in central 600 pc region. 
Within the inner 600 pc region, and in the region between the ionised cone walls and the disk, the velocity dispersion $\sigma_{\rm gas}$ is increased by a factor of 3-4 for both the cold molecular and the warm ionised phase with respect than in SF galaxies, suggesting that a disk-wind interaction locally boosts the gas turbulence. 
In the ionisation cones on large scales the wind velocity is of the order of 200 \kms, in agreement with previous findings \citep{veilleux2001, friedrich2010}, but a large range of wind velocity is detected at any position within the cones, owing to line of sight effects. The electron density in the cones, $n_{\rm e}$, spans a range of 100-$10^4$ cm$^{-3}$, and implies a total ionised wind mass of $M_{\rm of,ion}= 3.2\times 10^7$ \msun, and a total ionised outflow rate of $\dot M_{\rm of,ion}=13.5\pm1$ \sfr.

On kpc scales we also detected a slower, clumpy cold molecular wind, with velocity $v_{98}$ up to 200 \kms, which carries a mass of $M_{\rm of,mol}=4.3\times 10^7$ \msun, that is about twice the ionised wind mass on the same physical scales. On scales above the $\sim$kpc, the wind is detected in only the ionised phase across the ionisation cones. 

In the central kpc, the multiphase wind is associated with the radio bubbles, as the cold molecular outflowing clumps and turbulent outflowing ionised gas are detected across the radio bubbles and at their limbs. As proposed by \citet{xu2022}, it is possible that the hot gas probed by soft X-ray emission filling the radio bubbles in NGC 2992 is heated by wind shocks, which could be produced either by the fast [O III] wind or by the UFO, or both \citep{zakamska2014}. The [OI]/\ha ratio and the gas velocity dispersion support the presence of shocks in and around the radio bubbles. 
The overall picture suggests a shocked wind heating the gas on scales up to 700 pc from the AGN, which in turn emits the soft X-rays forming the hot bubble. The ionised wind entrains the cold molecular gas and dust from the disk and lifts it up to $\sim$ kpc distance.
A compact $<150$ pc radio outflow, unresolved in current data, could also be the origin of the radio bubbles and of the multiphase wind. Higher resolution VLA data could shed light on the presence of a nuclear radio jet and its power. 

We detected a cold dust reservoir co-spatial with the cold molecular disk, with a dust mass $M_{\rm dust} = (4.04 \pm 0.03) \times \, 10^{6} \, M_{\odot}$ and a gas-to-dust ratio $\approx$ 80. We proposed that this dust reservoir is responsible for the extended Fe K$\alpha$ emission seen on 200 pc scales around the nucleus in hard X-rays and interpreted as reflection by cold dust.

\begin{acknowledgements}
We thank the anonymous referee for their insightful report that helped us improving the paper.
We thank A. Marconi and G. Venturi for helping in data analysis.
This paper makes use of the ALMA data from projects: ADS/JAO.ALMA$\#$2017.1.01439.S and 2017.1.00236.S. ALMA is a partnership of ESO (representing its member states), NSF (USA) and NINS (Japan), together with NRC (Canada), MOST and ASIAA (Taiwan), and KASI (Republic of Korea), in cooperation with the Republic of Chile. The Joint ALMA Observatory is operated by ESO, AUI/NRAO and NAOJ. The National Radio Astronomy Observatory is a facility of the National Science Foundation operated under cooperative agreement by Associated Universities, Inc. Based on observations collected at the European Southern Observatory under ESO program 094.B-0321. 
This research made use of observations made with the NASA/ESA Hubble Space Telescope, and obtained from the Hubble Legacy Archive, which is a collaboration between the Space Telescope Science Institute (STScI/NASA), the Space Telescope European Coordinating Facility (ST-ECF/ESAC/ESA) and the Canadian Astronomy Data Centre (CADC/NRC/CSA).
We acknowledge financial support from PRIN MIUR contract 2017PH3WAT, and PRIN MAIN STREAM INAF "Black hole winds and the baryon cycle".
Sebastiano Cantalupo and Andrea Travascio gratefully acknowledge support from the European Research Council (ERC) under the European Union’s Horizon 2020 research and innovation program grant agreement No 86436
\end{acknowledgements}

\bibliographystyle{aa} % style aa.bst
\bibliography{bibliografia} % your references Yourfile.bib
\begin{appendix}

\section{CO emission over different range of spatial frequencies }\label{appendixa}

\begin{figure*}[ht]
 \resizebox{\hsize}{!}{\includegraphics{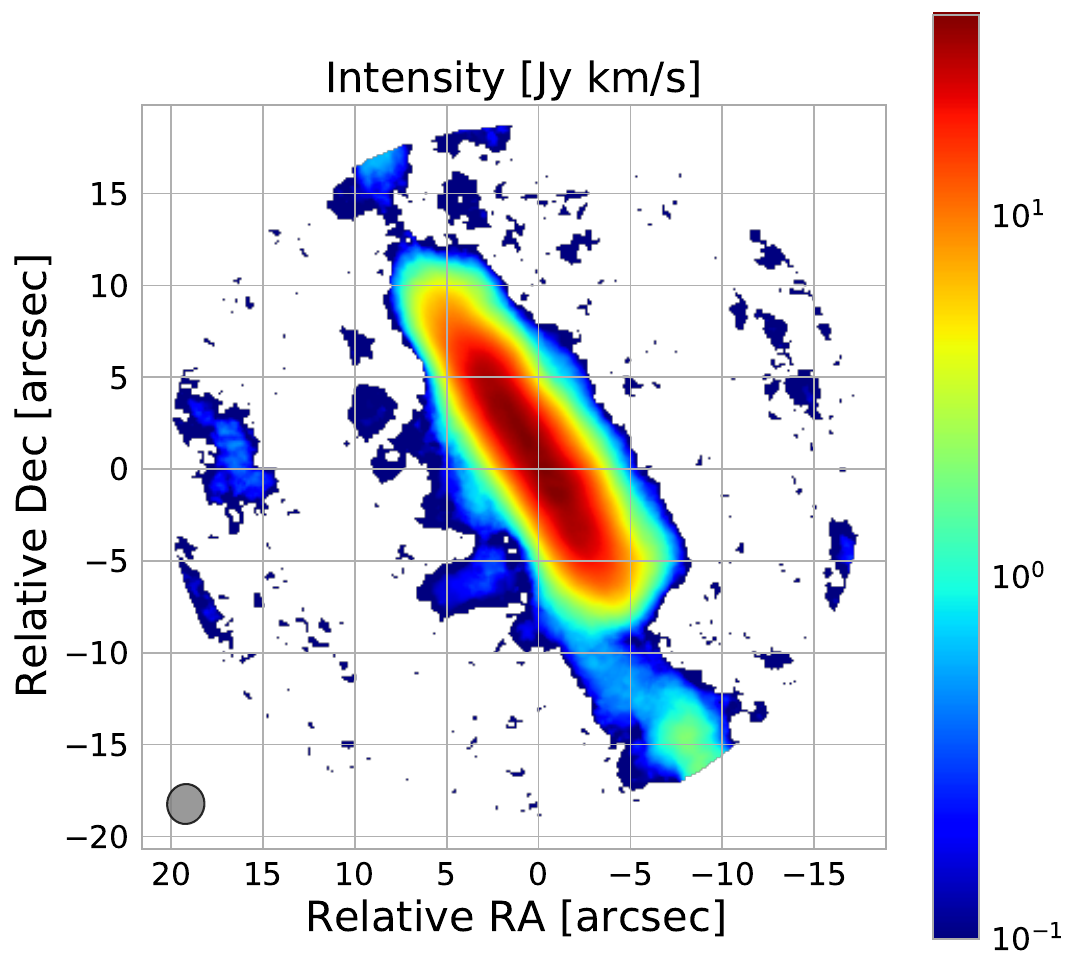} \includegraphics{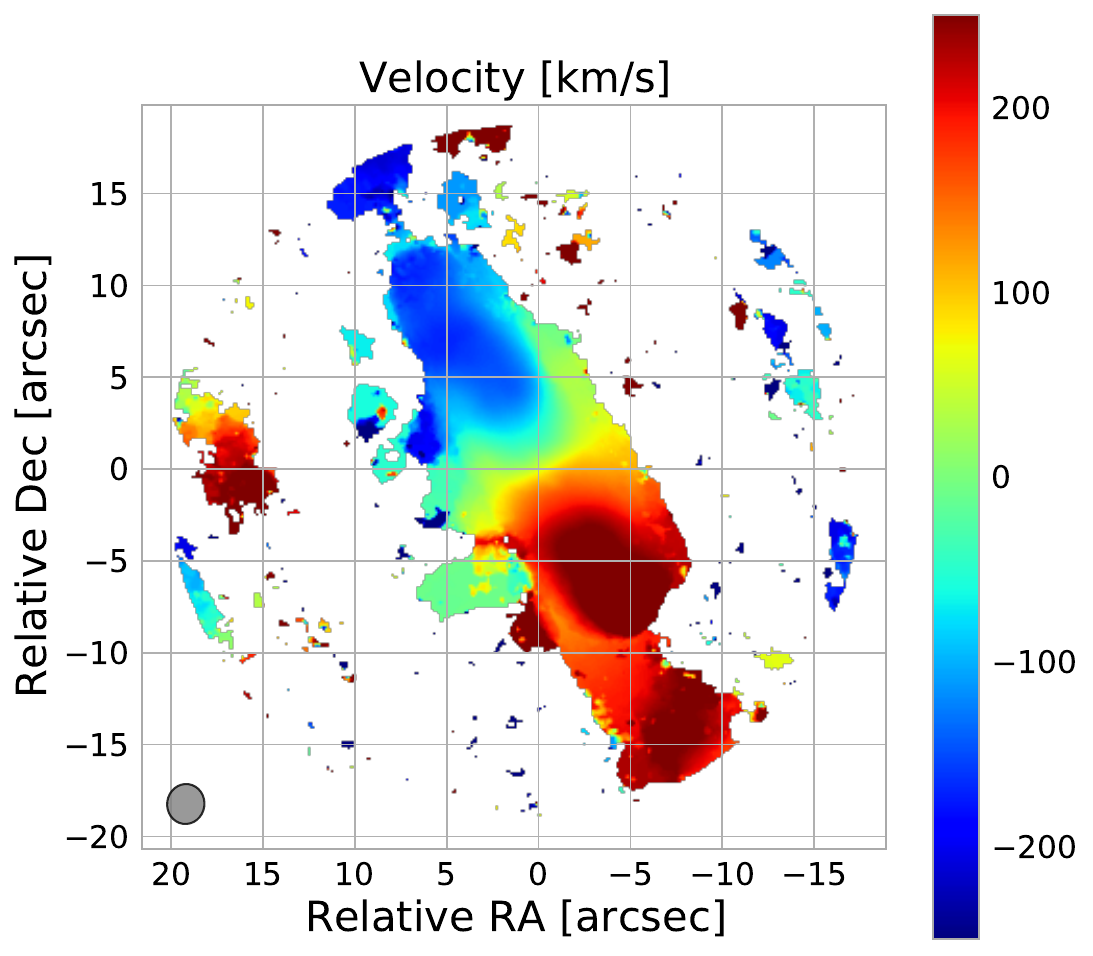} \includegraphics{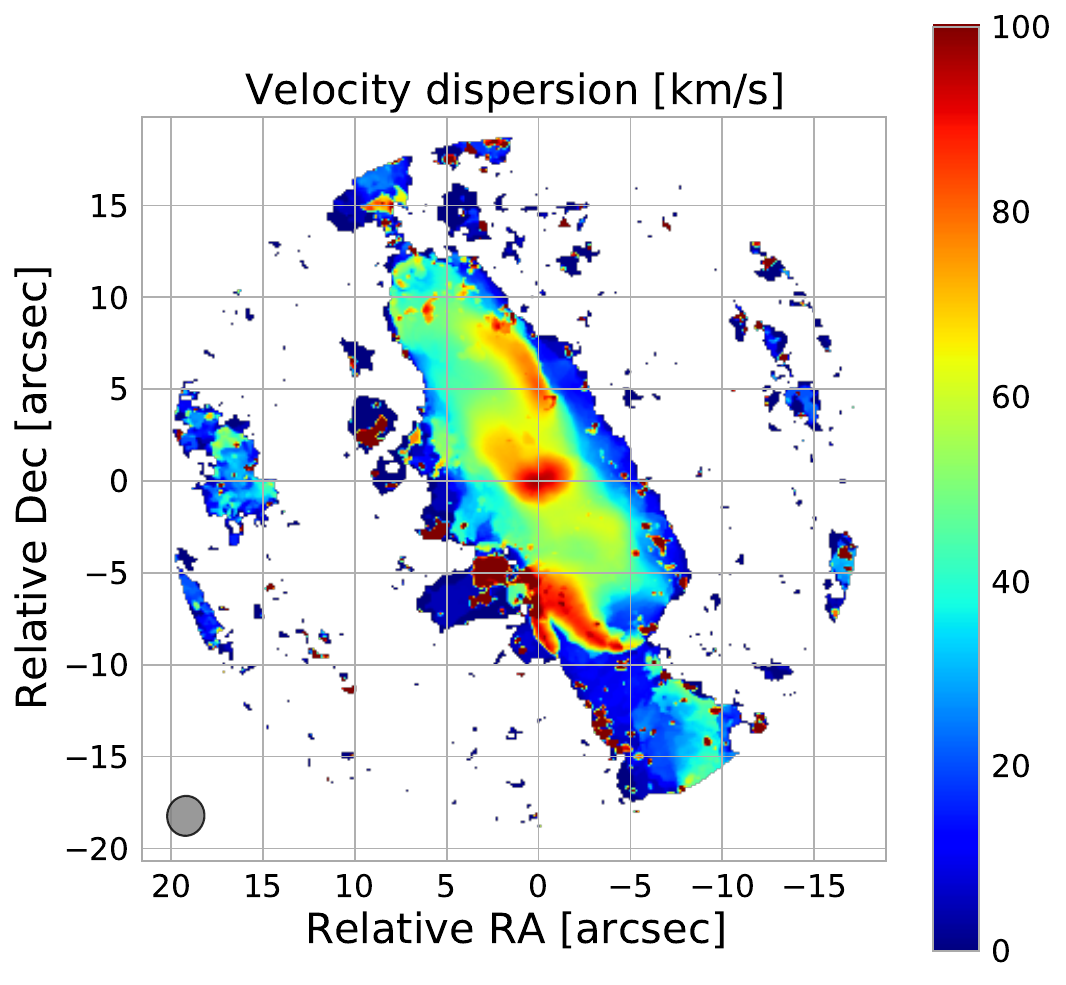}}
 \resizebox{\hsize}{!}{\includegraphics{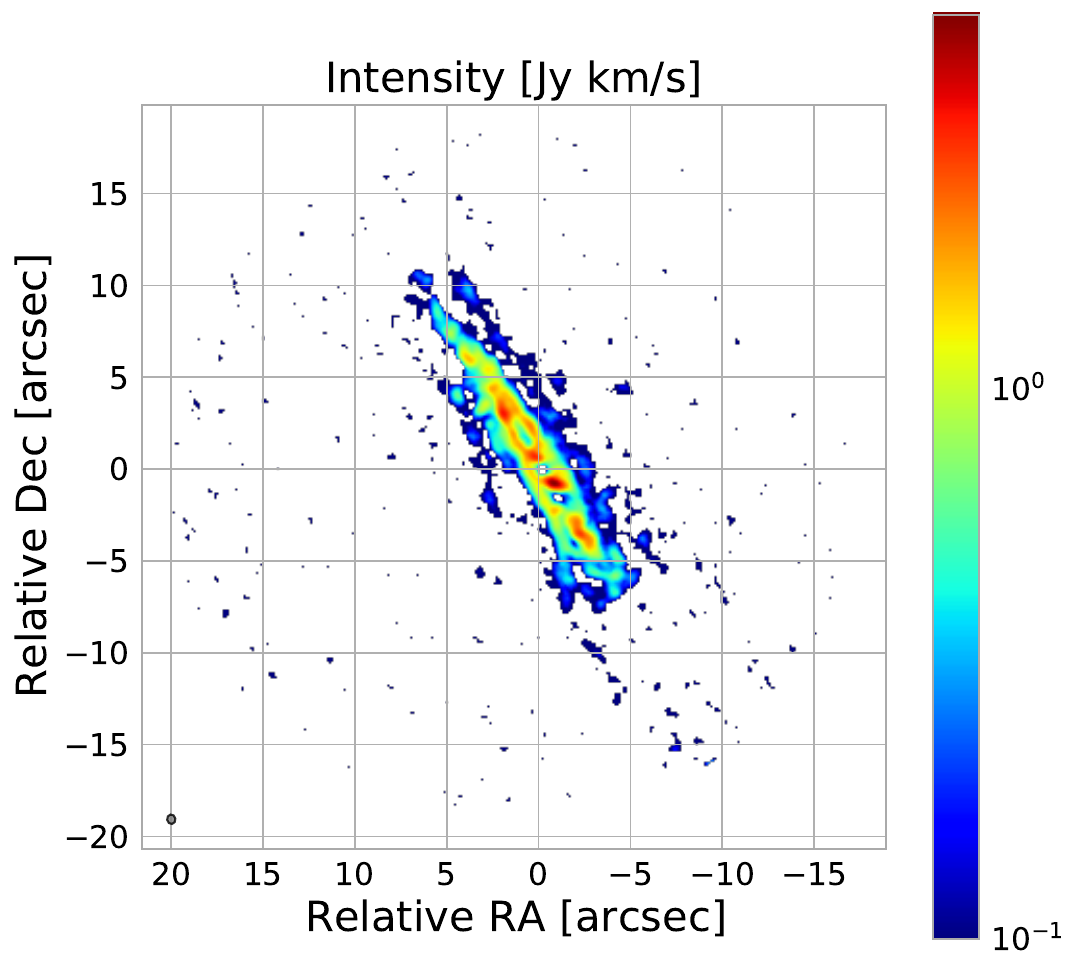} \includegraphics{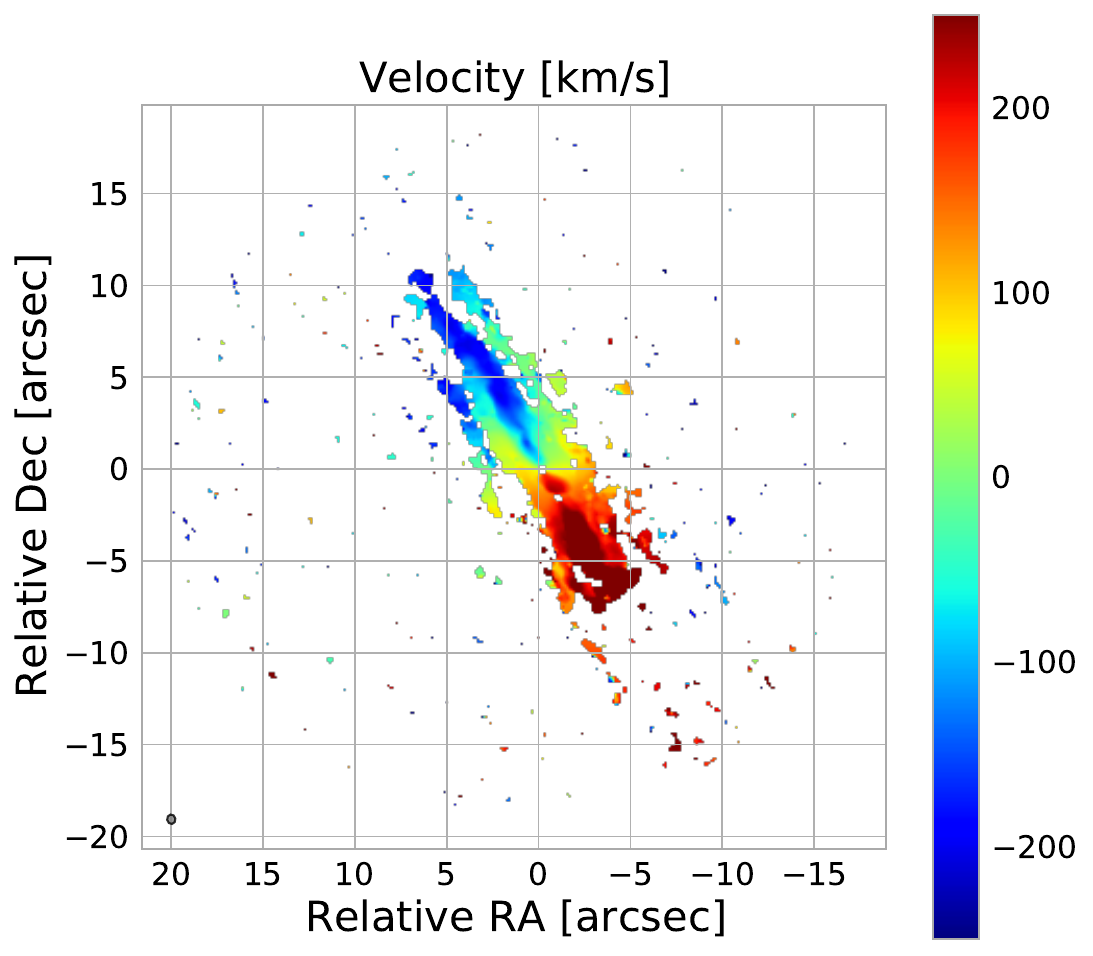} \includegraphics{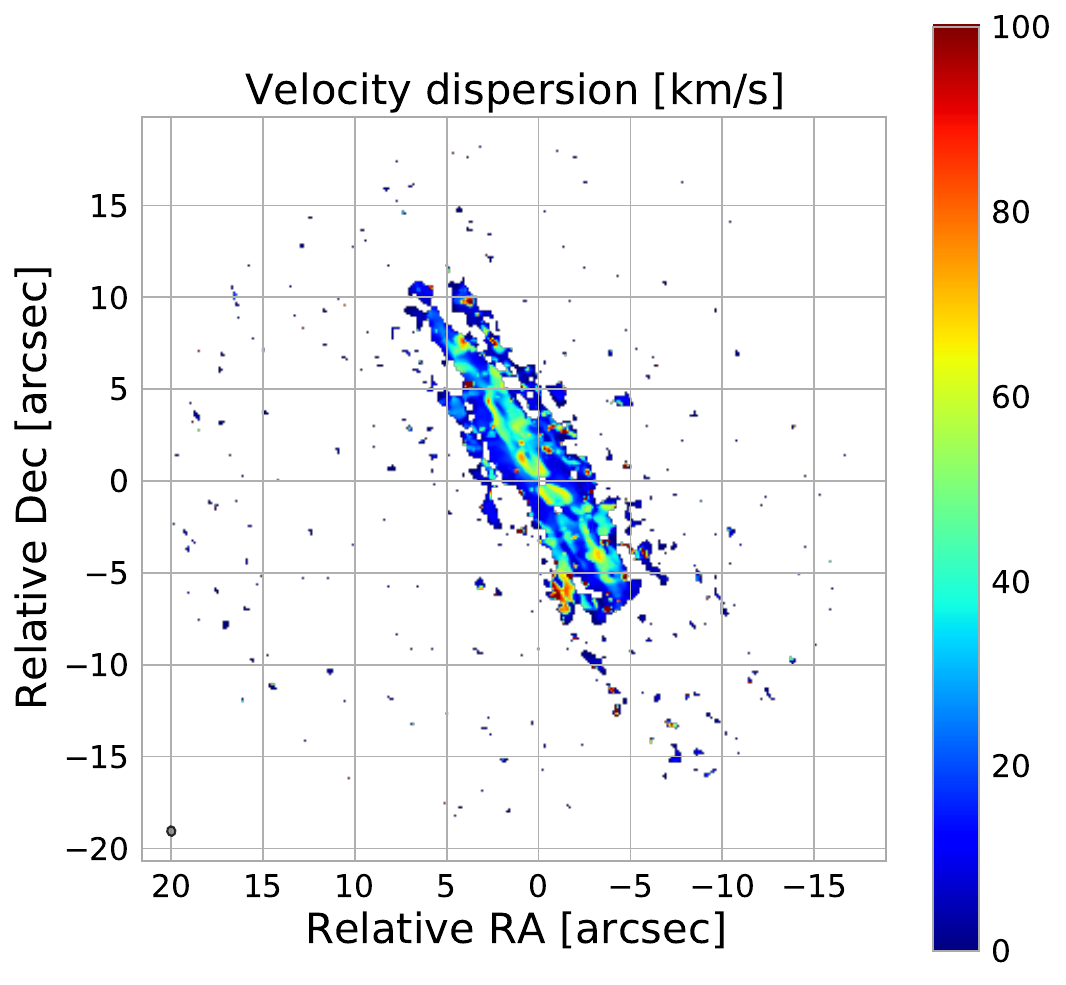}}
 \caption{ALMA first three moment maps obtained from the 2017.1.01439.S data-set selecting the baselines in order to separate the CO(2-1) emission over extended (top panels) and compact angular scales (bottom panels). From left to right: integrated flux (moment-0), mean velocity (moment-1) and velocity dispersion (moment-2). The grey ellipses in lower-left corner show the synthesized beam size (i.e. the angular resolution): 2.18 $\times$ 2.03 $\rm arcsec^2$ at PA = -82 deg and 0.52 $\times$ 0.45 $\rm arcsec^2$ at PA = 81 deg (top and bottom panels respectively).}
  \label{app:momentmaps_extended_compact}
\end{figure*}

Using the \texttt{immoments} task of CASA \citep{mcmullin} and applying a 3$\sigma$ threshold, we derived the first three moment maps of the extended and compact CO(2-1) emission of the ALMA data-set with project ID 2017.1.01439.S.
To recover the emission on compact scales, we selected the baselines with \texttt{uvrange} above 100m, to recover the emission on extended scales we imaged the visibilities with \texttt{uvrange} below 100m. To remove the compact scales contribution, we used the \texttt{immath} CASA task to subtract to the resulting image the image of the emission on compact scales. The resulting maps are reported in Figure \ref{app:momentmaps_extended_compact}. The high resolution data and the data-cube with \texttt{uvrange}$>$100 m trace the CO emission on comparable spatial scales and we find comparable cold molecular gas masses. On the other hand the data-cube with \texttt{uvrange}$<$100m is sensible to the bulk of the CO emission, mainly diffuse on extended spatial scales.

\begin{figure}[ht] 
\resizebox{\hsize}{!}{\includegraphics{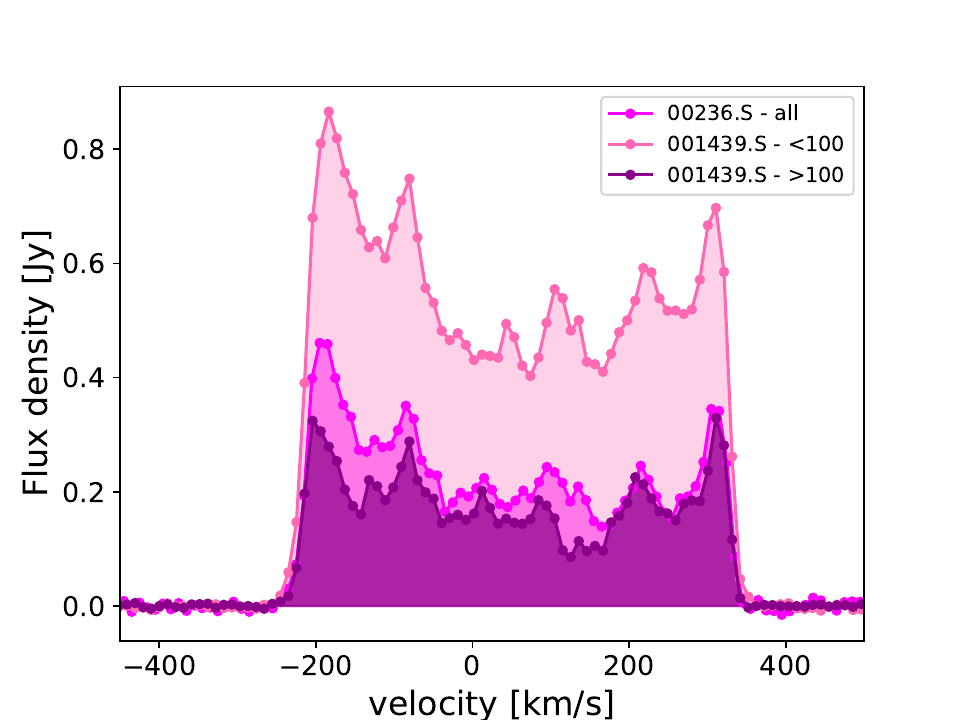}}
 \caption{CO(2-1) emission line spectrum extracted
from the ALMA data-cubes selecting the region above a 3$\sigma$ threshold.}
  \label{app:spectraCO}
\end{figure}

\section{Ionisation parameter}\label{appendixc}

\begin{figure}[ht] 
\resizebox{\hsize}{!}{\includegraphics{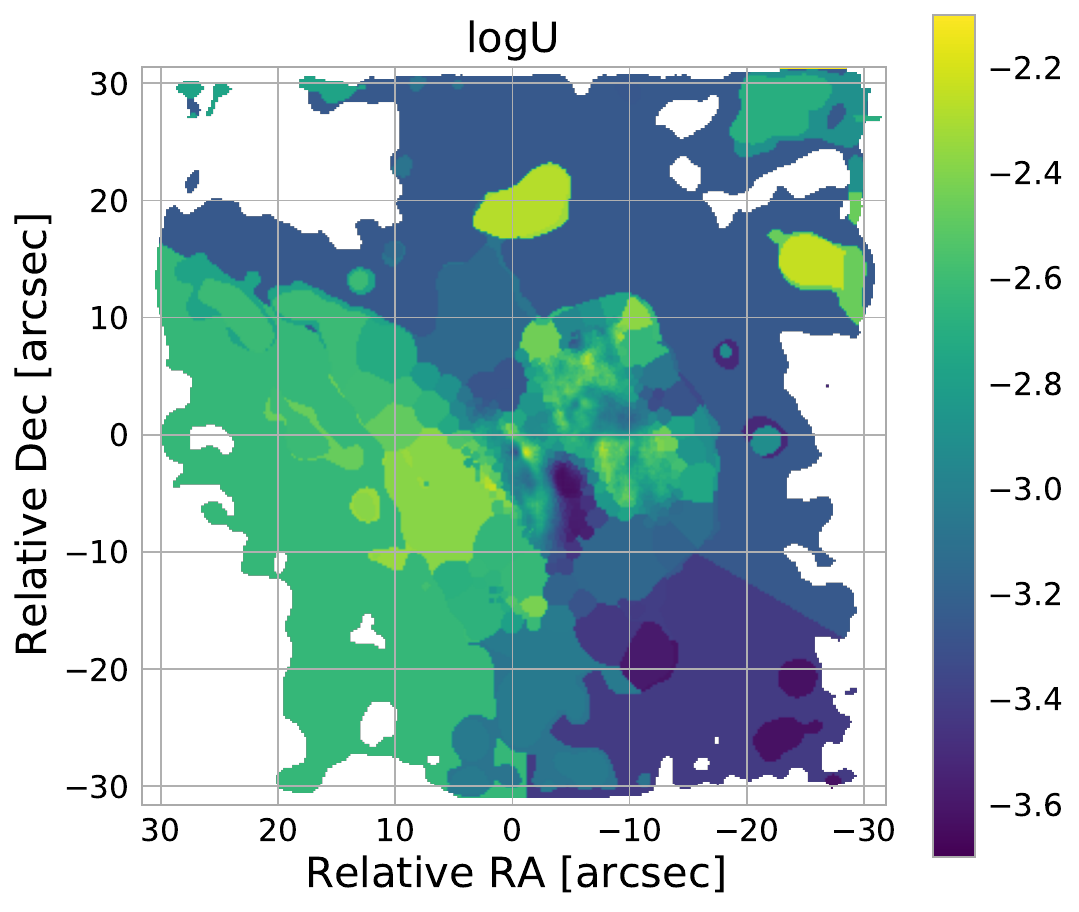}}
 \caption{Map of the ionisation parameter obtained by applying the Voronoi tassellation.}
  \label{fig:logU}
\end{figure}

We used the [O III]/H${\beta}$ and [NII]/\ha peak line ratios to compute the ionisation parameter by adopting the Voronoi tassellation and the analytical expression by \citet{baronnetezer2019} reported below:
\begin{multline*}
    \log U = a_1+ a_2log\biggl(\frac{\rm [O III]}{\rm H{\beta}}\biggr)+ a_3\log\biggl(\frac{\rm [O III]}{\rm H{\beta}}\biggr)^2+ \\ a_4\log\biggl(\frac{\rm [NII]}{\rm H{\alpha}}\biggr)+ a_5\log\biggl(\frac{\rm [NII]}{\rm H{\alpha}}\biggr)^2 ,
\end{multline*}
where $a_1 = -3.766$, $a_2 = 0.191$, $a_3 = 0.778$, $a_4 = -0.251$ and $a_5 = 0-342$.
This expression is only valid for ionization parameters in the range $\log U$ = -2 to $\log U = -3.8$ with a typical uncertainty of 0.11 dex \citep{baronnetezer2019}. The resulting map is reported in Figure \ref{fig:logU}. Higher values of the ionization parameter are found at the nucleus and in the cones, where U decreases with increasing distance from the AGN. U is on average low within the host galaxy disk.

\section{The [OI]$\lambda 6003 \AA$/\ha ratio}\label{appendix_OI}

\begin{figure}[ht] 
\resizebox{\hsize}{!}{\includegraphics{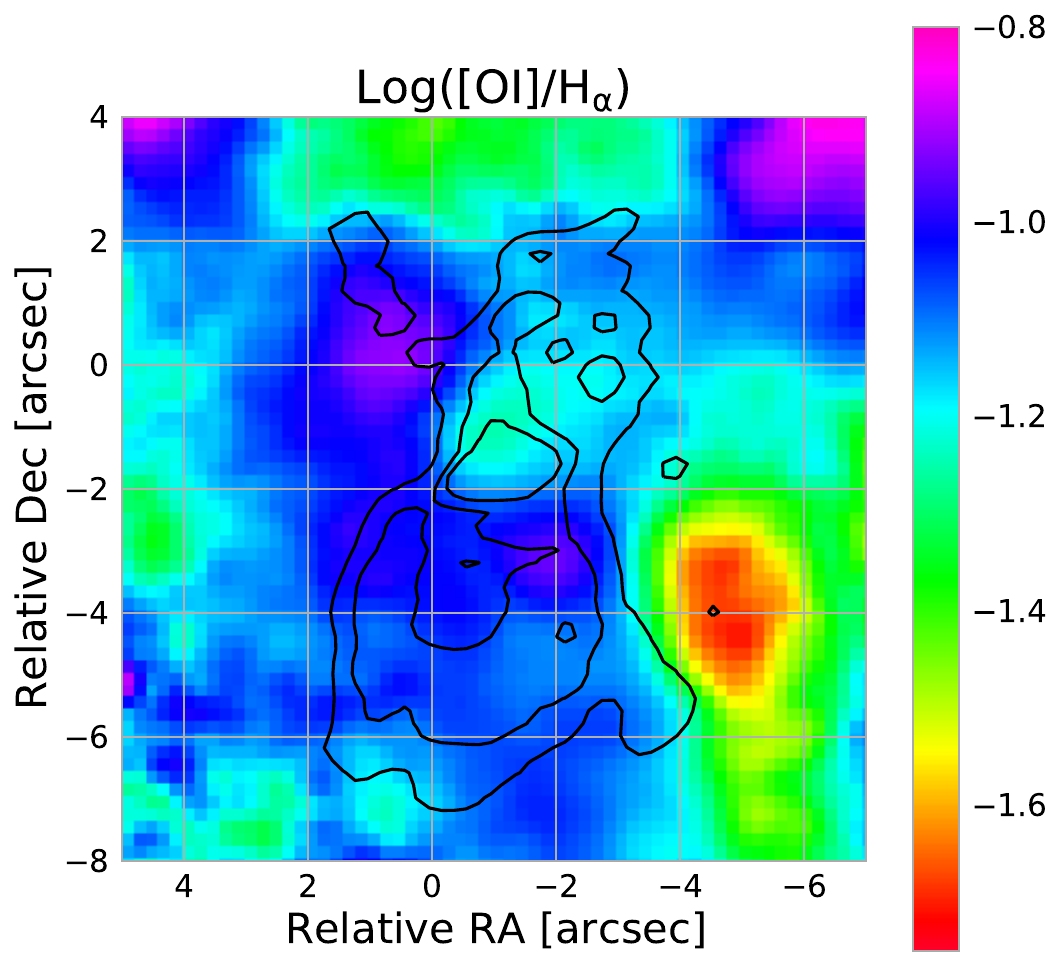}}
 \caption{Map of the [OI]$\lambda 6003 \AA$/\ha ratio in log scale. Black contours are from EVLA radio continuum map at 6cm (Figure \ref{continuo} left panel), drawn at (1, 5, 10)$\sigma$.}
  \label{fig:OIHa}
\end{figure}

By applying the the same procedure described in Section \ref{sec:ionisedgas} we produced the continuum subtracted [OI]$\lambda 6003 \AA$ emission line data-cube. We used the \texttt{Cube2Im} task to derive the surface brightness map of the [OI] emission line by applying a signal-to-noise threshold equal to 3. We used this maps the surface brightness \ha maps to derive the [OI]$\lambda 6003 \AA$/\ha ratio map reported in Figure \ref{fig:OIHa}. 

\end{appendix}

\end{document}